\documentclass[twocolumn]{aastex62}

\graphicspath{{./}{figures/}}
\usepackage{longtable}
\usepackage{amsmath}

\def\cgmsq {CGM$^2$~}
\def\lya{Ly$\alpha$~}
\def\lyb{Ly$\beta$~}
\def\NHI{$N_{\rm HI}$~}

\usepackage{hyperref}
\hypersetup{
    pdfnewwindow=true,
    colorlinks=true,
    citecolor=cyan,
    linkcolor=blue,
    urlcolor=blue
}

\begin{document}
\title{\cgmsq I: The Extent of the Circumgalactic Medium Traced by Neutral Hydrogen}

\correspondingauthor{Matthew C. Wilde}
\email{mwilde@uw.edu}

\author[0000-0003-1980-364X]{Matthew C. Wilde}
\affil{University of Washington, Department of Astronomy, Seattle, WA 98195, USA}

\author[0000-0002-0355-0134]{Jessica K. Werk}
\affil{University of Washington, Department of Astronomy, Seattle, WA 98195, USA}

\author[0000-0002-1979-2197]{Joseph N. Burchett}
\affil{University of California, Santa Cruz; 1156 High St., Santa Cruz, CA 95064, USA}
\affil{Department of Astronomy
, New Mexico State University, PO Box 30001, MSC 4500, Las Cruces, NM 88001}

\author[0000-0002-7738-6875]{J. Xavier Prochaska}
\affil{University of California, Santa Cruz; 1156 High St., Santa Cruz, CA 95064, USA}
\affil{Kavli Institute for the Physics and Mathematics of the Universe (Kavli IPMU) The University of Tokyo; 5-1-5 Kashiwanoha, Kashiwa, 277-8583, Japan}

\author[0000-0003-0789-9939]{Kirill Tchernyshyov}
\affil{University of Washington, Department of Astronomy, Seattle, WA 98195, USA}

\author[0000-0002-1218-640X]{Todd M. Tripp}
\affil{Department of Astronomy, University of Massachusetts, 710 North Pleasant Street, Amherst, MA 01003-9305, USA}

\author[0000-0002-1883-4252]{Nicolas Tejos}
\affil{Instituto de F\'{i}sica, Pontificia Universidad Cat\'{o}lica de Valpara\'{i}so, Casilla 4059, Valpara\'{i}so, Chile}

\author[0000-0001-9158-0829]{Nicolas Lehner}
\affil{Department of Physics, University of Notre Dame, Notre Dame, IN 46556}

\author[0000-0002-3120-7173]{Rongmon Bordoloi}
\affil{North Carolina State University, Department of Physics, Raleigh, NC 27695-8202}

\author[0000-0002-7893-1054]{John M. O'Meara}
\affil{W. M. Keck Observatory, 65-1120 Mamalahoa Hwy., Kamuela, HI 96743, USA}

\author[0000-0002-7982-412X]{Jason Tumlinson}
\affil{Space Telescope Science Institute, Baltimore, MD, USA}

\begin{abstract}
We present initial results from the  \textit{COS and Gemini Mapping the Circumgalactic Medium} (\mbox{CGMCGM} $\equiv$ \cgmsq) survey.  The \cgmsq survey consists of 1689 galaxies, all with high-quality Gemini GMOS spectra, within 1 Mpc of twenty-two $z \lesssim 1$ quasars, all with S/N$\sim$10 {\emph{HST/COS}} G130M$+$G160M spectra.  For 572 of these galaxies having stellar masses $10^{7} M_{\odot} < M_{\star} < 10^{11} M_{\odot}$ and $z \lesssim 0.5$, we show that the \ion{H}{1} covering fraction above a threshold of \NHI$>10^{14} $cm$^{-2}$ is $\gtrsim 0.5$ within 1.5 virial radii ($R_{\rm vir} \sim R_{200m}$). We examine the \ion{H}{1} kinematics and find that the majority of absorption lies within $\pm$ 250 km s$^{-1}$ of the galaxy systemic velocity. We examine \ion{H}{1} covering fractions over a range of impact parameters to infer a characteristic size of the CGM, $R^{14}_{\rm CGM}$, as a function of galaxy mass. $R^{14}_{\rm CGM}$ is the impact parameter at which the probability of observing an absorber with \NHI $>$ 10$^{14}$ cm$^{-2}$ is $>$ 50\%. In this framework, the radial extent of the CGM of $M_{\star} > 10^{9.9} M_{\odot}$ galaxies is $R^{14}_{\rm CGM} = 346^{+57}_{-53}$ kpc or $R^{14}_{\rm CGM} \simeq 1.2R_{\rm vir}$. Intermediate-mass galaxies with $10^{9.2} < M_{\star}/M_{\odot} < 10^{9.9}$ have an extent of $R^{14}_{\rm CGM} = 353^{+64}_{-50}$ kpc or $R^{14}_{\rm CGM} \simeq 2.4R_{\rm vir}$. Low-mass galaxies, $M_{\star} < 10^{9.2} M_{\odot}$, show a smaller physical scale $R^{14}_{\rm CGM} = 177_{-65}^{+70}$ kpc and extend to $R^{14}_{\rm CGM} \simeq 1.6R_{\rm vir}$. Our analysis suggests that using $R_{\rm vir}$ as a proxy for the characteristic radius of the CGM likely underestimates its extent.
\end{abstract}

\section{Introduction} \label{introduction}
    The Circumgalactic Medium (CGM) represents the complex interface between stellar evolution, feedback from super-massive massive black holes, and the cosmic web, dictated by the large scale cosmology of the universe (e.g., $\Lambda$CDM) \cite{tumlinson17}. These phenomena span many orders of magnitude in relevant spatial scales and involve several different sub-disciplines of astrophysics. In the canonical cosmological picture of galaxy evolution of \cite{white78}, gas accretes onto galactic halos from the intergalactic medium (IGM), cools and condenses to form the interstellar medium, and then eventually stars. At the same time, mass and energy is returned to the CGM via supernovae and AGN feedback, with some fraction potentially deposited back into the IGM. 
    
    In spite of its role as the harbor for a significant fraction of galactic baryons \citep[e.g.][]{dave10, shull12, werk14}, the CGM is predicted to have low gas densities (n $\ll$ 0.01 cm$^{-3}$) due to its vast volume. Such low-density gas is most efficiently observed using absorption lines found in the spectra of bright background objects, usually quasars (QSOs).  Generally, QSO absorption spectroscopy provides very high sensitivity to extremely low column densities in various ions, down to  $N \sim$ 10$^{12}$ cm$^{-2}$, and access to a broad range of ionized metal transitions that trace gas at both high and low densities in the CGM. 
            
    \cite{bahcall69} first posited that galaxies are responsible for the high-column density absorption lines in QSO spectra. In the 40 years since the prediction of \cite{bahcall69}, there has been much effort to link QSO absorption lines to both the IGM and CGM. In particular, the IGM traces the large scale structure of the universe, and is observed mainly through the \lya forest (see \citealp{rauch98} and \citealp{mcquinn16} for reviews). Previous works have found that the \ion{H}{1} \lya forest is associated with two distinct populations based on the strength of absorption systems; a low-density population that traces the cosmic-web, and a high-column density population that lies within dark matter halos \citep[e.g.,][]{morris93, tripp98, chen05, tejos14,  burchett20}. Linking sets of absorption lines arising at the same redshift (absorber systems) with redshifts of foreground galaxies in close proximity to the QSO sightline has essentially defined the last 20 years of CGM work \citep[e.g.][]{chen08, prochaska11b, tumlinson13, rudie19}. \cite{boksenberg78} and \cite{bergeron86} reported some of the first identifications of intervening galaxies responsible for intermediate redshift absorption line systems in quasar spectra. Subsequently, \cite{lanzetta95} found that all the luminous galaxies in their sample exhibit extended gaseous envelopes (not yet called CGM), out to at least $\sim$160 kpc.
    
    Most of the early studies either searched for galaxies associated with known absorbers or they conducted blind surveys that started with no previously known information about either the absorbers or the galaxies at the outset of the survey. However, \cite{bowen91, bowen95, bowen02} used an inverted approach in which they searched for specific absorption features (e.g., \ion{Ca}{2} \ion{Mg}{2} , or \ion{H}{1}) affiliated with galaxies with redshifts and properties that were known prior to the absorption survey.  The advantage offered by this approach is that the CGM of galaxies with specific properties of interest could be accumulated more efficiently than in blind or absorption-selected surveys.  Unfortunately, while many QSOs can be found behind low-z galaxies at interesting impact parameters, the first generation of HST spectrographs could not go deep enough to access most of those QSOs with a practical allocation of telescope time.
    
    In 2009, the installation of the {\emph{Cosmic Origins Spectrograph}} (COS) on HST, 20$-$30 times more sensitive than its UV spectrograph predecessors,  allowed for targeted CGM studies of statistical samples of galaxies nearby in projection to background QSOs. The studies carried out with HST/COS over the last decade have revealed complex connections between the bulk properties of galaxies and their CGM. For example, the presence of \ion{O}{6}  in the CGM, 10$-$150 kpc from the star-forming host galaxy's disk, is somehow linked to the present-day star-forming properties of the disk \citep{tumlinson11}. The \ion{H}{1} content of the CGM scales with that of the host galaxies' \ion{H}{1} disks, perhaps implying that the disk is being fed by accretion from the IGM and CGM \citep{borthakur15}.  In addition, large quiescent galaxies that have long since ceased star formation were observed to contain rich reservoirs of cool ($\sim$10$^{4}$ K) gas \citep[e.g.][]{thom12, chen18} which should provide fuel for new star formation.
    
    Thus far, the empirical picture of the CGM remains somewhat piecemeal, assembled from a variety of targeted observations of small-moderate sized samples carried out by different teams. Generally, various surveys of the CGM at low-z have revealed a multi-phase medium extending as far as 0.5 $-$ 1 $R_{\rm vir}$ ($R_{200m}$), which exhibits absorption from \ion{H}{1} and a range of metal ions with ionization potential energies ranging from 10 - 239 eV \citep[e.g.][]{lanzetta95, stocke13, tripp11, muzahid11, tumlinson13, werk13, burchett19}. Absorption signatures of low-ionization state metals like \ion{Mg}{2} and \ion{Si}{2} are rarely detected greater than $\sim 100$ kpc away from their host galaxies \citep[][]{bordoloi11, nielsen13, werk13, bordoloi14}. By comparison, highly ionized metal lines such as \ion{O}{6}  are detected out to $\sim$ 200 kpc \citep{tumlinson11, johnson15, johnson17, prochaska19} and in M31, out to $\sim$ 500 kpc \citep{lehner20}. This progress at low-$z$ has been mirrored at $z \sim$2$-$3 with the Keck Baryonic Structure Survey (KBSS; \cite{rudie12}), which found typical scales of the CGM to be $\sim 300$ kpc for $L^*$ galaxies. 

    Observational progress enabled by HST/COS has presented several challenges to our current models of galaxy evolution. Photoionization modeling of the 10$^{4}$~K cool CGM gas around L* galaxies has revealed it to be an order of magnitude under-pressurized with respect to the envisioned 10$^{6}$~K virialized, ambient hot halo \citep[e.g.][]{werk14, prochaska17, haislmaier21}, and thus likely not in thermal or pressure equilibrium as expected from simulations or simple theoretical arguments. Complex, non-equilibrium physical processes may include precipitation \citep[e.g.][]{voit19}, shocks \citep[e.g.][]{mcquinn18}, cosmic rays \citep[e.g.][]{ji19}, kpc-scale fountain cycles \citep[e.g.][]{fraternali08, kim18}, multi-filament gas flows fueling star formation \citep{martin19}, galactic winds \citep{bordoloi11, burchett20b, huang20} and turbulent mixing layers \citep{fielding20b}. Furthermore, simulations have generally under-predicted observed column densities of various ions, of both high-and-low ionization potential, compared to observations of galactic halos \citep[e.g.][]{stinson12, hummels13, oppenheimer16, liang16, fielding17}. Newer simulations \citep[e.g.][]{oppenheimer18, lehner20, nelson20} have more faithfully reproduced observations. 
    
    Recent simulations have shown that the CGM content, thermal structure,  and kinematic properties are highly dependent on the mass and spatial resolution of simulations \citep[e.g.][]{vandeVoort19}. High-resolution simulations that focus explicitly on the CGM, with resolved spatial scales of $\sim$100~pc and mass scales of $\sim$10$^{3}$ M$_{\odot}$ out to 250 kpc, have helped to ease the discrepancies with the observed column densities \citep{peeples19, hummels19}, more naturally producing cool, denser clumps. However, it remains important to characterize the full extent of the CGM, as simulations still struggle to recreate all key aspects of the extended CGM.
    
    For the above reasons, it has been challenging to put quantitative, physically-motivated bounds on the radial extent of the CGM. Generally, statistical QSO-galaxy samples are limited to single sightlines per galaxy due to the rarity of QSOs \citep[but see][]{chen14, bowen16} and thus assembling large enough samples to place constraints on the extent of the CGM is challenging work. A comprehensive $z < 1$ CGM survey requires both space and ground-based spectroscopy. The former are required to measure key far-ultraviolet transitions such as Ly$\alpha$, \lyb and metal species (e.g., \ion{C}{2}, \ion{Si}{2},  \ion{C}{4}, \ion{O}{6}). At low redshift, only \ion{Ca}{2}, \ion{Na}{1}, \ion{Fe}{2}, and \ion{Mg}{1} and \ion{Mg}{2} can be observed from the ground. However, these species are confused by ionization and dust depletion and are hard to interpret by themselves. On the other hand, large samples of galaxies \textit{are} accessible to ground based spectrographs to measure precise ($z_{\rm err} < 10^{-4}$) redshifts of potentially associated galaxies. 
        
    To fill this need for a comprehensive, uniform CGM study, we have completed a deep spectroscopic survey of $>$1000 galaxies around $z > 0.5$ QSOs, all within a few arcminutes of the background QSO sightline. The galaxies and absorption systems are blindly selected such that no preference to galaxy type or absorber optical depth is explicitly imposed.  Our survey, \textit{COS and Gemini Mapping the Circumgalactic Medium} (\mbox{CGMCGM} $\equiv$ \cgmsq), was designed to address many puzzles related to the CGM, such as: (i) What is the physical state of the gaseous halo when galaxies quench their star formation? (ii) What effects do environment and merger history have on the physical state and content gaseous halo? (iii) How do galaxy properties relate to the metal content and kinematic structure of their gaseous halos and vice versa? (iv) What is the physical extent of the CGM?  In this first presentation of the \cgmsq survey,  we address the last question and in particular, how its spatial extent may depend upon halo mass and galaxy type. Ultimately, this study will inform how galaxy evolution and the large scale structure of the universe are connected.
    
    \cgmsq includes a significant investment of effort from a team of University of Washington Undergraduate Astronomy majors led by \cgmsq PI Jessica Werk, the Student Quasar Absorption Diagnosticians (aka the \textit{Werk SQuAD}) whose work is described in detail in the relevant sections. The outline of this paper is as follows. Section \ref{data} describes the observations and data reduction for the \cgmsq spectroscopic galaxy survey. Section \ref{analysis} describes an analysis of the galaxies' spectroscopic and photometric properties.  In Section \ref{survey}, we discuss our initial empirical results from associating galaxies with absorbers. We then derive a characteristic physical scale for the CGM in Section \ref{section:clustering} by examining the \ion{H}{1}-galaxy clustering. Section \ref{discussion} compares our results to previous surveys of similar design and hydrodynamical cosmological simulations. We summarize and conclude this work in Section \ref{summary}. Throughout this analysis we adopt the Planck15 \citep{planck16} cosmology as encoded in the \texttt{ASTROPY} package \citep{astropy13, astropy18}. All distances are in physical space (not co-moving) unless otherwise noted.
    
    \section{Observations and Data Reduction}
    \label{data}
    \subsection{QSO Sample Selection}
    
    The \cgmsq survey is built upon the COS-Halos (GO11598, GO13033; \citealp{tumlinson13}) and COS-Dwarfs (GO12248;
    \citealp{bordoloi14}) surveys, which used 263 orbits of COS time to observe $\sim$ 80 QSOs with $z = 0.2-1.0$. These surveys were designed to probe the halo(s) of one or two foreground galaxies well inside $R_{\rm vir}$ using suitable QSO-galaxy pairs. The COS-Halos QSO catalog was selected from the SDSS DR5 quasar catalog for QSOs that are UV-bright (GALEX FUV $\lesssim 18$) and lie at $z \lesssim 1$. The target QSOs were further selected to avoid Mg II absorbers at $z > 0.4$ to avoid losing a large range of the FUV QSO spectrum to LLSs. While this criterion did not affect the original surveys, which typically targeted specific galaxies at $z < 0.4$, it does affect the \cgmsq sample because it selects against high column density systems at $z \gtrsim 0.4$. 
    
    COS-Halos galaxies were selected via photometric redshifts to target stellar masses of $M_{\star} \simeq 10^{10-11} M_{\odot}$ at $0.1 < z < 0.3$ with  projected separations from a nearby QSO $\rho < 150$ kpc. The precise redshifts (and other galaxy properties) were subsequently constrained with follow-up spectroscopy \citep{werk12}. The COS-Dwarfs survey was designed to probe the CGM of galaxies with $\log M_{\star}/M_{\odot} \lesssim 10^{10}$ at z $\simeq$ 0.01 $-$ 0.05 with masses and redshifts derived via SDSS photometry and spectroscopy. Both surveys yield galaxies that are the closest spectroscopically-identified galaxy to each QSO sightline and were designed to avoid biases with respect to galaxy neighbors, large-scale environment, or status as a satellite of a larger halo (in cases where neighbors were known). 
    
    For the \cgmsq multislit spectroscopic follow-up, we targeted a subset of ``high-value" QSO fields with our Gemini program, either those with $z_{\rm QSO} >$ 0.6 or those with HST imaging available, which allows us to obtain morphology of any absorption-hosting galaxies with $z < 0.5$. To-date, we have obtained multislit galaxy spectra in 22 QSO fields as part of \cgmsq. The fields we have surveyed, and some basic properties of the background QSOs, are tabulated in Table~\ref{tab:qsodata}.  
    
    \begin{deluxetable*}{cccccc}
    \tablecaption{ QSO Fields in \cgmsq \label{tab:qsodata}}
    \tablehead{\colhead{QSO} & \colhead{QSOLong} & \colhead{RA} & \colhead{Dec} & \colhead{$z_{\rm QSO}$} & \colhead{m$_g$}}
    \colnumbers
    \startdata
    J0226+0015 & J022614.46+001529.7 & 36.56028 & 0.25827 & 0.615 & 17.15 \\
    J0809+4619 & J080908.13+461925.6 & 122.2839 & 46.3238 & 0.657 & 16.54 \\
    J0843+4117 & J084349.49+411741.6 & 130.9562 & 41.2949 & 0.99 & 17.31 \\
    J0914+2823 & J091440.38+282330.6 & 138.66829 & 28.39184 & 0.735 & 17.79 \\
    J0935+0204 & J093518.19+020415.5 & 143.82581 & 2.07098 & 0.649 & 16.99 \\
    J0943+0531 & J094331.61+053131.4 & 145.88173 & 5.52541 & 0.564 & 17.16 \\
    J1001+5944 & J100102.55+594414.3 & 150.2606 & 59.7373 & 0.746 & 16.08 \\
    J1016+4706 & J101622.60+470643.3 & 154.09418 & 47.11204 & 0.822 & 17.12 \\
    J1022+0132 & J102218.99+013218.8 & 155.57913 & 1.53856 & 0.789 & 16.75 \\
    J1059+1441 & J105945.23+144142.9 & 164.9385 & 14.6953 & 0.631 & 16.93 \\
    J1059+2517 & J105958.82+251708.8 & 164.9951 & 25.2858 & 0.662 & 17.39 \\
    J1112+3539 & J111239.11+353928.2 & 168.16296 & 35.65784 & 0.636 & 17.73 \\
    J1133+0327 & J113327.78+032719.1 & 173.36578 & 3.45533 & 0.525 & 17.54 \\
    J1134+2555 & J113457.62+255527.9 & 173.7401 & 25.9244 & 0.71 & 16.8 \\
    J1233-0031 & J123304.05-003134.1 & 188.26688 & -0.52616 & 0.471 & 17.76 \\
    J1241+5721 & J124154.02+572107.3 & 190.4751 & 57.35205 & 0.583 & 17.58 \\
    J1342-0053 & J134251.60-005345.3 & 205.71503 & -0.89592 & 0.326 & 16.92 \\
    J1419+4207 & J141910.20+420746.9 & 214.79251 & 42.1297 & 0.873 & 17.04 \\
    J1437+5045 & J143726.14+504555.8 & 219.35892 & 50.76551 & 0.783 & 17.57 \\
    J1553+3548 & J155304.92+354828.6 & 238.27052 & 35.80795 & 0.722 & 16.46 \\
    J1555+3628 & J155504.39+362848.0 & 238.76833 & 36.48001 & 0.714 & 17.76 \\
    J2345-0059 & J234500.43-005936.0 & 356.2518 & -0.99335 & 0.789 & 16.8 
    \enddata
    \tablecomments{The full sample of QSOs included in the CGM$^{2}$ Survey: (1) QSO Short Name; (2) QSO Long Name, RA in hms and Dec in dms, all J2000; (3 \& 4) RA, Dec in decimal degrees; (5) QSO redshift; (6) SDSS $g$-band magnitude of QSO.}
    \end{deluxetable*}
    
    \begin{deluxetable*}{cccccc}
    \tablecaption{\cgmsq Multislit Mask Observations \label{tab:maskdata}}
    \tablehead{\colhead{QSO} & \colhead{Maskname} & \colhead{Date} & \colhead{Instrument} & \colhead{N(slits)} & \colhead{N(z)}}
    \colnumbers
    \startdata
    J0809+4619 & GN2014AQ001-01 & 2014-11-19 & GMOS-N & 31 & 24 \\
    J0809+4619 & GN2014AQ001-02 & 2014-11-20 & GMOS-N & 27 & 21 \\
    J0809+4619 & GN2014AQ001-03 & 2014-11-21 & GMOS-N & 29 & 21 \\
    J0809+4619 & GN2014AQ001-04 & 2014-11-23 & GMOS-N & 17 & 15 \\
    J1134+2555 & GN2014AQ001-05 & 2014-06-21 & GMOS-N & 29 & 26 \\
    J1134+2555 & GN2014AQ001-06 & 2014-06-22 & GMOS-N & 29 & 26 \\
    J1134+2555 & GN2014AQ001-07 & 2014-06-25 & GMOS-N & 24 & 20 \\
    J1134+2555 & GN2014AQ001-08 & 2014-12-21 & GMOS-N & 16 & 13 \\
    J1241+5721 & GN2014AQ001-09 & 2014-06-25 & GMOS-N & 27 & 18 \\
    J1241+5721 & GN2014AQ001-10 & 2014-06-30 & GMOS-N & 24 & 23 \\
    J1555+3628 & GN2014AQ001-11 & 2014-06-24 & GMOS-N & 33 & 29 \\
    J1555+3628 & GN2014AQ001-12 & 2014-06-24 & GMOS-N & 26 & 24 \\
    J0914+2823 & GN2014BLP003-01 & 2015-01-17 & GMOS-N & 45 & 29 \\
    J0914+2823 & GN2014BLP003-02 & 2015-01-17 & GMOS-N & 44 & 27 \\
    J0914+2823 & GN2014BLP003-03 & 2015-01-17 & GMOS-N & 45 & 26 \\
    J0843+4117 & GN2014BLP003-04 & 2015-01-16 & GMOS-N & 39 & 28 \\
    J0843+4117 & GN2014BLP003-05 & 2015-01-16 & GMOS-N & 39 & 31 \\
    J0843+4117 & GN2014BLP003-06 & 2015-01-16 & GMOS-N & 39 & 31 \\
    J1059+1441 & GN2014BLP003-08 & 2015-01-16 & GMOS-N & 45 & 38 \\
    J1001+5944 & GN2014BLP003-10 & 2015-01-18 & GMOS-N & 45 & 36 \\
    J1001+5944 & GN2014BLP003-11 & 2015-01-18 & GMOS-N & 45 & 29 \\
    J1001+5944 & GN2014BLP003-12 & 2015-01-18 & GMOS-N & 41 & 33 \\
    J1016+4706 & GN2014BLP003-13 & 2015-01-19 & GMOS-N & 42 & 34 \\
    J1016+4706 & GN2014BLP003-14 & 2015-01-19 & GMOS-N & 40 & 30 \\
    J1016+4706 & GN2014BLP003-15 & 2015-01-19 & GMOS-N & 42 & 33 \\
    J1112+3539 & GN2014BLP003-19 & 2015-01-17 & GMOS-N & 49 & 33 \\
    J1112+3539 & GN2014BLP003-20 & 2015-01-17 & GMOS-N & 46 & 31 \\
    J1112+3539 & GN2014BLP003-21 & 2015-01-17 & GMOS-N & 43 & 37 \\
    J1059+2517 & GN2015ALP003-01 & 2015-05-18 & GMOS-N & 47 & 31 \\
    J1059+2517 & GN2015ALP003-02 & 2015-05-20 & GMOS-N & 43 & 29 \\
    J1059+2517 & GN2015ALP003-03 & 2015-06-08 & GMOS-N & 42 & 25 \\
    J1419+4207 & GN2015ALP003-04 & 2015-05-17 & GMOS-N & 45 & 27 \\
    J1419+4207 & GN2015ALP003-05 & 2015-05-17 & GMOS-N & 46 & 33 \\
    J1419+4207 & GN2015ALP003-06 & 2015-05-17 & GMOS-N & 43 & 32 \\
    J1419+4207 & GN2015ALP003-07 & 2015-05-17 & GMOS-N & 41 & 25 \\
    J1437+5045 & GN2015ALP003-08 & 2015-05-24 & GMOS-N & 40 & 33 \\
    J1437+5045 & GN2015ALP003-09 & 2015-05-22 & GMOS-N & 43 & 31 \\
    J1437+5045 & GN2015ALP003-10 & 2015-06-19 & GMOS-N & 42 & 30 \\
    J1553+3548 & GN2015ALP003-11 & 2015-05-17 & GMOS-N & 45 & 32 \\
    J1553+3548 & GN2015ALP003-12 & 2015-05-18 & GMOS-N & 45 & 35 \\
    J1553+3548 & GN2015ALP003-13 & 2015-05-18 & GMOS-N & 41 & 28
    \enddata
    \end{deluxetable*}
    
    \begin{deluxetable*}{cccccc}
    \tablenum{2}
    \caption{Continued}
    \tablehead{\colhead{QSO} & \colhead{Maskname} & \colhead{Date} & \colhead{Instrument} & \colhead{N(slits)} & \colhead{N(z)}}
    \colnumbers
    \startdata
    J0943+0531 & GS2014AQ002-01 & 2015-04-19 & GMOS-S & 19 & 13 \\
    J0943+0531 & GS2014AQ002-02 & 2015-04-18 & GMOS-S & 7 & 3 \\
    J0943+0531 & GS2014AQ002-03 & 2015-02-18 & GMOS-S & 28 & 22 \\
    J0943+0531 & GS2014AQ002-04 & 2015-02-19 & GMOS-S & 18 & 9 \\
    J1133+0327 & GS2014AQ002-05 & 2015-03-15 & GMOS-S & 32 & 27 \\
    J1133+0327 & GS2014AQ002-06 & 2015-04-25 & GMOS-S & 24 & 18 \\
    J1133+0327 & GS2014AQ002-07 & 2015-04-25 & GMOS-S & 26 & 21 \\
    J1133+0327 & GS2014AQ002-08 & 2015-05-14 & GMOS-S & 22 & 17 \\
    J1233-0031 & GS2014AQ002-09 & 2015-04-19 & GMOS-S & 30 & 19 \\
    J1233-0031 & GS2014AQ002-10 & 2015-05-14 & GMOS-S & 29 & 24 \\
    J1342-0053 & GS2014AQ002-11 & 2015-01-04 & GMOS-S & 31 & 18 \\
    J1342-0053 & GS2014AQ002-12 & 2015-06-06 & GMOS-S & 29 & 22 \\
    J0226+0015 & GS2014BLP004-01 & 2014-10-26 & GMOS-S & 47 & 40 \\
    J0226+0015 & GS2014BLP004-02 & 2014-10-26 & GMOS-S & 44 & 32 \\
    J0226+0015 & GS2014BLP004-03 & 2014-10-26 & GMOS-S & 43 & 29 \\
    J2345-0059 & GS2014BLP004-04 & 2015-06-19 & GMOS-S & 41 & 25 \\
    J0935+0204 & GS2014BLP004-07 & 2015-02-13 & GMOS-S & 45 & 26 \\
    J0935+0204 & GS2014BLP004-08 & 2015-02-13 & GMOS-S & 32 & 20 \\
    J0935+0204 & GS2014BLP004-09 & 2015-02-17 & GMOS-S & 38 & 25 \\
    J1022+0132 & GS2015ALP004-01 & 2015-02-17 & GMOS-S & 40 & 26 \\
    J1022+0132 & GS2015ALP004-02 & 2015-02-17 & GMOS-S & 38 & 30 \\
    J1022+0132 & GS2015ALP004-03 & 2015-02-17 & GMOS-S & 36 & 27 \\
    J1022+0132 & GS2015ALP004-04 & 2015-02-17 & GMOS-S & 34 & 23 
    \enddata
    \tablecomments{Multislit Observation with GMOS taken as part of the CGM$^{2}$ Survey: (1) QSO short name; (2) Unique program mask name with project ID; (3) Date of mask observation; (4) Instrument, either GMOS-N or GMOS-S; (5) Number of slits placed on each mask; (6) Number of slits that yielded a reliable redshift.}
    \end{deluxetable*}
    
    \subsubsection{COS Data Reduction}
    The COS spectra were taken using both the G130M and G160M gratings. The balance between the time allocated to G130M and G160M was designed to achieve S/N $\simeq 10-12$ per resolution element (FWHM $\simeq$ 16-18 km s$^{-1}$) or better over $1150$-$1800$~\AA. The reduction of the COS spectra is explained in detail in \cite{tumlinson13} and \cite{bordoloi14} and follows the same method used by \cite{tripp11}, \cite{meiring11}, \cite{tumlinson11} and \cite{thom12} but we provide a brief description here. The COS data were obtained from MAST\footnote{http://archive.stsci.edu} and processed via CALCOS (v2.12) \citep{holland12} with standard parameters and reference files. First, exposures taken at the same grating and CENWAVE were combined. These coadded spectra were then coadded with exposures in the same grating at different CENWAVEs. This was followed by a combination of the two grating spectra to produce a single one-dimensional (1D) trace from $1150$-$1800$~\AA. At each coadd, exposures were combined by aligning common Milky Way interstellar absorption lines. Features related to the design of COS show up in the raw data and must be removed. The photocathode grid wires cast shadows on the detector as well as other fixed-pattern noise features and were removed with a flatfield and moreover are mitigated by the use of FP SPLITS. Flat-field reference files, prepared and communicated to us by D. Massa at STScI and filtered for high-frequency noise by E. Jenkins, were used to correct these fixed pattern features. These flats do not however, correct for gain-sag depressions in the spectra created by prolonged exposure to bright geocoronal emission lines \citep{sahnow11}. The affected regions are flagged by the CALCOS pipeline and are rejected in our coaddition process. The 1D spectra are binned to Nyquist sampling resulting in a 1D, flat-corrected spectra with two bins per COS resolution element (FWHM $\simeq 18$ km s$^{-1}$) and a S/N $\sim$ 8-10. Errors arising from counting statistics (Poisson) are propagated through each step in the calibration. 
    
    \subsection{Spectroscopic Galaxy Survey}
    \label{sec:survey}
        \subsubsection{Survey Design}
        All galaxy spectra were obtained at the Gemini North and South Observatories using the GMOS instrument \citep[]{hook04, gmoss16} in multi-object spectrograph mode. The GMOS observations enabled us to obtain accurate, precise redshifts for low-redshift ($0.1 \lesssim z \lesssim 1$) galaxies as faint as $g \sim 24.5$ for star-forming galaxies and $g \sim 24$ for early-type galaxies. Over 3 observing semesters from 2014$-$2015 at both GMOS-N and GMOS-S, we surveyed galaxies in the high-quality QSO fields listed in Table~\ref{tab:qsodata}.
        
        GMOS uses slit masks that must be cut for each field based on highly-accurate astrometry from either a `pre-image' taken in GMOS imaging mode or derived from an accurate catalog. For most of our fields, we used Gemini GMOS pre-imaging to create the masks.  However, for fields observed in 2014A, specifically J0809$+$4619, J0943$+$0531,  J1133$+$0327, J1134$+$2555, J1241$+$5721, and J1555$+$3628, we used a combination of HST-WFC3 imaging for angular separations less than 1$'$ from the QSO along with SDSS imaging for galaxies further from the sightline to construct a slit mask target catalog. We refer to the observational program that observed these fields as the ``Rollup" program and are differentiated in Table \ref{tab:maskdata} with a `Q' in the Maskname column. Subsequent observations used Gemini-GMOS $g$ and $i$ band pre-imaging to create the slit mask target lists and are referred to as ``Large Program" (LP) observations. From the pre-imaging, we derived astrometry as well as magnitudes from \texttt{SExtractor} \citep{sextractor96} using Gemini-calibrated photometric zero points and color corrections. We then optimized to get as many $g < 24$ galaxies into each slit mask as possible (see below). 
        
        In our GMOS programs, we used the R400 grating with $1''$ slits to balance wavelength coverage with spectral resolution. The fields from the Rollup programs listed above were dithered across two exposures per mask at central wavelengths of $\lambda\lambda 6000, 6900$\AA~ in order to avoid losing information to the chip gaps. The LP fields have spectra consisting of three exposures per mask at three separate central wavelengths, $\lambda\lambda 6900, 7000, 7100$\AA. Despite these differences, both programs yielded similar depths with total exposure times of 1 hour per mask. Our final spectra achieved a S/N of at least a few per pixel at $\lambda_{obs} = 4800$\AA, the approximate wavelength of the 4000\AA~break at $z \sim 0.2$. 
        
        Because precise redshift determination from our spectra was the primary goal, we chose a grating to cover 4400 to 9000 \AA~ in order to detect emission from $z \lesssim 0.5$ [OII], H$\beta$ (for both redshifts and star formation metrics) and H$\alpha$ as well as absorption from Ca H$+$K and Na D in passively-evolving galaxies. The R400 grating provided a spectral resolving power of $R = \lambda / \Delta\lambda \simeq 950$, which corresponds to a velocity resolution of  300 km s$^{-1}$ per resolution element. Ultimately, this setup allowed for determining the velocity centroid to $\sim 50$ km s$^{-1}$ for precise redshift determination. We note that every spectrum does not exhibit uniform wavelength coverage given the design of GMOS. Galaxies that are not placed near the center of the field have redder or bluer coverage than the range quoted above.
        
\begin{figure}[t!]
        \begin{centering}
        \hspace{-0.35in}
        \includegraphics[width=1.2\linewidth]{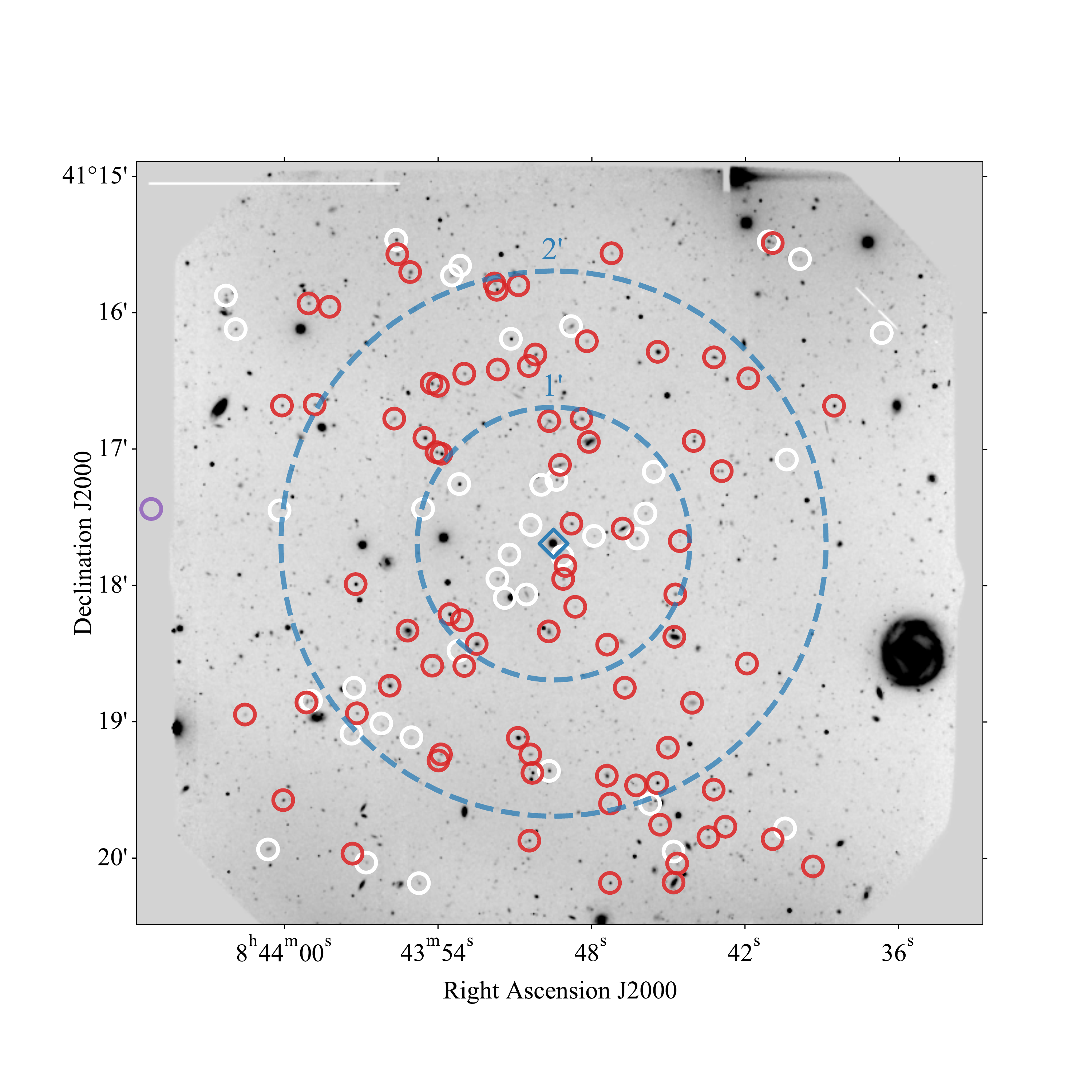}
        \hspace{-0.15in}
        \end{centering}
        \caption{An example of the survey design and targeting strategy of \cgmsq showing the slit placements centered around QSO J0843+4117 (blue diamond) overlayed on the $g$-band pre-image from the Gemini-GMOS detector. The circles denote where slits were placed on the slitmasks. White circles indicate galaxies with slits whose final spectra did not yield a reliable redshift,  while the red circles indicate galaxies that produced reliable redshifts. Large blue dashed circles show the one and two arcminute radii from the QSO. The purple circle just off the left of the detector is the COS-Dwarfs \citep{bordoloi14} galaxy target for this field whose spectra was obtained by SDSS with an impact parameter of $\rho = 113$~kpc at $z = 0.0300$.
        \label{fig:fov}}
        \end{figure}

        The slit masks were designed such that slits were placed on $\sim 85\%$ of objects within 1 arcminute of the QSO, with additional slits placed to fill the 5.5$\times$5.5 arcminute area of the detector. Slit placement constraints meant that $\sim 40-50$ slits could be placed on one mask. We aimed to obtain spectra of  $\sim 80$-$120$ unique galaxies per QSO field. Some fields in the earlier Rollup programs (those observed in 2014A) had as few as two and as many as four masks per field. In the later observing campaigns, all fields (except J1059$+$1441, which was only observed with one mask due to problems in the mask making) have three masks per field. This is shown in Table \ref{tab:maskdata}. An example of the targeting strategy can be seen in Figure~\ref{fig:fov} for the field J0843$+$4117. The red circles highlight the galaxies with  $z < z_{\rm QSO}$ for which we obtained reliable redshifts while the white circles were targeted for spectra but for which we did not recover a reliable redshift.
        
        All masks were observed between June 2014 and June 2015 on either GMOS-N or GMOS-S.   Table \ref{tab:maskdata} provides details of each multislit mask observed. Columns 5 and 6 list the number of slits placed and the number of slits that yielded reliable redshifts, respectively. A summary of the completeness of our galaxy spectroscopic survey is shown Table \ref{tab:completeness} for each field as a function of angular separation from the central QSO and limiting magnitude. The completeness is given as the fraction of reliable, redshift-yielding spectra, with `ZQ' $> 2$ of galaxies (\texttt{CLASSSTAR} $< 0.5$) by \texttt{SExtractor} within some angular separation to the QSO. The fields J2345-0059 and J1059+1441 have only one mask per field and thus have a lower completeness relative to the rest of the sample. In general, we are nearly 100\% complete out to 2 arcminutes from the QSO for galaxies with an $i$-band magnitude of $ <22$. The completeness drops to $\sim$50\% for a limiting magnitude of $i<23$. 68\% of the galaxies used in this analysis lie within 2 arcminutes of the central quasar. Details of the survey design, completeness and their impact on scientific results will be discussed in the full presentation of the \cgmsq survey, currently in preparation (Werk et al. 2021).
        
        \begin{deluxetable*}{cCCCCCC}
        \tablecaption{Completeness \label{tab:completeness}}
        \tablenum{3}
        \tablehead{
              QSO &  C_{22,1'} &  C_{22,2'} &  C_{22,\rm FOV} &  C_{23,1'} &  C_{23,2'} &  C_{23,\rm FOV}
              }
        \colnumbers
        \startdata
         J0226+0015 &      0.57 &      0.79 &        0.47 &      0.34 &      0.41 &        0.23 \\
         J0809+4619 &      1.00 &      1.00 &        1.00 &      0.77 &      0.70 &        0.37 \\
         J0843+4117 &      1.00 &      1.00 &        0.63 &      0.70 &      0.51 &        0.31 \\
         J0914+2823 &      1.00 &      1.00 &        1.00 &      0.80 &      0.63 &        0.40 \\
         J0935+0204 &      0.93 &      0.86 &        0.42 &      0.33 &      0.40 &        0.20 \\
         J0943+0531 &      1.00 &      1.00 &        0.54 &      0.61 &      0.42 &        0.20 \\
         J1001+5944 &      1.00 &      1.00 &        0.65 &      0.87 &      0.53 &        0.32 \\
         J1016+4706 &      1.00 &      1.00 &        0.55 &      0.81 &      0.62 &        0.31 \\
         J1022+0132 &      1.00 &      0.97 &        0.79 &      0.59 &      0.51 &        0.35 \\
         J1059+1441 &      0.60 &      0.47 &        0.32 &      0.24 &      0.24 &        0.15 \\
         J1059+2517 &      1.00 &      1.00 &        0.68 &      0.72 &      0.69 &        0.36 \\
         J1112+3539 &      1.00 &      1.00 &        0.49 &      0.54 &      0.43 &        0.22 \\
         J1133+0327 &      1.00 &      0.95 &        0.48 &      0.68 &      0.46 &        0.23 \\
         J1134+2555 &      1.00 &      0.99 &        0.66 &      0.70 &      0.57 &        0.33 \\
         J1233-0031 &      1.00 &      0.44 &        0.25 &      0.43 &      0.19 &        0.11 \\
         J1241+5721 &      1.00 &      0.79 &        0.35 &      0.90 &      0.36 &        0.14 \\
         J1342-0053 &      1.00 &      0.58 &        0.29 &      0.54 &      0.30 &        0.14 \\
         J1419+4207 &      1.00 &      1.00 &        1.00 &      1.00 &      1.00 &        0.76 \\
         J1437+5045 &      1.00 &      1.00 &        0.66 &      1.00 &      0.63 &        0.33 \\
         J1553+3548 &      1.00 &      1.00 &        0.82 &      1.00 &      0.74 &        0.40 \\
         J1555+3628 &      1.00 &      1.00 &        0.62 &      0.67 &      0.53 &        0.24 \\
         J2345-0059 &      0.28 &      0.40 &        0.16 &      0.16 &      0.19 &        0.09 \\
        \hline
         Median     &      1.00 &      1.00 &        0.57 &       0.69 &     0.51 &       0.27 \\
        \enddata
        \tablecomments{Completeness of the \cgmsq galaxy catalog along with the median in each radial and magnitude limited bin. The completeness is defined as the fraction of reliable spectra, with `ZQ' $> 2$, to objects designated as a galaxy (\texttt{CLASSSTAR} $< 0.5$) by \texttt{SExtractor} (1) QSO Field; (2) Completeness within 1' of the QSO at a limiting $i$-band magnitude of 22; (3) Completeness within 2' of the QSO at a limiting $i$-band magnitude of 22; (4) Completeness within the GMOS field of view of the QSO at a limiting $i$-band magnitude of 22; (5) Completeness within 1' of the QSO at a limiting $i$-band magnitude of 23; (6) Completeness within 2' of the QSO at a limiting $i$-band magnitude of 23; (7) Completeness within the GMOS field of view of the QSO at a limiting $i$ -band magnitude of 23. The fields J2345-0059 and J1059+1441 have only one mask per field and thus have a low completeness relative to the rest of the sample. }
        \end{deluxetable*}

        \subsubsection{Gemini Data Reduction}
        The spectra were reduced using a combination of Gemini's \texttt{PyRAF} package and \texttt{PypeIt}\footnote{https://github.com/pypeit/PypeIt} \citep{pypeit20}.
        The initial reduction closely follows the GMOS Data Reduction Cookbook\footnote{Shaw, Richard A. 2016, GMOS Data Reduction Cookbook (Version 1.2; Tucson: National Optical Astronomy Observatory), available online at:http://ast.noao.edu/sites/default/files/GMOS\_Cookbook/}. After obtaining the raw spectra from Gemini Observatory Archive, the data were organized according to field and mask. Biases were created by downloading all bias exposures of the same $2\times2$ binning taken within $\sim 1$ month from the observations if they were the same detector. Flat field and NeAr wavelength calibration exposures were taken along with each mask and were prepared using the standard methods in the cookbook. This procedure performs the bias subtraction, performs automatic slit edge finding, cuts the slits out of the image and isolates them, flat-fields each slit with the flats taken at the same central wavelength, and then performs a wavelength transformation to each slit. This results in multiple 2-D spectra. In general, the typical RMS of the wavelength solution is 0.1 pixel, which corresponds to 0.04\AA~ given the dispersion of GMOS of 0.4\AA~ per pixel. We then turned to \texttt{PypeIt} to perform sky subtraction, spectral continuum tracing, and combination of the wavelength-dithered. The spectra were then flux calibrated using a sensitivity function based on a selection of spectro-photometric flux calibration standards, BD28+4211, EG21, EG131, and Wolf1346, choosing whichever was closest on the sky. The final 1D spectra were co-added in vacuum wavelength space weighted by the inverse variance of the individual exposures. 

\section{Analysis} \label{analysis}
    \subsection{Galaxy Redshift Determination}
    The process of determining galaxy redshifts was done in two stages. First,  each 1D extracted spectrum was passed through an automated redshift fitting code, \texttt{REDROCK} \footnote{https://github.com/desihub/redrock} (v0.14). \texttt{REDROCK} was developed by the DESI team and uses a template fitting algorithm to generate a set of ranked best-fitting models, identifying the object's type (QSO, Star, Galaxy) and corresponding redshift.

\begin{figure}[ht!]
    \hspace{-0.15in}
    \includegraphics[width=1.1\linewidth]{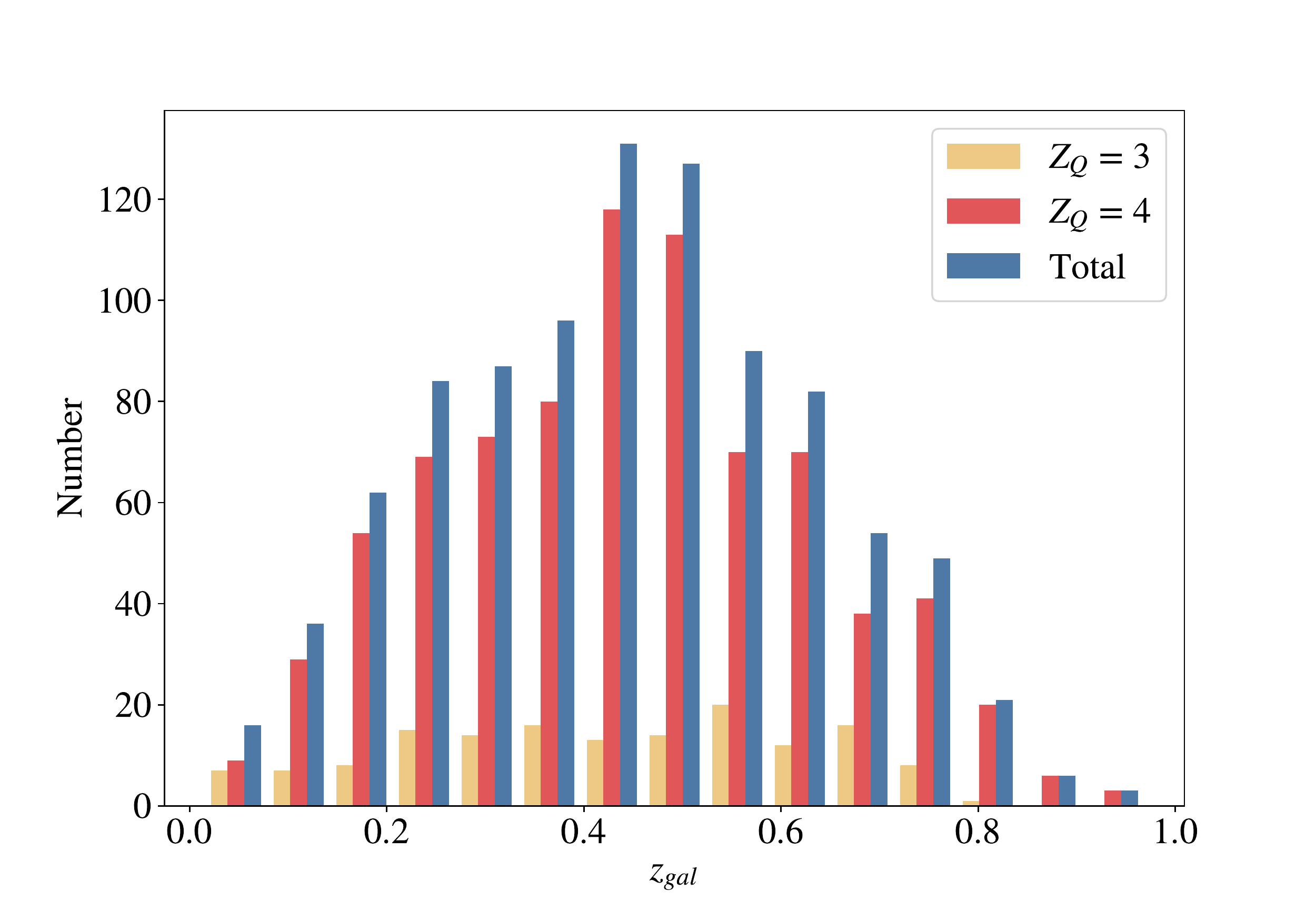}
    \caption{The redshift distribution of the \cgmsq galaxy catalog for galaxies with $z < z_{\rm QSO}$. The redshift reliability is  encoded in yellow and red.  Red represents our most reliable redshift quality flag of `4' with spectra displaying more than one strong absorption or emission line. A quality flag of `3' was reserved for spectra with only one strong emission line and thus a less reliable redshift designation. Approximately 85\% (820 out of 971) of our spectra were given the highest reliability flag. The typical statistical uncertainly of our redshifts is $\sigma_z \sim 50$-$100$ km s$^{-1}$ ($z \simeq 0.00016$-$0.00030$). \label{fig:zhist}}
    \end{figure}

    However, much of our galaxy sample has moderate to poor S/N, and the automated \texttt{REDROCK} redshift guesses can thus fail catastrophically by fixating on spurious features. We constructed a method of manually vetting the \texttt{REDROCK} redshifts by eye using a custom GUI, \texttt{VETRR}\footnote{https://github.com/mattcwilde/vetrr}. Each redshift was visually assessed and assigned a quality flag, $Z_{Q}$, by one of the authors, and by at least two members of the Werk SQuAD. A $Z_{Q}$ of either 0, 1, 3, 4 were assigned to each spectrum, where a $Z_{Q} = 0$ indicates the spectrum has a S/N that is too low to be useful. $Z_{Q} = 1$ denotes a good spectrum but we are not confident in identifying a redshift (i.e. has no clear absorption or emission lines). $Z_{Q} = 3$ are spectra with  one absorption or emission line that was confidently identified. These are usually strong [OII], H$\alpha$ emission or weak CaII absorption. A solitary H$\alpha$ emission only falls into this category if H$\beta$ is off the detector and the strong emission line is too narrow to be [OII], which is a marginally unresolved doublet in these data. $Z_{Q} = 4$ represents a spectrum for which we are most confident in the redshift, with at least two absorption or emission lines identified. $Z_{Q} =  2$ was not used. The fully vetted galaxy survey database that we use for our analysis contains only spectra with $Z_{Q} > 2$. The statistical uncertainly of our redshift identification was determined by computing the standard deviation of redshift identifications from at least three humans for a sample of 50 galaxies and is typically in the range of $\sigma_z \sim 50$-$100$ km s$^{-1}$ ($z \sim 0.00016$-$0.00030$). The redshift distribution of the vetted galaxy database containing 971 galaxies with redshifts less than that of the field quasar is shown in Figure \ref{fig:zhist}.
    
\begin{figure}$
    \begin{array}{c}
    \includegraphics[scale=0.36]{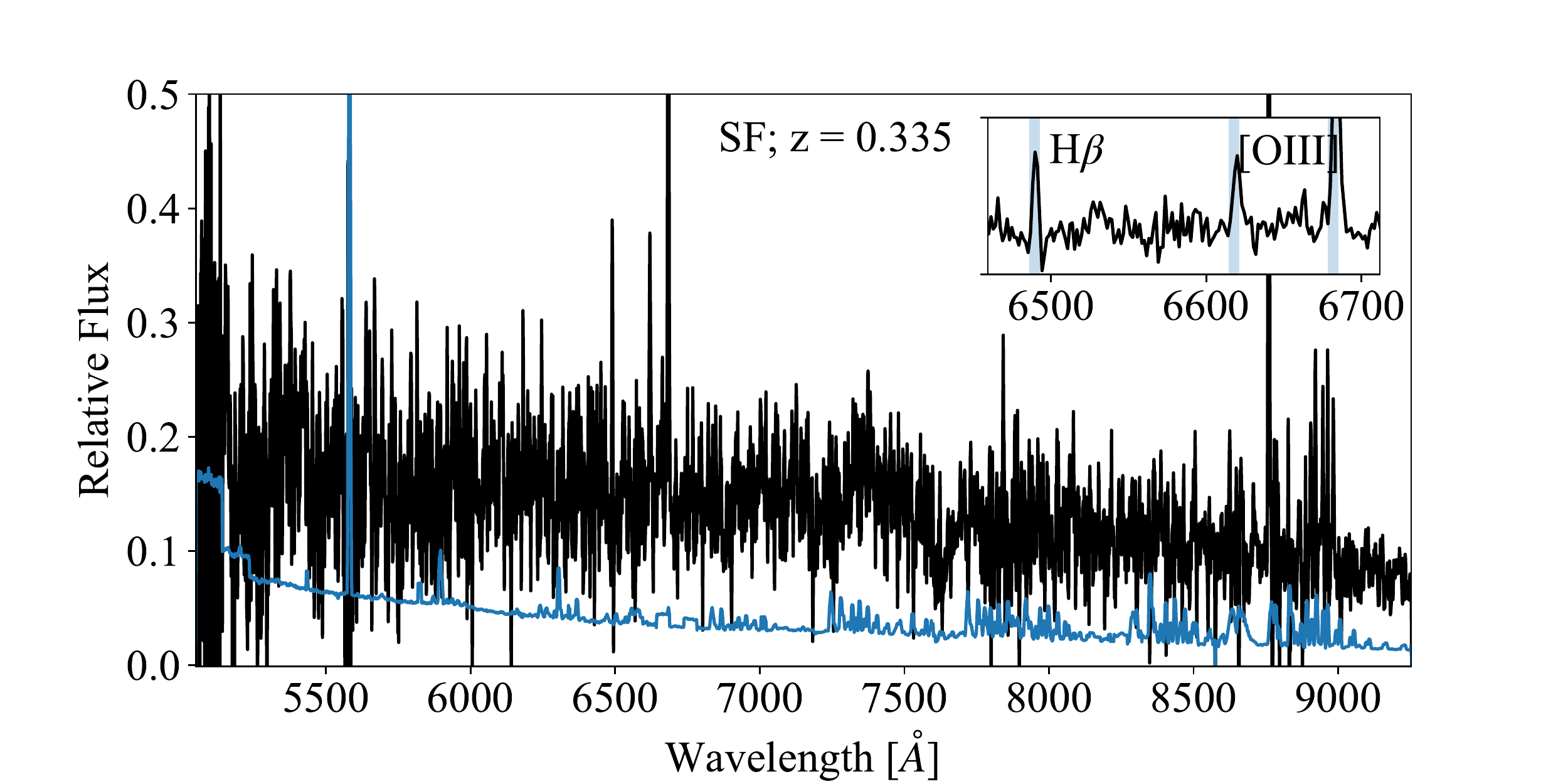} \\
    \includegraphics[scale=0.36]{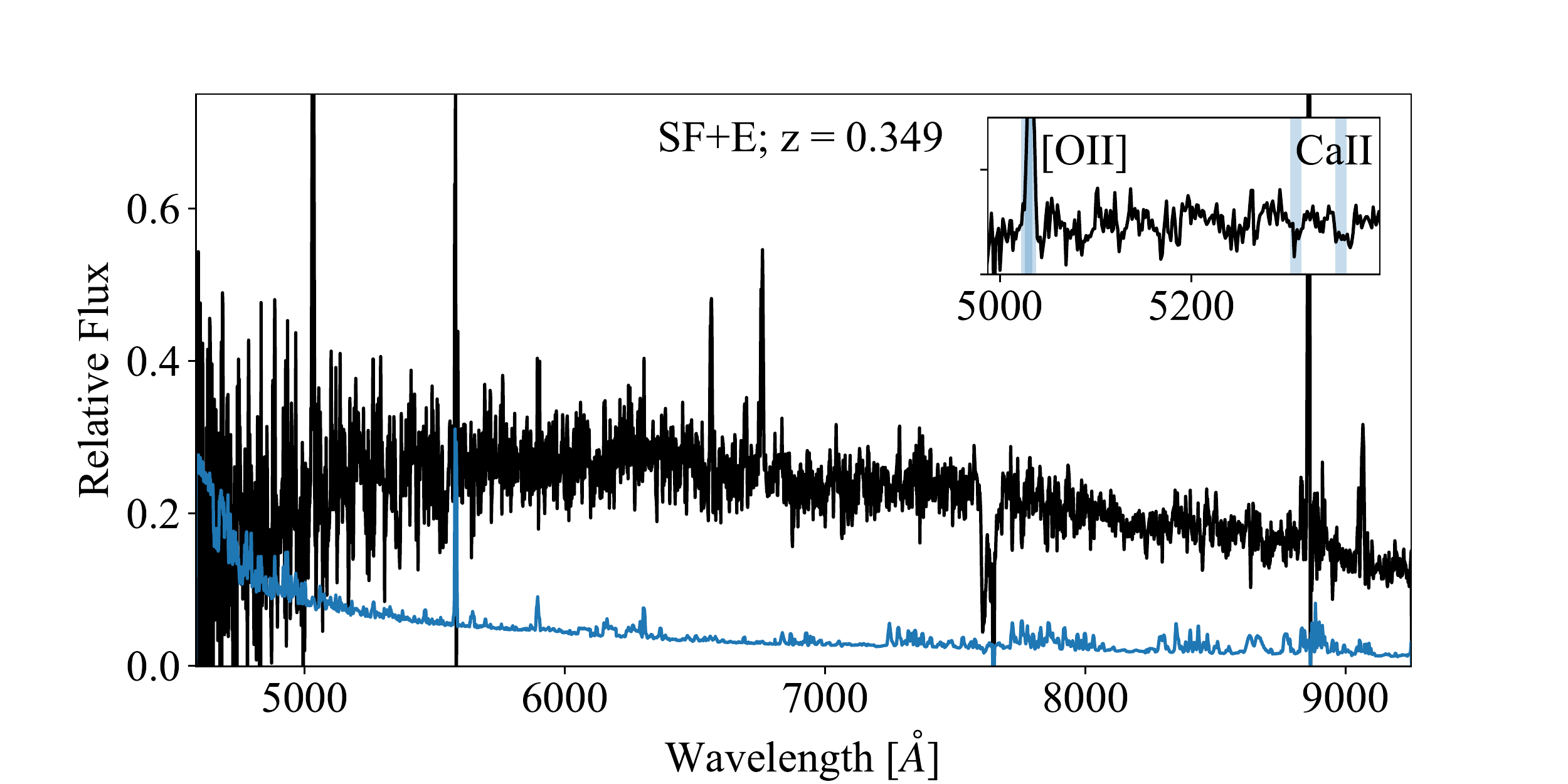} \\
    \includegraphics[scale=0.36]{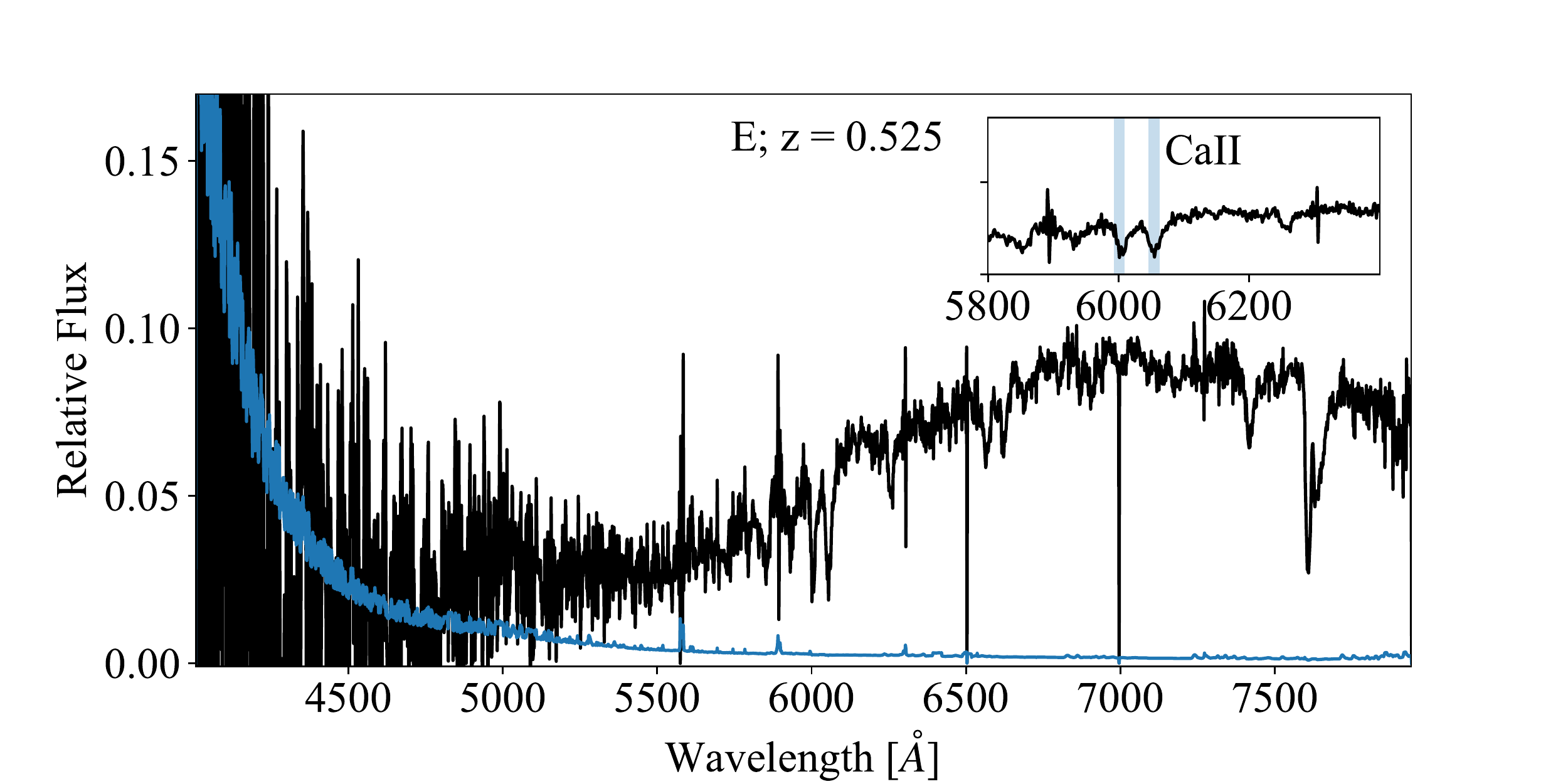} 
    \end{array}$
    \vspace{+5pt}
    \caption{Examples of typical Gemini-GMOS spectra with a quality flag of $Z_Q=4$, along with the error shown in blue.  These spectra highlight our method of visual galaxy spectral typing. The cut out insert in each panel shows an example of the key spectral features used in redshift determination. The top spectrum is classified as star-forming galaxy, the middle panel shows a galaxy with both emission lines and absorption lines, and the bottom spectrum is an example of galaxy with an older stellar population with strong Ca H+K absorption. 
    }
    \label{fig:galspec}
    \end{figure}

    In addition to manually vetting the galaxy redshifts, we also visually identified a galaxy spectral type during the process of examining each 1D spectrum.  If strong emission lines and weak continuum were present,  we classified the galaxy as ``star forming" or ``SF". If only absorption lines along a detected continuum were found,  we classified the galaxy as ``quiescent" or ``E". If both emission and absorption lines were identified,  we classified the galaxy as a combination of the two, ``SF+E." In several cases, stars were mistakenly targeted in our slit masks, and they have the identification as star. Stars are never included as part of our vetted galaxy database.  Examples of  galaxy spectra of each type are shown in the three panels of Figure \ref{fig:galspec}.  These spectra contain some poorly-subtracted sky lines, and telluric absorption lines at $\sim$7600\AA, which are the sorts of spurious spectral features that can cause \texttt{REDROCK} to fail. The top panel displays an emission-line `SF' galaxy that shows emission from several highlighted strong emission lines, the middle panel shows a combination-type, `SF+E' spectrum with both [OII] emission and strong CaII absorption against a bright stellar continuum, and the bottom panel shows an example of an absorption-only spectrum, type `E.' 

    \subsection{Galaxy Photometry and Spectral Energy Distribution Fitting}

    In addition to the galaxy spectroscopic catalog, we constructed a photometric galaxy catalog to derive stellar masses and galaxy star formation rates (SFRs). Our spectra are generally insufficient in signal and flux calibration to analyze them via spectral fitting codes \citep[e.g.][]{cappellari17}. To estimate stellar masses, we used CIGALE \citep[]{cigale11, boquien19} to fit the spectral energy distribution and retrieve stellar mass and SFRs. We note that while there are myriad other SED fitting codes available (as well as direct color-mass relations) we chose CIGALE for its options for stellar population models, dust models, etc. across large swaths of parameter space. We also chose it to compare directly to complementary surveys such as CASBaH \citep{burchett19, prochaska19}. 
    
    \subsubsection{Galaxy Photometry}

    One major challenge in constructing a photometric catalog for the \cgmsq survey was that the spectroscopic target catalog is generally fainter than available public all-sky surveys and thus the photometric coverage is not uniform in all fields. We chose to gather photometric data for every spectroscopic target in the galaxy catalog, totalling 2310 unique targets.  
    
    We created the photometric catalog by cross-matching our target to the DESI Legacy Imaging Surveys Data Release 8 (DR8) \citep{dey19}.
    The imaging survey for DESI is composed of data from three telescopes covering $\sim14000$ deg$^2$ over $-18^{\circ} < \delta < +84^{\circ}$ ($|b| > 18^{\circ}$). These three Programs include The Beijing-Arizona Sky Survey (BASS), The DECam Legacy Survey (DECaLS), and the Mayall z-band Legacy Survey (MzLS) which provide $g$, $r$, and $z$ band photometry to $\sim 23.3$ mag.
    
    The photometry is corrected for Galactic extinction. This is the deepest publicly available optical survey and provided the bulk of the photometry. In cases of overlap between the North and South catalogs, we chose the Southern DECaLS observations. We limited matches to objects with S/N $>2$ and chose the closest match within 1.3 arcseconds of our targets in order to limit mismatches between our faint sources and the Legacy Survey catalogs. This gives us 1985 targets with at least one band of photometry. In addition to the $g$, $r$, and $z$ bands, DR8 provides cross-matched WISE \citep{cutri13} observations in 3.4, 4.6, 12, and 22 $\mu$m. 

    In order to cover a larger wavelength range to better estimate the SED, we also cross-matched our catalog with the Pan-STARRS Data Release 2 \citep{chambers16} with coverage of the $grizy$ bands, utilizing the MAST cross-match service, using a 1.3 arcsecond threshold. We limited photometry to those marked as extended objects with good stack photometry and $grizy <$ 23.3, 23.2, 23.1, 22.3, 21.3., giving 393 objects with photometry in at least one band.
    
    In addition we also queried the SDSS DR14 \citep{aboltathi18} survey with $ugriz$ coverage where we restricted matches to less than $ugriz <$ 22.15, 23.13, 22.70, 22.20, 20.71, totalling to 331 targets with photometry in at least one band. 
    
    To make sure we included photometry for every object in the \cgmsq survey that may be too faint or in crowded areas for the public surveys, we included photometry from the GMOS imaging. The target selection and slit mask design was based on Gemini GMOS $i$-band and $g$-band imaging except in the case of the Rollup programs, GN-2014A-Q1 and GS-2014A-Q2, which involved a combination of HST WFC-ACS and Gemini imaging.  This allowed us to use the Gemini images to get photometry for each field, the details of which are presented in Table~\ref{tab:imagingdata}. Two exposures in each band were combined and processed by Gemini. We obtained $g$ and $i$ band magnitudes of the Gemini sources using the \texttt{SExtractor} software. We used pre-calibrated photometric zero-points and color corrections for GMOS-N and GMOS-S for an initial pass on the Gemini photometry.  In order to further calibrate the Gemini photometry, we cross-matched our m$<$21 sources to the $g$ and $i$ band SDSS photometric sources, and then bootstrapped the Gemini photometry below the SDSS magnitude limit within each field. This consisted of applying a constant magnitude offset to the Gemini sources to match the SDSS photometry in the $g$ and $i$ bands.
    
    \begin{deluxetable}{l|c|c}
    \tablecaption{\cgmsq Gemini GMOS Imaging \label{tab:imagingdata}}
    \tablenum{4}
    \tablehead{\colhead{Project ID} & \colhead{$g$-$t_{\rm exp}$ [s]} & \colhead{$i$-$t_{\rm exp}$ [s]}}
    \colnumbers
    \startdata
    GN-2014A-Q1   &   X & 150 \\
    GS-2014A-Q2   &   X & 150 \\
    GN-2014B-LP-4 & 450 & 200 \\
    GS-2014B-LP-3 & 450 & 200 \\
    GN-2015A-LP-3 & 450 & 200 \\
    GS-2015A-LP-4 & 450 & 200
    \enddata
    \tablecomments{Imaging observations with GMOS-N and GMOS-S taken as part of the CGM$^{2}$ Survey: (1) Project ID; (2) $g$-band exposure time in seconds; (3) $i$-band exposure time in seconds}
    \end{deluxetable}
    
    All photometry from Gemini, SDSS, and Pan-STARRS was corrected for Galactic reddening based on the values in \cite{schlafly11} provided by the NASA Extragalactic Database\footnote{The NASA/IPAC Extragalactic Database (NED) is funded by the National Aeronautics and Space Administration and operated by the California Institute of Technology.}  by querying our targets coordinates via \texttt{ASTROQUERY}\footnote{http://dx.doi.org/10.6084/m9.figshare.805208}. We employed the SVO Filter Profile Service \citep{rodrigo12} to obtain our filter transmission curves for the telescopes used in these surveys as required input to CIGALE.
   
    CIGALE includes many models as options to include in fitting. For stellar populations, we used the \cite{bruzual03} models, assuming a \cite{chabrier03} initial mass function (IMF). We chose a grid of metallicities ranging from $0.001$-$2.5 Z_{\odot}$. We used a delayed star formation history (SFH) model with an exponential burst. The e-folding time of the main stellar population models ranged from 0.1-8~Gyr. We varied the age of the oldest stars in the galaxy from 2-12~Gyr. We included an optional late burst with an e-folding time of 50~Myr and an age of 20 Myr. We varied the burst mass fraction from 0.0 or 0.1 to turn this feature on or off. Nebular emission and reprocessed dust models \citep{dale14} were also included with the default values. The dust models have slopes ranging from $1-2.5$ and the nebular models include no active galactic nuclei.

    We employed the \cite{calzetti94} dust attenuation law, but we also included a ``bump" in the UV (see discussion in \cite{prochaska19}) at 217.5~nm with a FWHM of 35.6~nm. The bump amplitude is set at 1.3 and the power law slope is -0.13 \citep{lofaro17}. We varied the color excess of the stellar continuum from the young population, E(B-V), from 0.12-1.98. Finally, we used a reduction factor of 0.44 to the color excess for the old population compared to the young stars. 
    
    \subsection{Galaxy Stellar Masses and Derived Properties}\label{derived}
        
\begin{figure}[ht!]
    \hspace{-0.15in} 
    \includegraphics[width=1\linewidth]{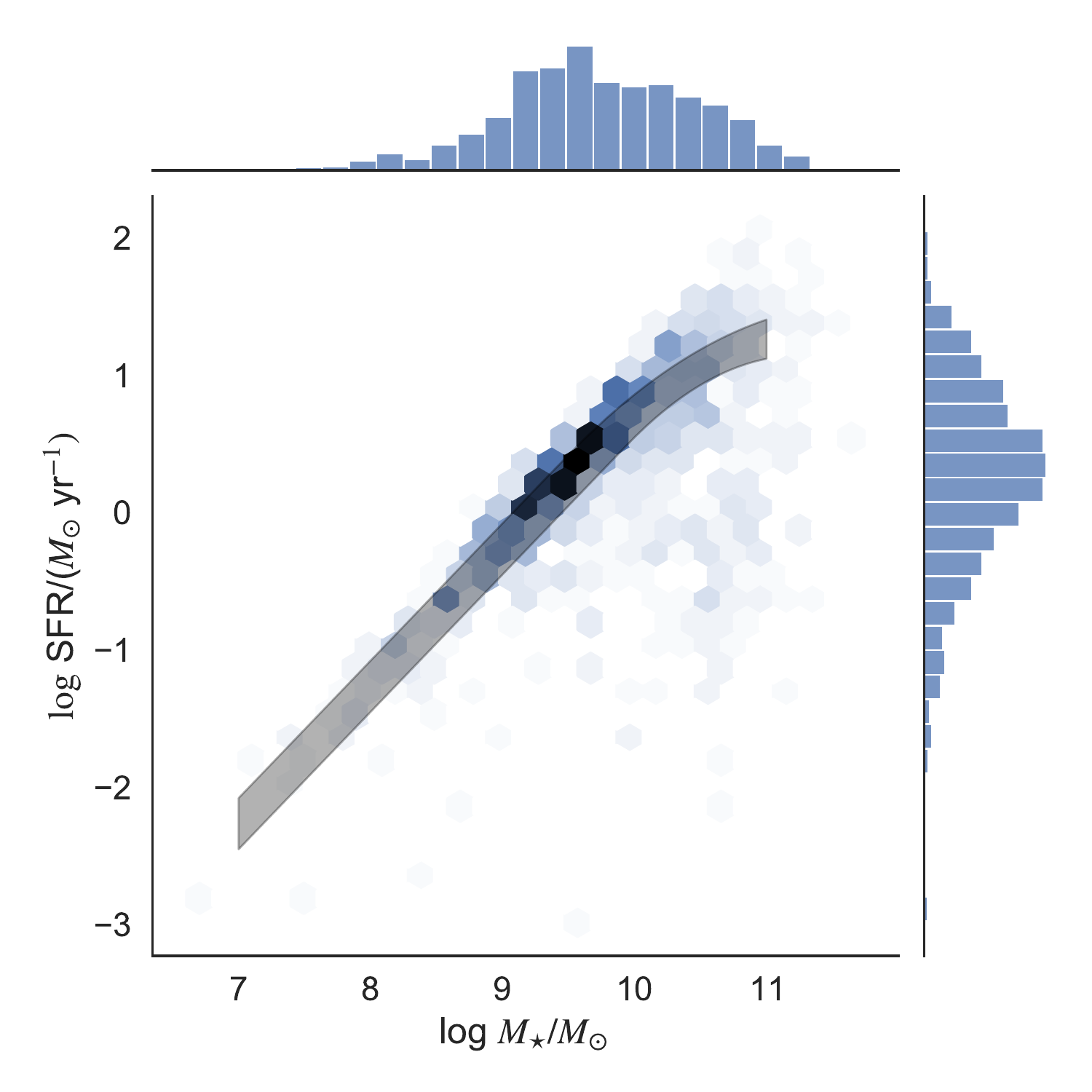}
    \caption{Star formation rate (SFR) vs. stellar mass for the \cgmsq galaxy sample as estimated by CIGALE. The locus of galaxies tracks a monotonic increase of SFR with stellar mass, known informally as the ``star-forming main-sequence" (SFMS). The density of galaxies in this space is indicated via shading of the hexagonal bins. The grey shaded region corresponds to the redshift dependent fit of the SFMS from \cite{schreiber15} of galaxies spanning $z = [0.23, 0.63]$. This range represents the 16th and 84th percentiles of the galaxy catalogs redshift distribution}.
    \label{fig:SFR}
    \end{figure}

\begin{figure}[ht!]
    \hspace{-0.15in} 
    \includegraphics[width=1.1\linewidth]{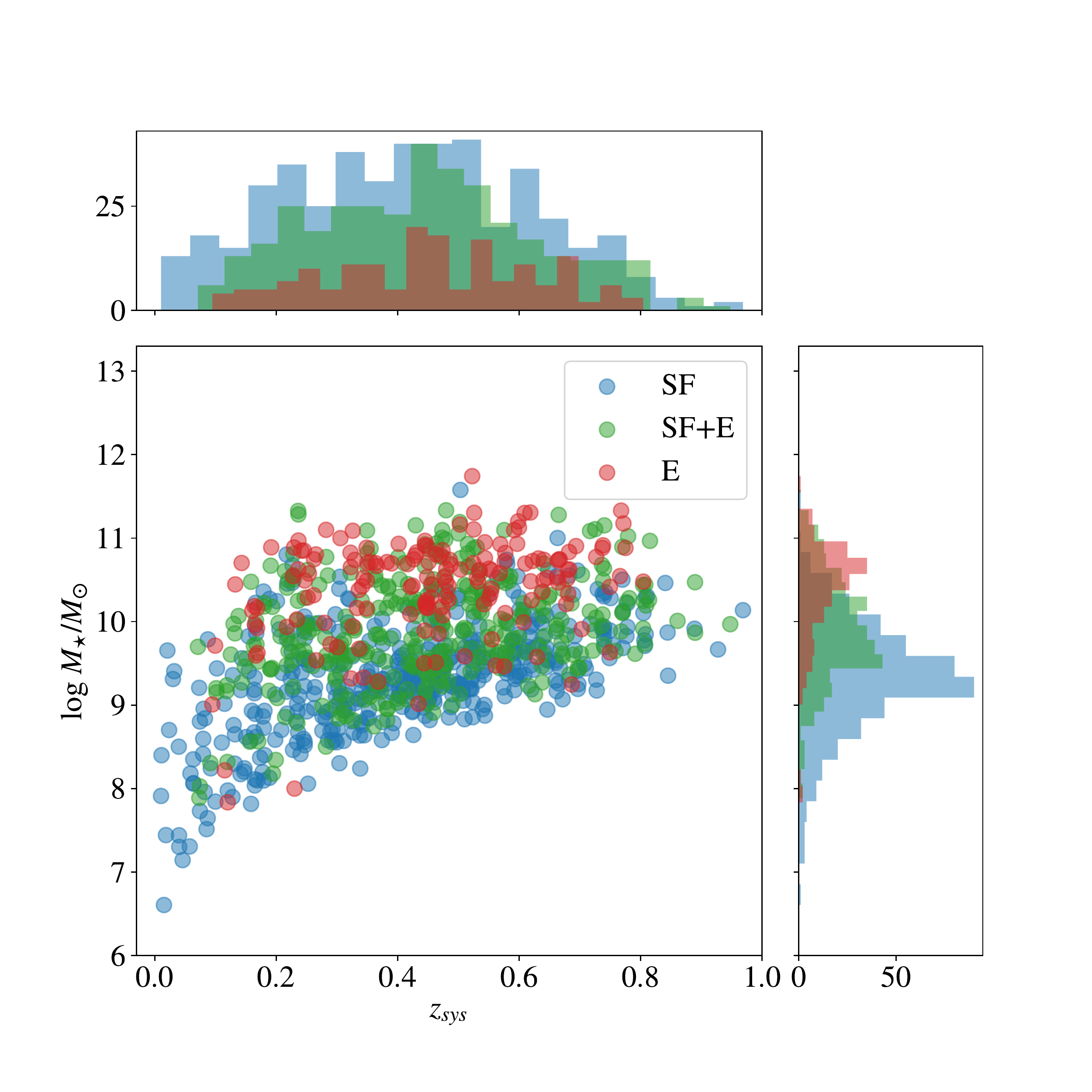}
    \caption{The distribution of galaxy stellar mass as a function of galaxy systemic redshift with marginal distributions on the right and top.  The red, green, blue colors correspond to the galaxy spectral type determined from visual inspection of GMOS spectra. Red circles show absorption-line only, or elliptical (E) type galaxies, green circles show galaxies displaying a combination of absorption and emission lines associated with star formation (SF+E), and blue circles show emission-line only,  or star forming (SF) galaxies. \label{fig:mstars_vs_z}}
    \end{figure}

    After fitting each galaxy's SED, CIGALE then outputs several useful parameters including stellar mass, SFR, and the rest-frame absolute luminosity in each band. Figures \ref{fig:SFR} and \ref{fig:mstars_vs_z} show the resultant mass distributions with masses spanning $M_{\star} \approx 10^{6} - 10^{11} M_{\odot}$ and $\bar{M}_{\star} = 10^{9.3} M_{\odot}$ at $\bar{z} = 0.44$. In order to calculate the virial radius of the galaxies, we first calculate the halo mass using the abundance matching method of \cite{moster13} with the modifications used in \cite{burchett16}. We adopt $R_{200m}$, the radius within which the average mass density is 200 times the mean matter density of the universe, as the virial radius ($R_{\rm vir}$) of a galaxy halo.

    Figure \ref{fig:SFR} shows the CIGALE-derived SFRs vs stellar masses, which exhibit, as expected, a positive correlation known as the ``star-forming main-sequence" (SFMS). We compare our models to the redshift dependent fit from \cite{schreiber15} over a range of galaxies spanning $z = [0.23, 0.63]$. This range represents the 16th and 84th percentiles of the \cgmsq galaxy catalogs redshift distribution. Typical uncertainties on photometry-derived stellar masses range from a factor of $3-5$ \citep[e.g.][]{blanton07, werk12} corresponding to $\sim 0.5$ dex, and errors on CIGALE-derived SFRs are similar.

\begin{figure}[ht!]
    \hspace{-0.15in} 
    \includegraphics[width=1.1\linewidth]{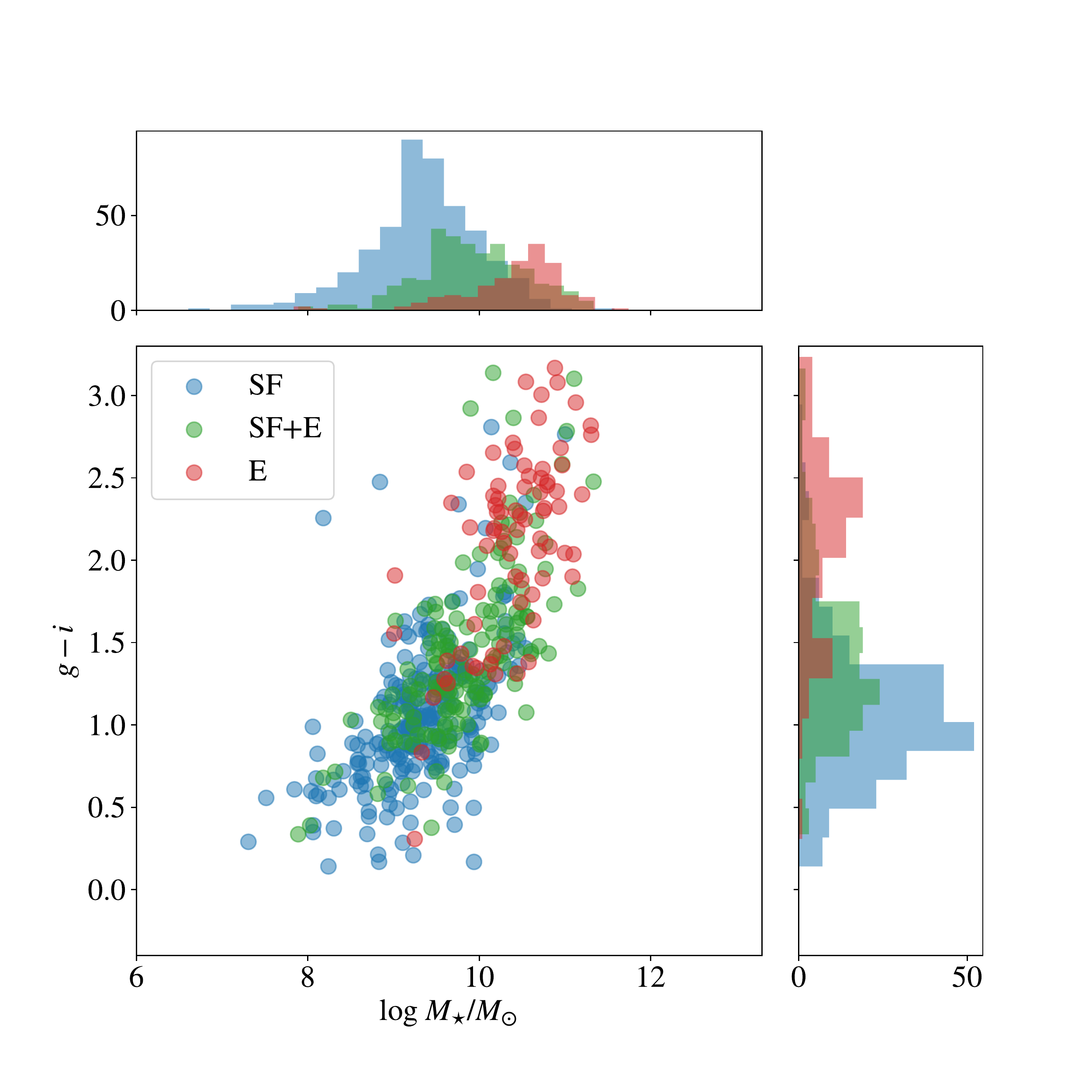}
    \caption{\cgmsq sample in a color-mass diagram using the $g-i$ color and the galaxy stellar mass, $M_{\star}$. Multi-band} photometry was not available for all of the galaxy targets, only objects with both bands are shown here. The bi-modal populations of the star-forming and passive galaxies are evident. Due to the nature of our survey, we are slightly biased against faint, passive galaxies since retrieving a redshift in the case of absorption lines requires large continuum flux.
    \label{fig:colormag_gr_mstars}
    \end{figure}

    Figure \ref{fig:mstars_vs_z} shows galaxy stellar masses vs. galaxy systemic redshift, and differentiates among our three galaxy spectral types visually derived from the GMOS spectra. Above $z\sim$0.5, we are no longer detecting galaxies with M$_{\star}$ $\lesssim$ 10$^{8}$ M$_{\odot}$. Both SF and SF$+$E galaxy spectral types are preferentially distributed among lower stellar masses as seen in the marginal distributions. We find that, as expected, there are more `E' type galaxies at higher galaxy stellar masses than a random distribution would predict. As shown in Figure \ref{fig:colormag_gr_mstars}, the CIGALE-derived color-mass diagram of our galaxy sample shows the bimodality of the star-forming and non-star-forming galaxies found in large galaxy surveys such as SDSS \citep[e.g.][]{chang15}. 

    \subsection{Absorber Catalog}\label{absorber_cat}
    
    To identify absorption features in the QSO spectra, the HST/COS UV spectra were visually inspected by members of the \textit{Werk SQuAD} in a multi-step process which closely follows the procedure described in \cite{tejos14} and is designed to leave no QSO absorption feature unidentified. This process includes identifying all of the most ubiquitous metal ions present in QSO spectra at the wavelengths covered by the HST/COS spectra, not just the \ion{H}{1} Lyman series.  These other metal ions will be explored in future \cgmsq papers. To identify the absorption features, the Werk SQuAD used a module from the \texttt{PYIGM}\footnote{https://github.com/pyigm/pyigm} software package, \texttt{IGMGUESSES} . The software allows for a straightforward comparison of multiple transitions from different elements, as multiple lines are displayed with their expected relative intensities given by their atomic parameters simultaneously for a given redshift.
    
    During the line identification process, the absorption lines are assigned a reliability score. `a' signifies a \textit{certain} feature. For example, the always-present Milky Way (MW) ISM lines at $z = 0$ fall into this reliability category. Other examples include absorption lines observed in two or more transitions of \ion{H}{1}, or multiple metal lines that align with hydrogen lines within $\pm$30 km s$^{-1}$, and metal doublets or multiplets that show the expected relative strengths as derived from their oscillator strengths and wavelengths \citep{morton03} and similar velocity profiles. A reliability score of `b', or \textit{possible}, includes single \ion{H}{1} lines with no associated metal lines, and metal ions having only one transition within the observed wavelength range. Other cases of assigning `b' values involve messy blends from absorption lines at different redshifts, weak or uncommon metal lines in an otherwise strong absorption system, and velocity offsets $> 30$ km s$^{-1}$ from other `a' lines. If a line did not fall into either of the previous categories we gave it a reliability score of `c', or \textit{unreliable}. In our analysis, we did not include any absorbers in the \textit{unreliable} 'c' category. The full catalog of identified absorbers will be presented in detail in future work.

   The end result of this line identification process for all 22 QSOs is a catalog of 2914 distinct absorption components, 2071 of which have a reliability rating of \textit{certain} or \textit{possible}. An absorption component is defined by an absorption line or lines with a distinct central velocity (or redshift). In practice, individual components offset by $< 20$~km s$^{-1}$ may not be separable in the HST/COS spectra. An absorber, or absorption system, is a set of absorption line components within $|\delta v| \approx 1000$ km s$^{-1}$. For example, Ly-$\alpha$ and Ly-$\beta$ are distinct lines but would be part of the same \ion{H}{1} component if aligned within the COS resolution velocity (redshift). This component may be grouped with other \ion{H}{1} or metal ion components to form an absorption system.
   
   Different absorbers lie at distinct redshifts and may physically correspond to clouds or layers of gas in the CGM of galaxies at their respective redshifts. Absorbers may also be clouds or filaments of gas in the IGM referred to as the \lya forest, not directly associated with a nearby galaxy. The term absorber is often used interchangeably with the term “system.” However, absorber distinctly does not imply an association with a galaxy and is a more empirical term. Absorption systems may have multiple components at different velocities within their assigned redshift ranges. The vast majority of our \ion{H}{1} absorption systems cover $>$ 3 Lyman series lines, many with multiple absorption components.  
   
   In order to retrieve more physical quantities such as column density, $N_{\rm HI}$, we used the apparent optical depth method (AODM) from \cite{savage91} as encoded in the \texttt{linetools}\footnote{https://github.com/linetools/linetools} package. Because we have column density measurements from several Lyman Series lines in most cases, the mean $1\sigma$ uncertainties on column density is 0.17 dex for unsaturated \ion{H}{1} lines to column densities $\simeq$ 10$^{17.5}$ cm$^{-2}$. 

    
    \begin{figure}[ht!]
    \begin{centering}
    \hspace{-0.15in} 
    \includegraphics[width=1.1\linewidth]{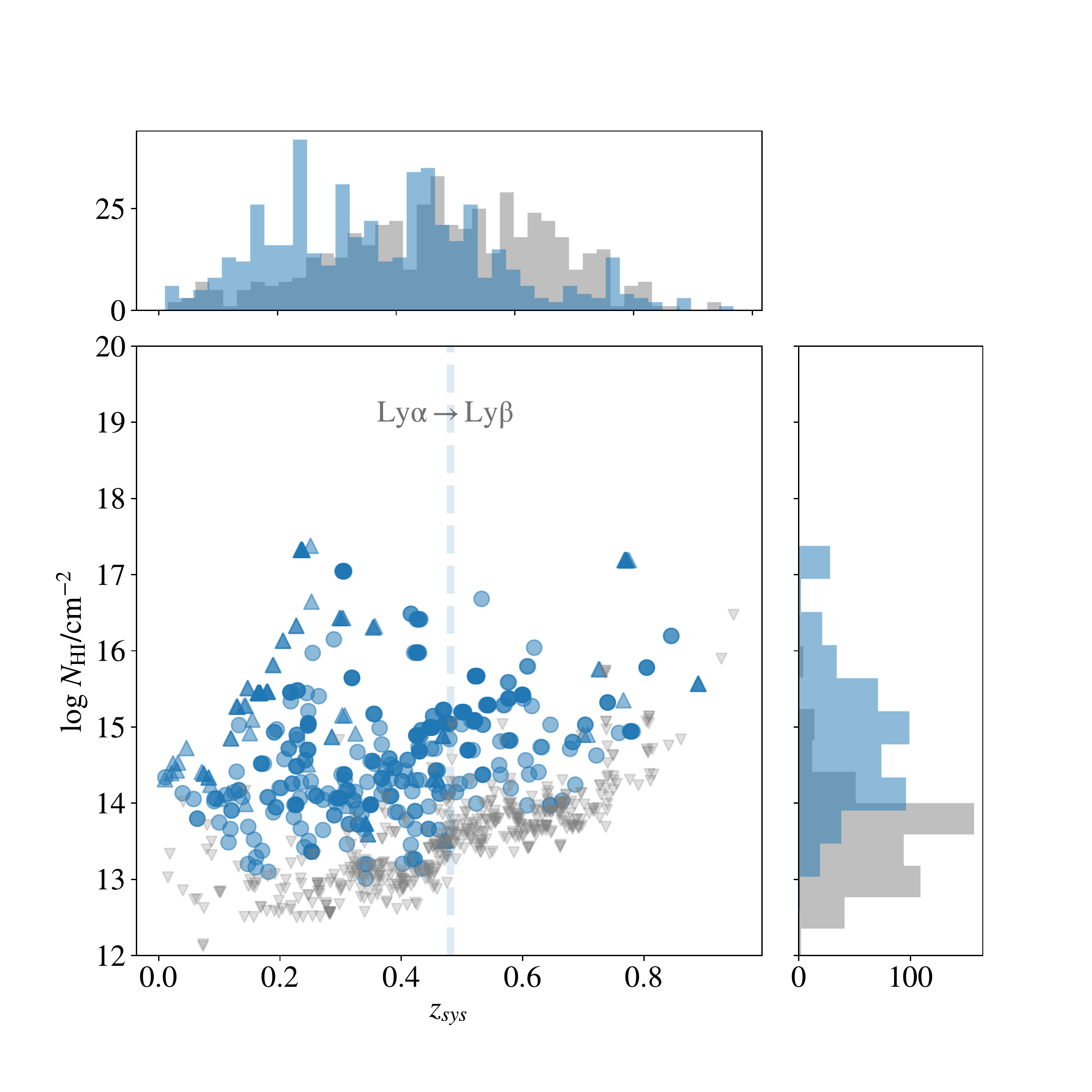}
    \end{centering}
    \vspace{-0.2in}
    \caption{Scatter plot and marginal distributions of column densities vs. redshift for the \ion{H}{1} systems detected in the \cgmsq survey. The mean $1\sigma$ uncertainties on column density is 0.17 dex for unsaturated \ion{H}{1} lines to column densities $\simeq$ 10$^{17.5}$ cm$^{-2}$, which is of order the size of the symbol (see Figure~\ref{fig:HI_mass} for the size of our uncertainties). The measurements designated with upward triangles are saturated absorption lines and are thus lower limits while circles represent detections. Smaller gray downward facing triangles are $2\sigma$ upper limits for galaxies where no corresponding absorption was measured. The visible break in the minimum \NHI at $\gtrsim$ 0.48 is shown by the vertical dashed line, and marks the redshift at which \lya shifts out of the COS G160M bandpass and thus becomes inaccessible. There can be multiple systems at the same column density and redshift which appear as darker points in this plot. This occurs when multiple galaxies lie within $|\delta v| < 500$ km s$^{-1}$ of the absorption systems. \label{fig:NHI_vs_z}}
    \end{figure}

\section{Connecting Galaxies and Absorbers: The CGM$^2$ \ion{H}{1} Survey} \label{survey}
\subsection{Defining CGM \ion{H}{1} Absorption Systems}
    With our separately completed galaxy and absorber databases, we can now begin to connect the two as a study of the CGM. In order to construct CGM systems, we first group the individual absorption component identifications into absorption systems, or absorbers. The grouping of absorption components was done using a clustering algorithm from \texttt{SKLEARN} \citep{pedregosa11}, \texttt{MEANSHIFT}. This algorithm groups individual absorption components together within a window function of 1000 km s$^{-1}$. The resultant absorber catalog consists of groups of components we call absorption systems. 
   
    We then cross-matched the galaxies and absorption systems if the relative velocity difference of the galaxy and the velocity centroid of at least one of the components of an absorption system exhibits $|\delta v| < 500$ km s$^{-1}$. We chose this velocity threshold to include absorption systems that could be at or above the escape velocity of the most massive galaxies in our sample. If no absorption system is found at the redshift of a galaxy, we measure the $2\sigma$ upper limit within $\delta v = \pm 30$ km s$^{-1}$ of the galaxies redshift using the normalized error of the quasar flux. If there was an interloping line at this redshift, we measure the AODM column density as a conservative upper limit. Our results are not sensitive to the choice in the velocity window. We find 181 systems consisting of 416 distinct components that exhibit HI column densities above our 2-$\sigma$ detection threshold, giving an average of 2.3 detected components per galaxy (absorption system) within our stated velocity window. We find 2.4 average components per galaxy for a smaller window of $|\delta v| < 250$ km s$^{-1}$, while for a larger velocity window, we find a small decrease to 2.2 detected components per galaxy for a window of $|\delta v| < 1000$ km s$^{-1}$. The average column density of detected, but not saturated, components remains $10^{14.9}$ cm$^{-2}$ in each case.
    
    Thus, we are left with a galaxy-centric CGM survey that consists of absorption line column density measurements (or limits) around 971 galaxies with reliable redshifts that lie $<$ 1000 km s$^{-2}$ from the quasar. We denote these galaxy-absorber pairs as CGM systems. Figure \ref{fig:NHI_vs_z} shows the AODM \ion{H}{1} column densities for all 971 CGM systems as a function of galaxy systemic redshift. Saturated lines provide only lower limits to the \ion{H}{1} column density and are shown as upward facing triangles. Non-detections are shown as 2$\sigma$ upper limits, and as downward facing triangles.  There is an obvious ``knee" of amplitude 0.5 dex in the lowest \ion{H}{1} column densities at a redshift of $z \sim 0.5$.  This decreased sensitivity to weak \ion{H}{1} absorption features is driven by the redshifting of \lya out of the wavelength range of the COS G160M grating. The column density measurements at $z > 0.481$ are derived from measurements of the \lyb absorption line and/or weaker Lyman series transitions, leading to a decrease in sensitivity for a fixed S/N. This shift in sensitivity motivates us to limit our \ion{H}{1} analysis to $z < 0.481$, corresponding to $\lambda_{\rm Ly\alpha}(1+z) = 1800$\AA, or the reddest end of the COS-G160M grating. By imposing this limit, we ensure nearly uniform sensitivity to \ion{H}{1} column density for our CGM sample.
     
    After limiting our CGM system survey to the aforementioned redshift, we construct the \cgmsq \ion{H}{1} survey, consisting of 572 total galaxy-absorber pairs. The survey includes all galaxies with reliable redshifts and \ion{H}{1} absorption systems (including the non-detections within$|\delta v| < 500$ km s$^{-1}$), collectively referred to as CGM systems. In the \ion{H}{1} covering fraction analysis that follows, it is possible to have multiple galaxies at similar redshifts but at differing impact parameters that match with individual absorption systems (74 systems total). For understanding the \ion{H}{1} extent of the CGM, we want to understand the correlation of galaxies and absorbers and thus do not limit our matches to the closest or most massive galaxy. However, in our complementary \ion{H}{1} velocity analysis, we limit the survey to the galaxy with the smallest impact parameter, thus leaving 522 CGM systems. Future studies, depending on their specific aims, will make independent choices about how to include galaxy-absorber pairs. 

\begin{figure}[t!]
    \begin{centering}
    \hspace{-0.15in}
    \includegraphics[width=1.0\linewidth]{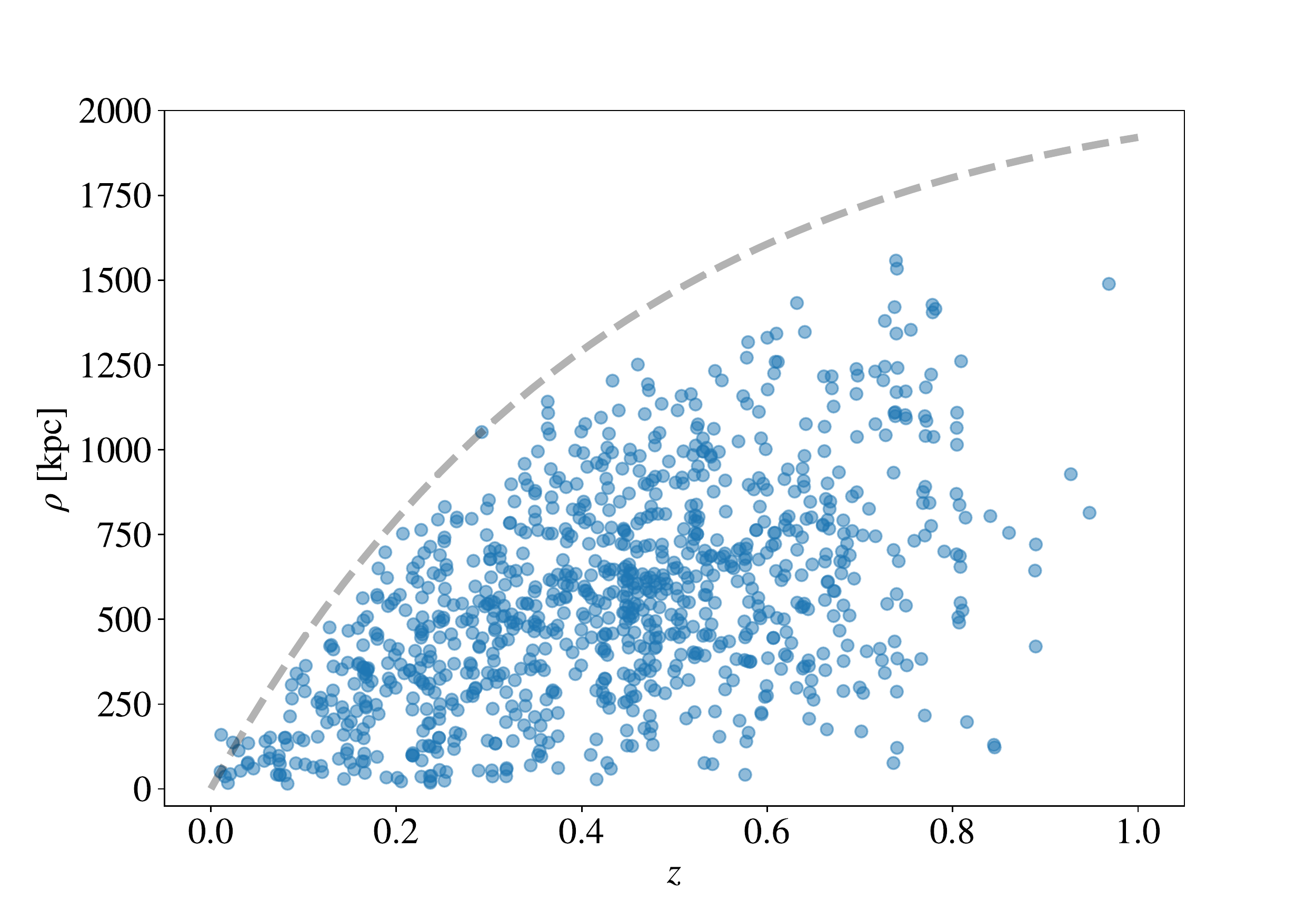}
    \end{centering}
    \vspace{0.0in}
    \caption{The distribution of impact parameters as a function of redshift. The grey curve approximates the distance to the edge of the detector in the 5.5$'$ GMOS FOV assuming the QSO is in the center to highlight the survey coverage as a function of redshift. Although most of the QSOs are centered in the FOV, in a few cases they had to be offset to avoid bright stars.  \label{fig:rho_z}}
    \end{figure}

    In the following analyses, we will examine trends with the impact parameter, $\rho$, which quantifies the projected distance between the QSO and the galaxy in the rest frame of the galaxy.  Figure \ref{fig:rho_z} shows the impact parameter--redshift distribution of CGM systems. The grey curve approximates the distance to the edge of the detector in the 5.5$'$ GMOS FOV, assuming the QSO is in the center, in order to highlight the survey coverage as a function of redshift. The QSO was slightly offset from the center in certain fields, either to better place guide stars or to avoid bright foreground stars; thus a few galaxies fall on or near this approximate field-size limit. 

    \subsection{\ion{H}{1} Covering Fraction: Definition and Threshold}
    To quantify the radial profile and extent of the CGM, we use the covering fraction $f_c$ as a measure of the probability of the presence of \ion{H}{1}. The covering fraction is the comparison of ``hits" (H) and ``misses" (M), with a hit being defined as a galaxy with a corresponding absorber at or above the detection threshold for the full ensemble (see below), while a miss occurs when the $2\sigma$ upper limit on a detection is below the threshold at the redshift of the galaxy. A system with a $2\sigma$ upper limit above the threshold is ignored altogether because this indicates that the S/N of the spectrum is not adequate for detection of the \ion{H}{1} lines. The covering fraction is the fraction of hits versus the total number of CGM systems above the threshold in a given bin, $f_c = H/(H+M)$. 

    We choose a threshold of \NHI$\geq 10^{14}$ cm$^{-2}$ for the covering fraction calculations which is supported by previous survey work. In particular, \cite{chen05} find that \ion{H}{1} column densities below 10$^{13.6}$ cm$^{-2}$ are consistent with being  randomly distributed with respect to known galaxies. They also show the correlation of galaxies and absorbers does not depend sensitively on \NHI for strong absorbers \NHI  $>$ 10$^{13.6}$ cm$^{-2}$.  Furthermore, a value of \NHI$\geq 10^{14}$ cm$^{-2}$ was shown in \cite{tejos14} to be more highly correlated with galaxies than gas at lower column densities. Adopting a threshold of \NHI$\geq 10^{14}$ cm$^{-2}$ additionally provides us with a sample against which we can compare to the cumulative column density distributions of \ion{H}{1} systems found in \cite{danforth16}, who also use this value.
    
    \subsection{The Empirical \ion{H}{1}-Galaxy connection}\label{empirical}
    
\begin{figure}$
    \begin{array}{c}
    \includegraphics[scale=0.33]{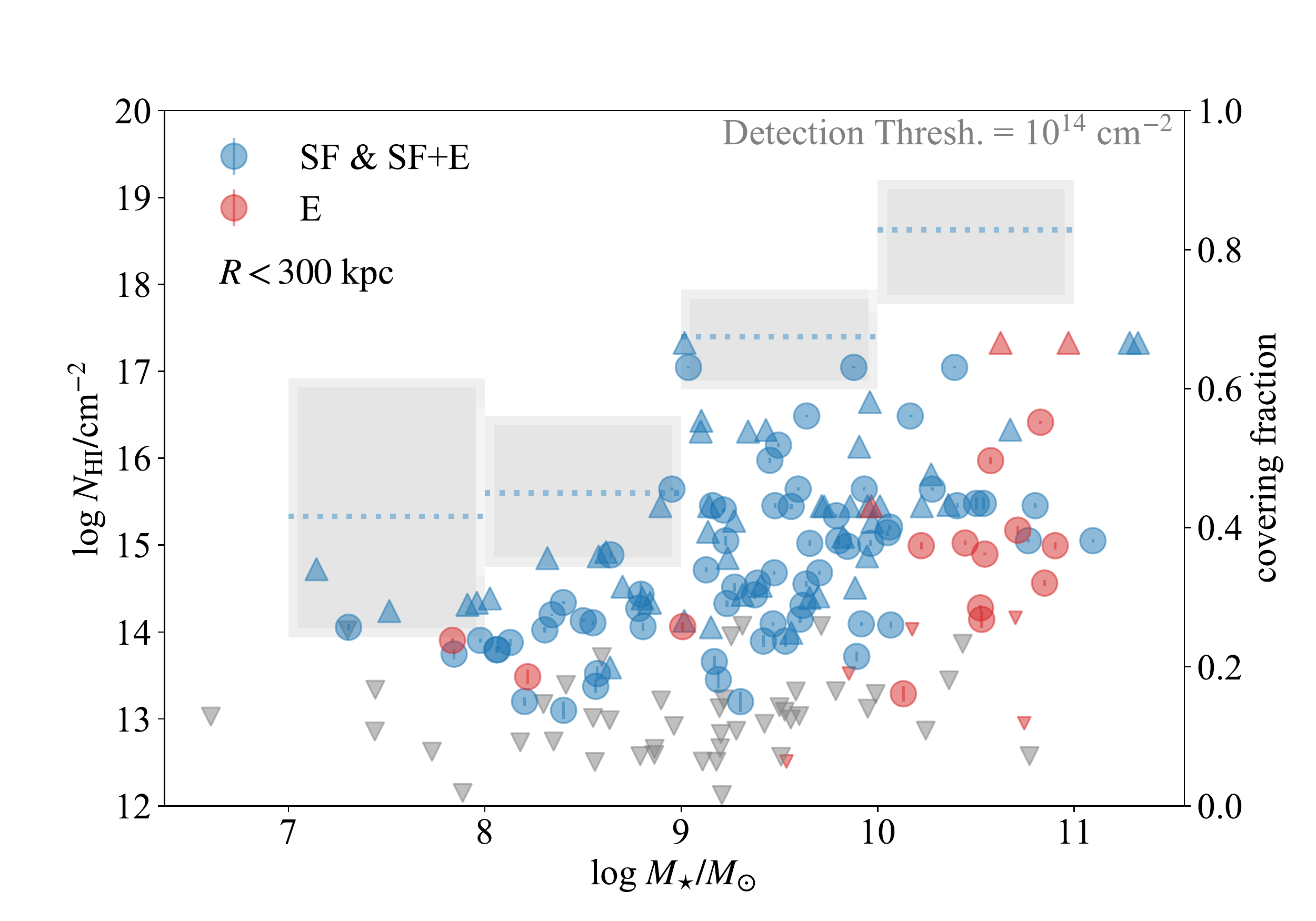} \\
    \includegraphics[scale=0.33]{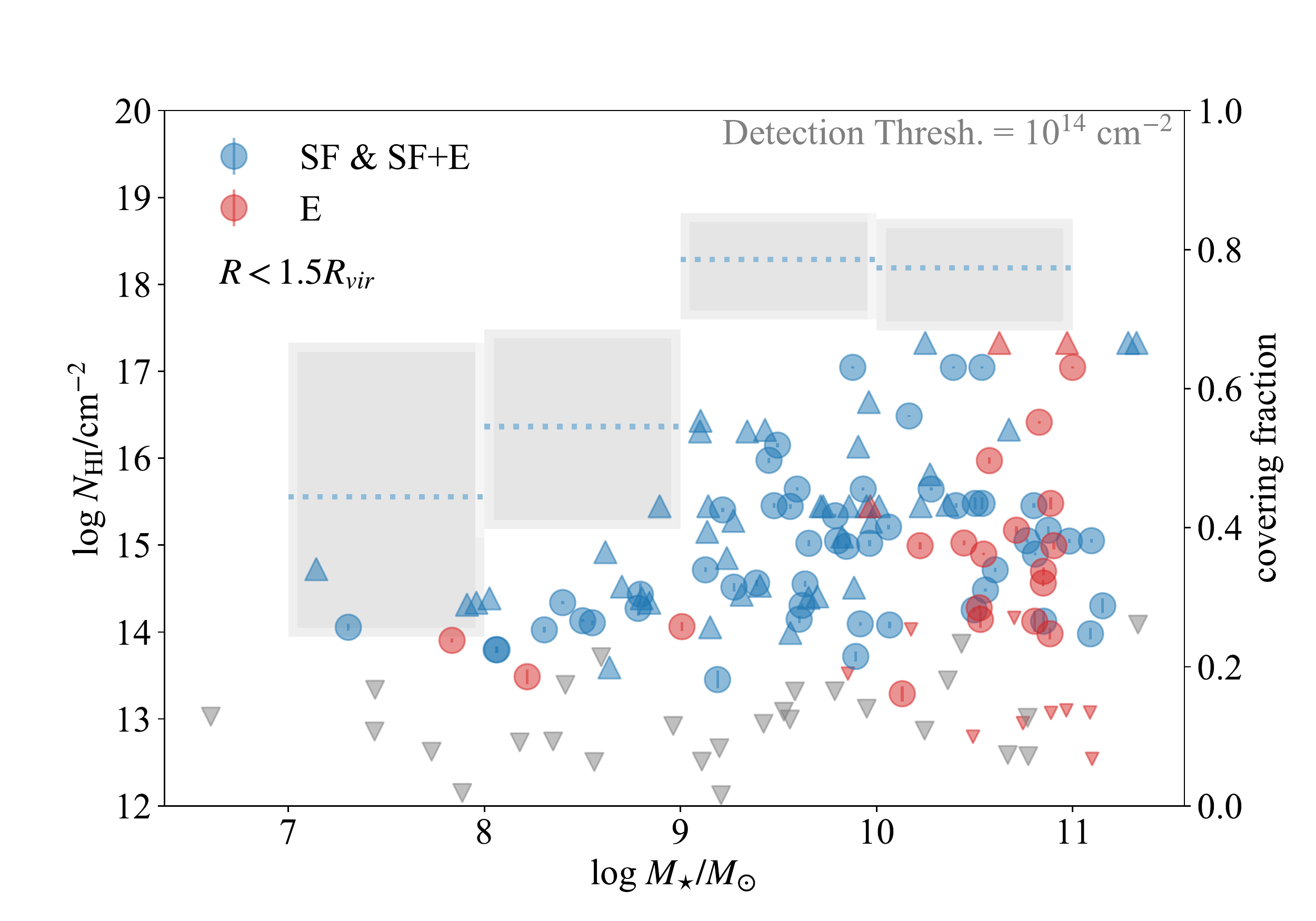} 
    \end{array}$
    \caption{Column density as a function of mass for $\rho < 300$ kpc (top) and $R < 1.5 R_{\rm vir}$ (bottom). CGM systems with saturated absorption are marked with upward facing triangles, while non-detections are displayed as lighter, downward facing triangles at their corresponding 2$\sigma$ upper limits. Circles represent CGM systems with measured $N_{\rm HI}$. Measured 1$\sigma$ uncertainties in the column density of the detected CGM systems are shown as lines inside the markers. Red markers indicate a spectroscopically-determined quiescent galaxy classification `E', while blue corresponds to those galaxies with emission lines present in their spectra, classified as  `SF' and `SF+E'. Our sample of quiescent galaxies predominately reside in the highest-mass bin. Covering fractions $f_c$ are plotted with respect to the right axes and are calculated without differentiating spectroscopic galaxy categories. The grey boxes correspond to the binomial confidence interval of the covering fraction (\NHI $> 10^{14}$ cm$^{-2}$) with the mean $f_c$ in each bin denoted with a dotted line. The column density increases as a function of mass while the covering fraction remains greater than $f_c > 0.5$ for galaxies with masses of $M_{\star} = 10^{8-11} M_{\odot}$.
    }
    \label{fig:HI_mass}
    \end{figure}

    This section presents an empirical analysis of the \ion{H}{1}-galaxy connection. In the following analysis we generally avoid differentiating the systems via their spectroscopic galaxy classifications due in part to the fact that we do not expect to see differences in covering fractions of \ion{H}{1} in star forming and quiescent galaxies \citep[]{thom12, keeney17} but include it in Figure \ref{fig:HI_mass} to illustrate this. Additionally, Figures \ref{fig:colormag_gr_mstars} and \ref{fig:mstars_vs_z} show that the galaxy spectral classification is correlated with the galaxy stellar mass, and thus we have relatively few galaxies with comparable masses but with differing SF classification. Future analyses that examine the metal ions present in the CGM will focus on galaxy spectroscopic type.
    
    In Figure \ref{fig:HI_mass}, we present the \ion{H}{1} column density (left axis) and covering fraction (right axis) as a function of stellar mass for galaxies within 300~kpc (top panel) and $1.5 R_{\rm vir}$ (bottom panel). The covering fraction for each bin is shown as a dotted line with the grey boxes corresponding to the 68\% binomial confidence intervals. At $R < 1.5 R_{\rm vir}$, $f_c$ remains consistent with $f_c \gtrsim 0.5$ at all ranges in mass. We also notice a trend of increasing covering fraction and column density as a function of galaxy stellar mass. 
    
    \begin{figure*}$
    \vspace{-0.1in}
    \begin{array}{c c}
    \includegraphics[scale=0.33]{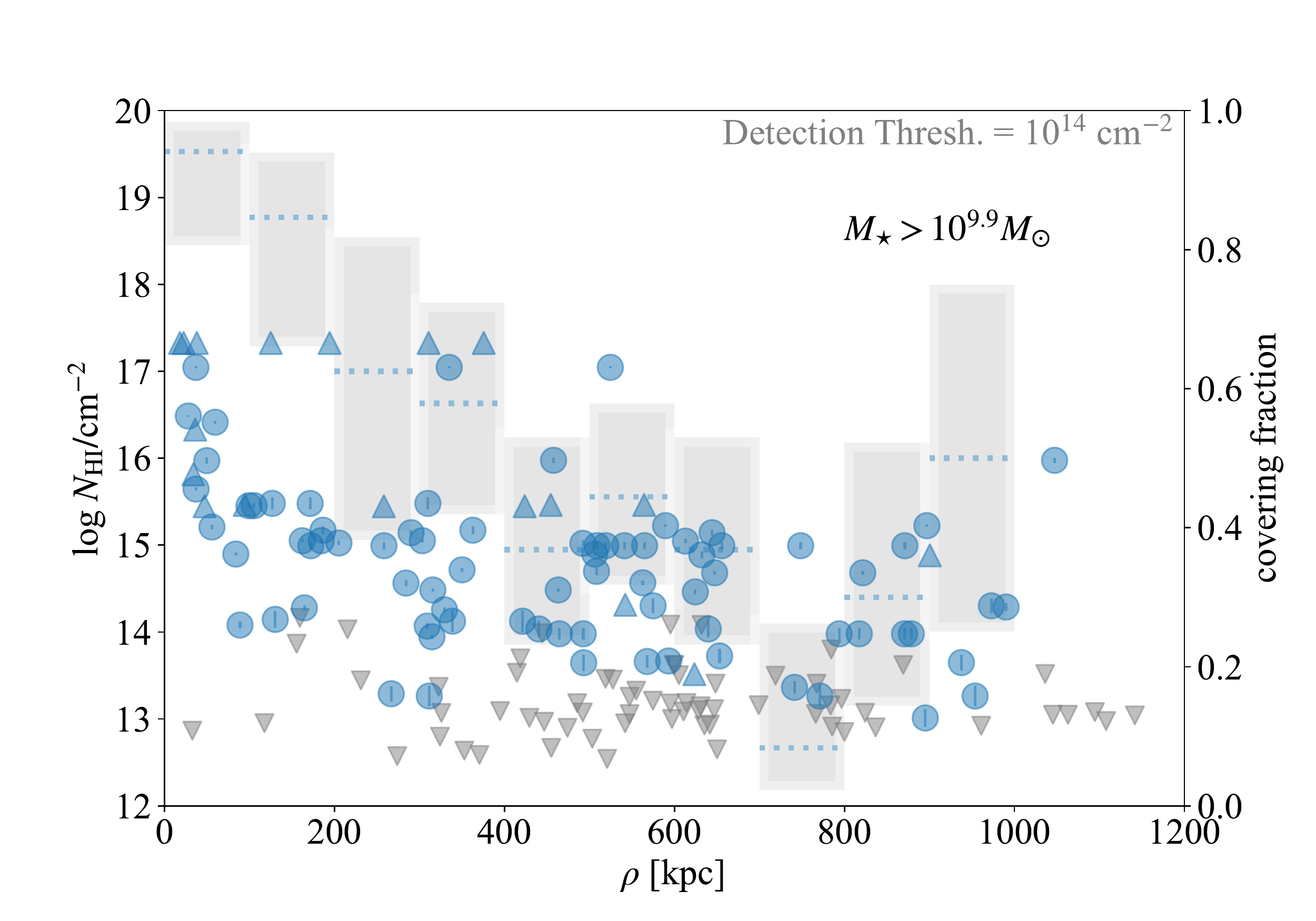} &
    \includegraphics[scale=0.33]{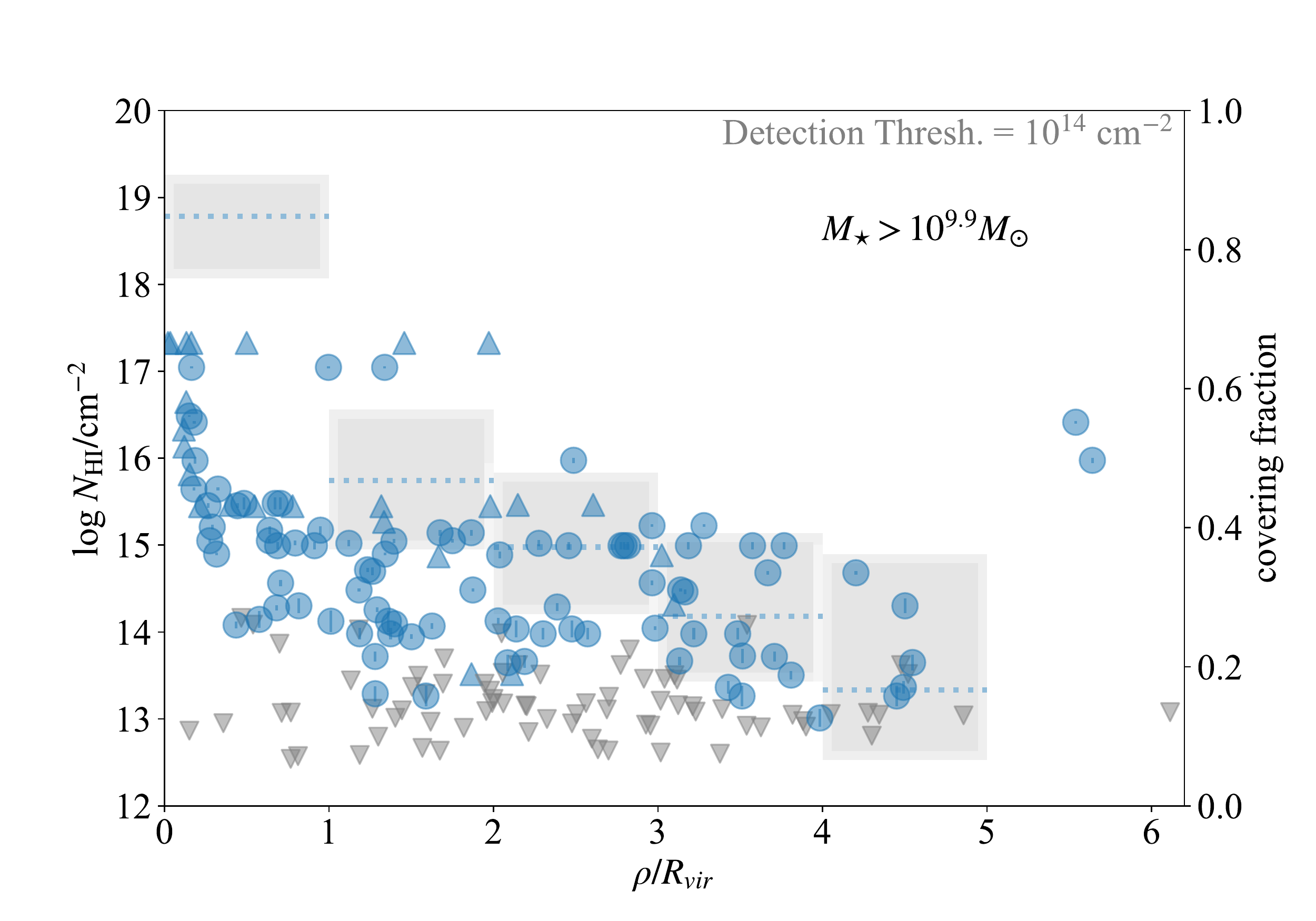} \\
    \includegraphics[scale=0.33]{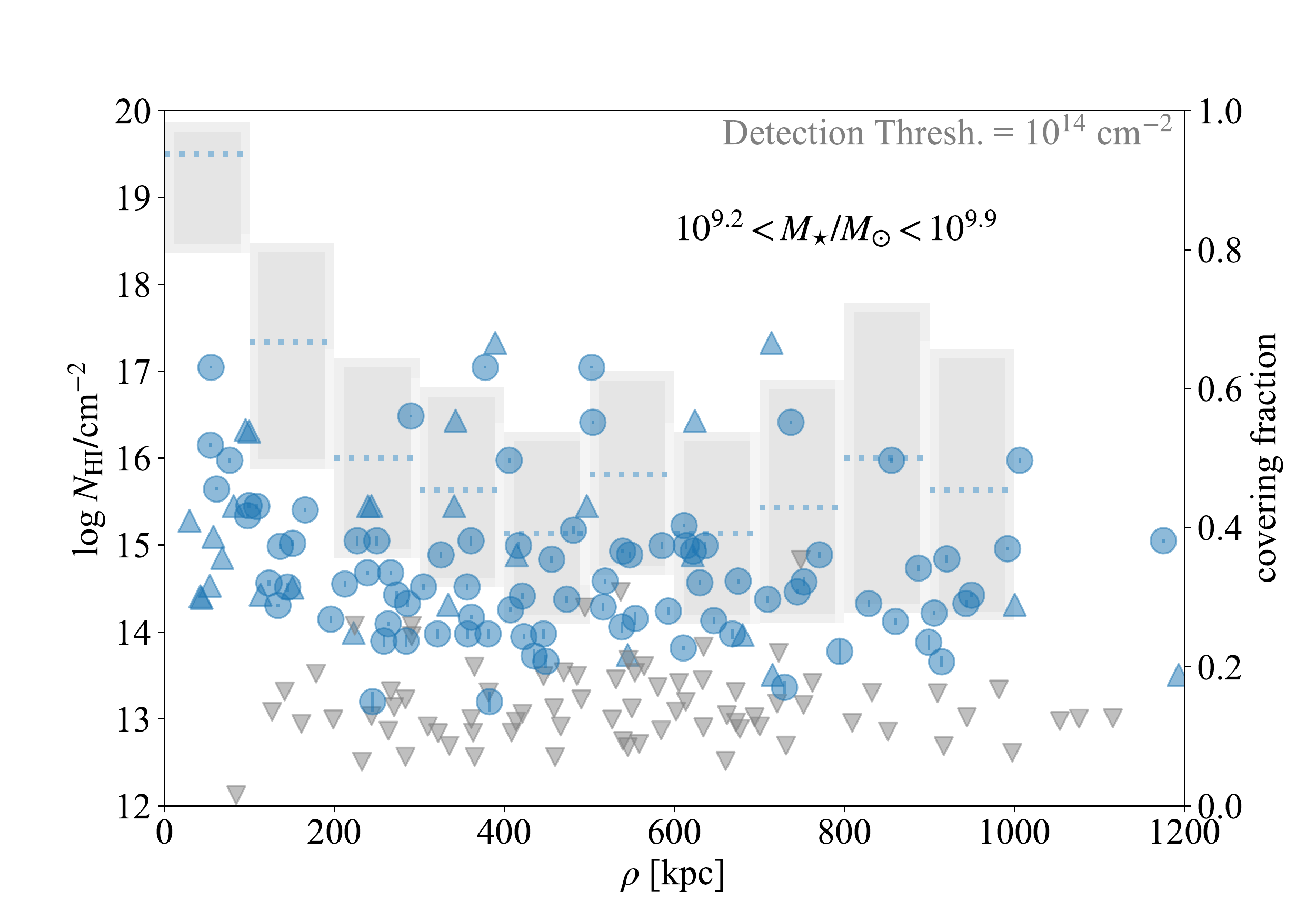} &
    \includegraphics[scale=0.33]{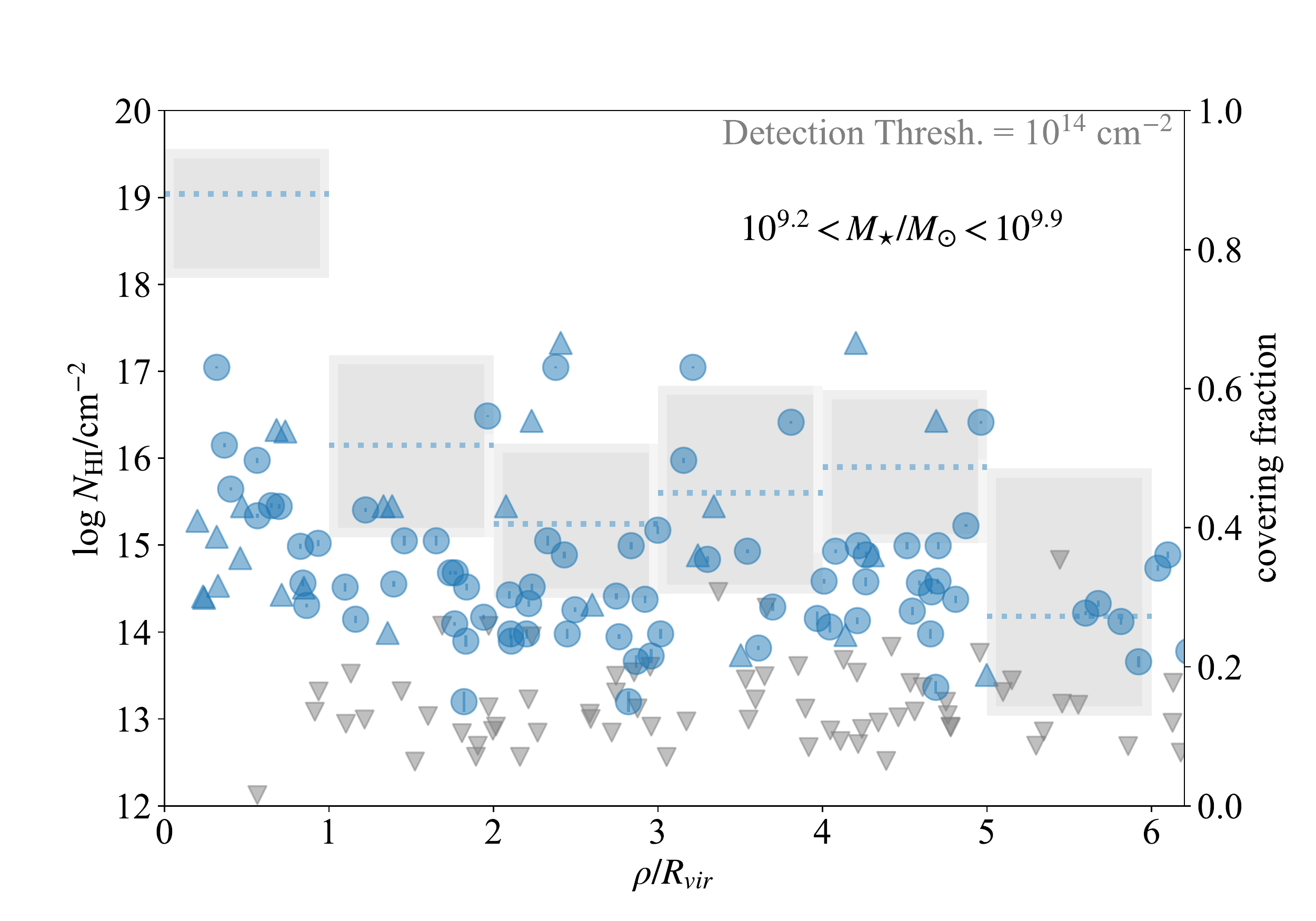} \\
    \includegraphics[scale=0.33]{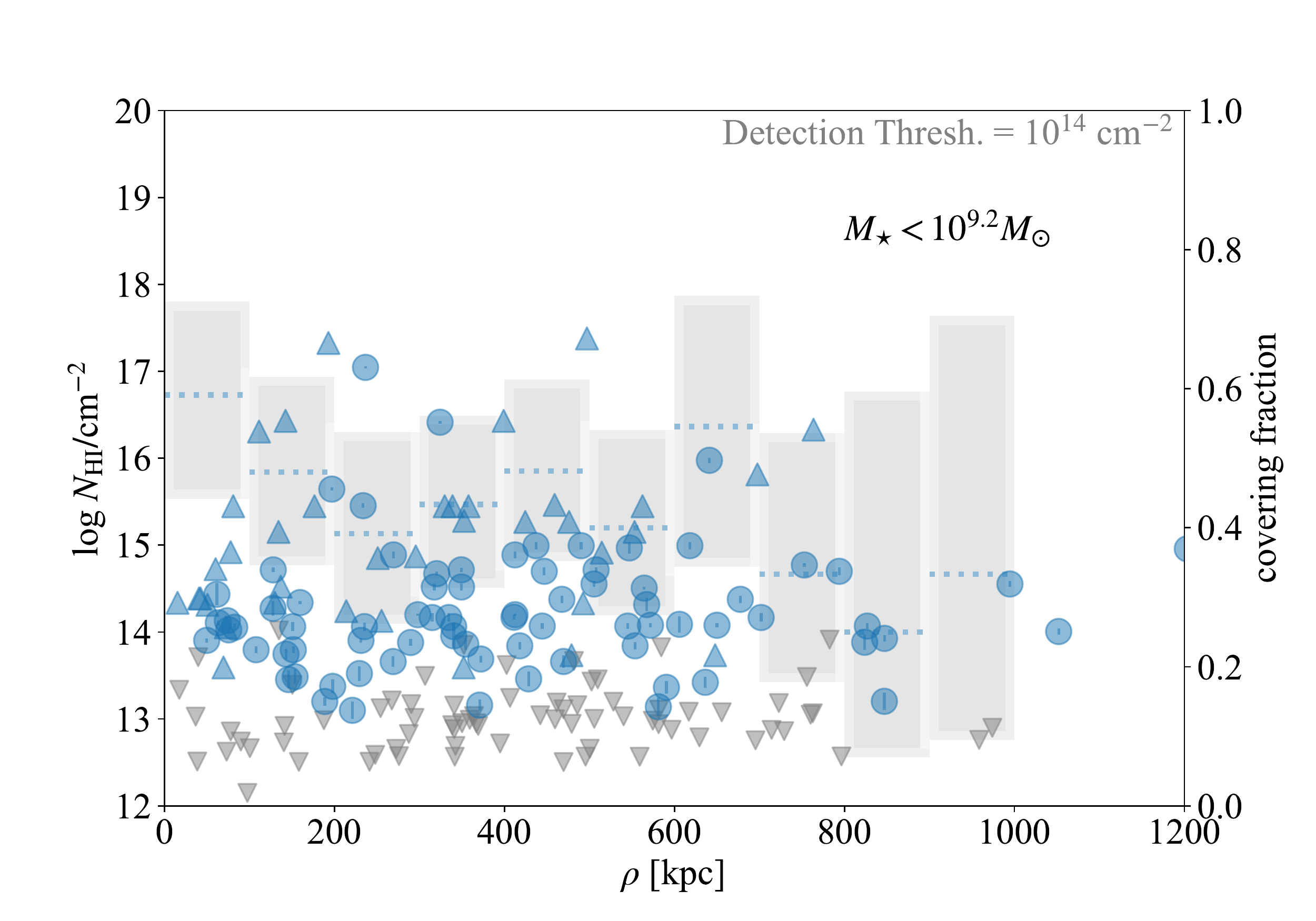}  &
    \includegraphics[scale=0.33]{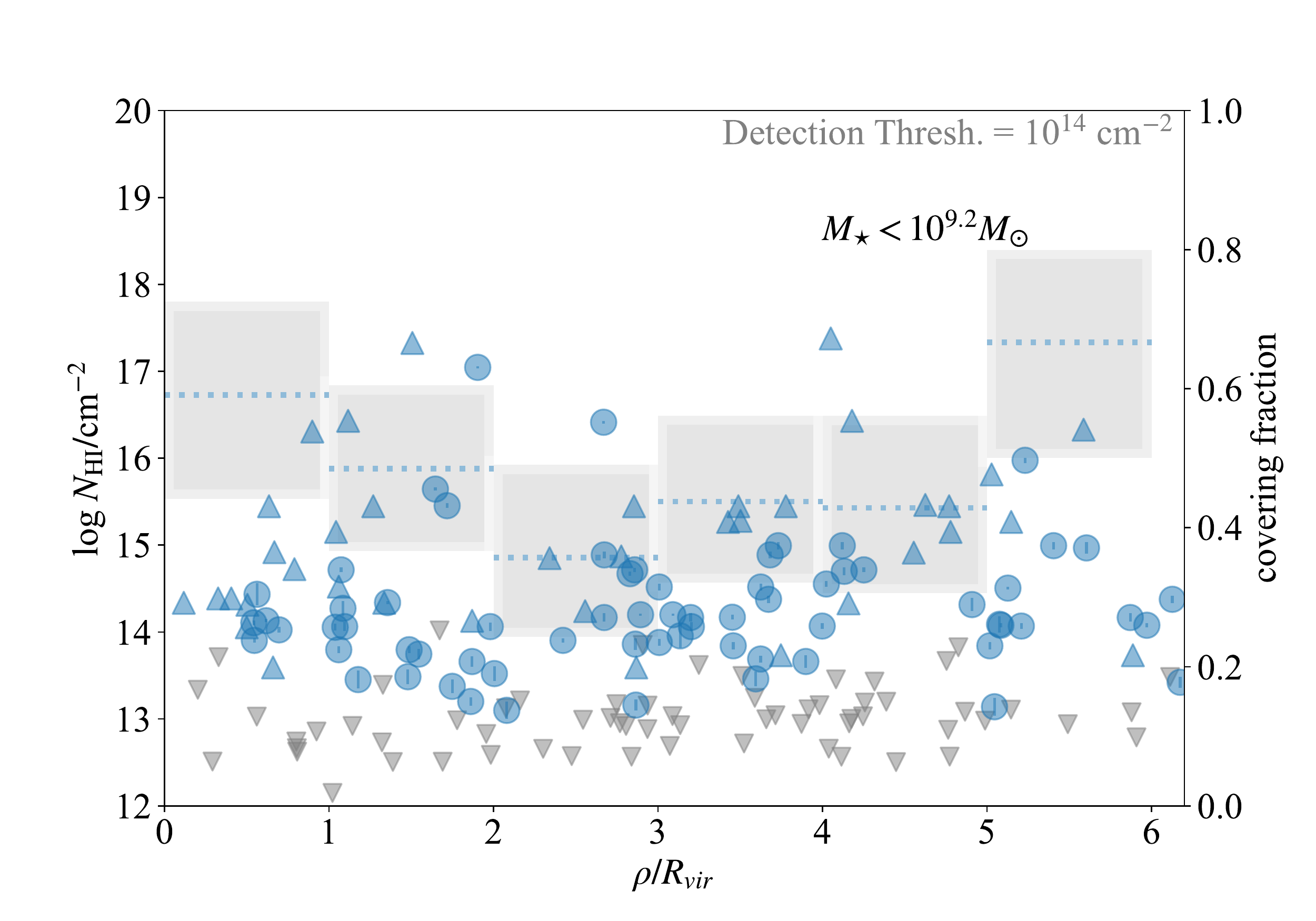}
    \end{array}$
    \caption{Column density \NHI as a function of impact parameter, $\rho$ (left) and $R_{\rm vir}$ (right). The panels are arranged vertically in order of decreasing mass. Galaxy-absorber systems with dark circles represent galaxies with \NHI detections in the corresponding QSO spectrum. Systems with dark, up-arrow symbols show our inferred lower limits due to saturation in $N_{\rm HI}$. Similarly, systems with gray, down-arrow symbols denote the $2\sigma$ upper limit on \NHI (non-detections). Measured $1\sigma$ uncertainties in the column density of the detected CGM systems are shown as grey lines but are smaller than the marker size in every case. \NHI covering fraction, $f_c$, corresponds to the right axes in bins of 100 kpc (left) and $1~R_{\rm vir}$ (right). The dotted lines represent $f_c$ assuming a detection threshold of \NHI $= 10^{14}$ cm$^{-2}$ with the 68\% binomial confidence interval shown as shaded gray regions about the mean $f_c$.
    %
    }
    \label{Fig:NHI_v_rho}
    \end{figure*}
    
    The \ion{H}{1} column densities and impact parameters are shown in Figure \ref{Fig:NHI_v_rho} as an anti-correlation of column density with increasing separation between the galaxy and QSO sightline. Figure \ref{Fig:NHI_v_rho} separates our galaxy sample into three mass bins, each containing approximately equal numbers of Galaxy-Absorber systems. The top panel contains 191 systems with $M_{\star} > 10^{9.9} M_{\odot}$. The middle panel also contains 191 systems with $10^{9.2} < 10^{10} M_{\star}/M_{\odot} < 10^{9.9}$ while the bottom panel corresponds to a sample of 190 systems with $M_{\star} < 10^{9.2} M_{\odot}$.
    
    We find that for the highest-and-intermediate mass galaxy samples, $f_c$ drops off monotonically but remains $\geq 50\%$ out to an impact parameter of $\rho \simeq $400~kpc, corresponding to $R\simeq 2R_{\rm vir}$ for these masses, before flattening out in the case of the intermediate mass sample. In the lowest-mass regime, we find lower ($f_c \sim 60$\%) covering fractions at the smallest separations. We also observe a shallower profile in covering fraction out to $R \simeq 2R_{\rm vir}$, beyond which the covering fraction remains elevated and consistent with $\sim 50\%$ out to 600~kpc, or $5R_{\rm vir}$.
    
    The extended distribution seen in the lower two panels, where the covering fraction remains elevated, could imply that the galaxies are located inside (or close to) the halos of other (massive) galaxies \citep[e.g.][]{burchett16}. In an upcoming \cgmsq~ paper, we plan to explore the impact of galaxies' environments on the properties of their CGM. Briefly, we find that 56\% (243/435) of galaxies with $M_{\star} < 10^{10} M_{\odot}$ have $M_{\star} > 10^{10} M_{\odot}$ neighbors within 300 kpc and 1000 km/s of the sightline. We observe 75\% (86/121) of galaxies with $M_{\star} > 10^{10} M_{\odot}$ have neighbors above this mass threshold within the same physical window, consistent with the fact that larger galaxies are more highly clustered. Furthermore, we generally see that low mass galaxies with nearby massive neighbors tend to have elevated HI covering fractions out to  $\sim$~2-3~$R_{\rm vir}$ compared to galaxies with no detected massive neighbors. In tandem, these effects support our conclusion that the elevated covering fractions out to large impact parameters for our $M_{\star} < 10^{10} M_{\odot}$ sample are consistent with environmental effects.
    
    Our results are consistent (within 2$\sigma$) with those presented by \cite{wakker09} who find that \lya absorbers at $z<$ 0.017 with equivalent width $>$ 50 m\AA~ (\NHI $\simeq$ $10^{13}$ cm$^{-2}$) have covering fractions of 100\% at $\rho<400$ kpc. Similarly, \cite{prochaska11b} find high covering fractions out to $\rho = 300 h^{-1}_{72}$ kpc for absorbers with equivalent widths $>$ 30 m\AA~. This observed enhancement of covering fraction in the low-mass sample could also be due in part to the fact that our CGM systems were defined to have separations $|\delta v| < 500$ km s$^{-1}$, which is around the expected escape velocities of high-mass galaxies but is more likely to encompass unassociated absorbers in the low mass sample that trace the cosmic web. However, low-mass galaxies have smaller escape velocities and shallower potential wells, and thus would likely exhibit gas being ejected at larger velocities.  There are clearly many competing effects in this mass range. 
    
\begin{figure}[h!]
    \begin{centering}
    \hspace{-0.15in}
    \includegraphics[width=1.1\linewidth]{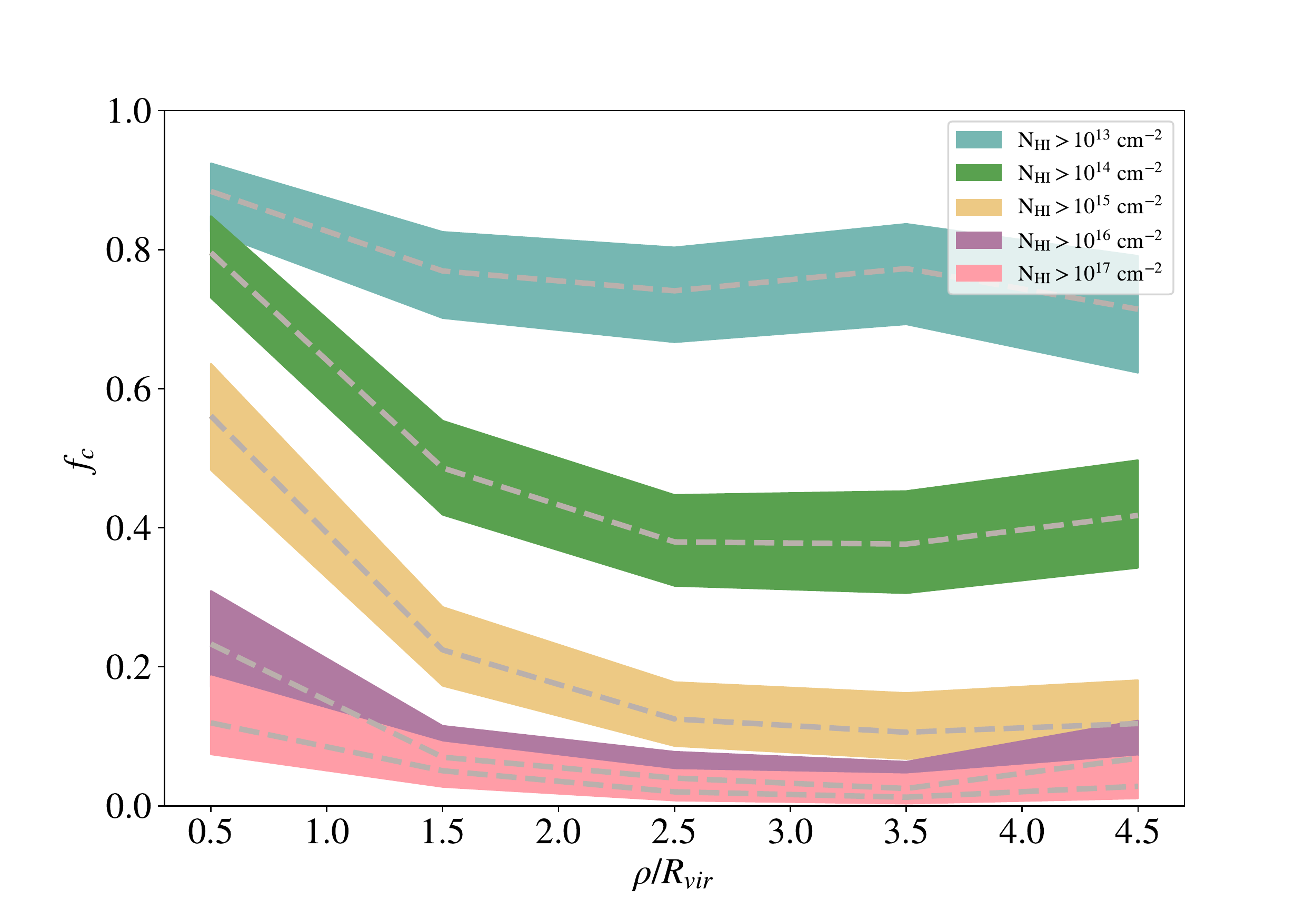}
    \end{centering}
  
    \caption{Covering fraction of \ion{H}{1} as a function of $\rho/R_{\rm vir}$ for column density thresholds of \NHI $> 10^{13}$ cm$^{-2}$ (blue), \NHI $> 10^{14}$ cm$^{-2}$ (green), \NHI $> 10^{15}$ cm$^{-2}$ (yellow), \NHI $> 10^{16}$ cm$^{-2}$ (purple), and \NHI $> 10^{17}$ cm$^{-2}$ (pink). Shaded regions represent the 1-$\sigma$ (68\%) binomial confidence intervals. Here we connect the center of the radial bins to highlight the difference in the distributions. We see that for column densities less than $10^{14}$ cm$^{-2}$ show little correlation with galaxies. The covering fraction at $R < R_{\rm vir}$ in for the highest column densities (\NHI $> 10^{15}$ cm$^{-2}$) never gets higher than 0.7.}
    \label{fig:covfrac_threshold}
    \end{figure}

    Figure~\ref{fig:covfrac_threshold} shows the covering fraction and confidence intervals for the total sample at \NHI $\in 10^{13-15}$ cm$^{-2}$ (blue, green, yellow). For the lowest column density threshold in the low-mass sample, the \ion{H}{1} covering fractions remain elevated at $\sim$80\% out to at least 4$R_{\rm vir}$. This signal must be dominated by galaxy-galaxy clustering as the $f_c$ of random incidence in a velocity window at the mean redshift is $f_c = 0.14$ for \NHI $> 10^{13}$ cm$^{-2}$. (see Section \ref{section:clustering}). In contrast, stronger absorbers preferentially occur at smaller impact parameters. For example, it is at log \NHI $\gtrsim$ 14.5 that the covering fraction drops below 50\%  by $R = R_{\rm vir}$. 
    
    \subsection{\ion{H}{1} velocity offsets}
    
    Here, we examine the kinematics of the \ion{H}{1} absorption in our \cgmsq sample of $z<$ 0.481 absorbers that are associated with galaxies. Velocity distributions of CGM absorbers quantify the amount of material gravitationally associated with the assumed host halo. For simplicity and clarity, we trim the sample to those galaxies closest to the absorber impact parameter ($\rho$), thus each absorber is associated with only one galaxy. We are left with a sample of 171 unique galaxy-absorber pairs at $z<$ 0.48 that were detected with a signal $> 2\sigma$. 
    
\begin{figure}
    $\begin{array}{cc}
        \includegraphics[scale=0.38]{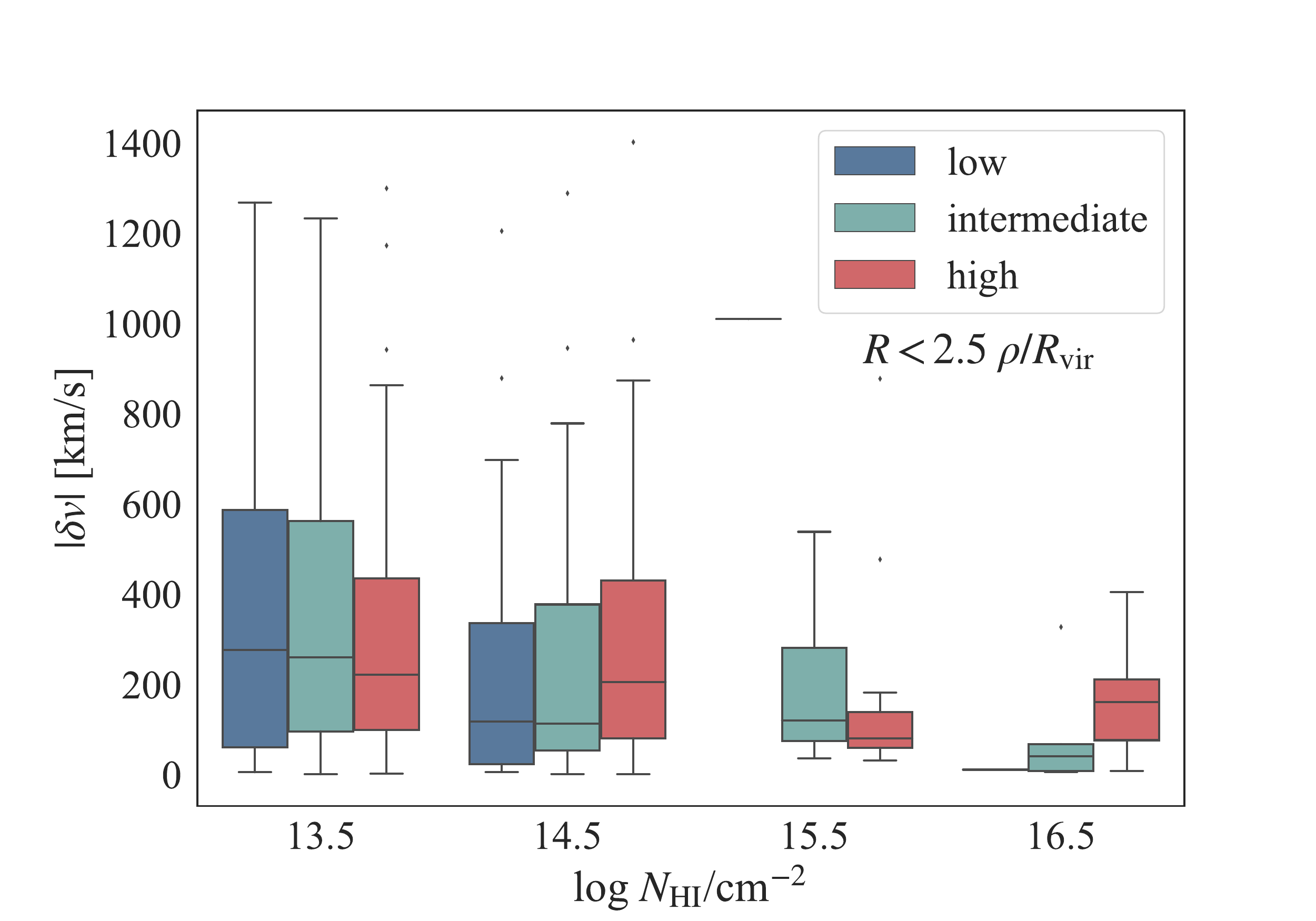} 
    \end{array}$
    
    \caption{The distribution of absolute velocities as a function of \NHI displayed in a box and whisker plot for each mass sample. The boxes display the quartiles of the distribution centered at each bin spanning 1 decade in column density while the whiskers extend to show the rest of the distribution of the bins. Outliers are defined as points that lie outside 1.5 times the innerquartile range and are displayed as small diamonds. Each bin is split into a high (blue), intermediate (green) and low-mass (red) sample. We see a strong anticorrelation between velocity spread and column density. The single line seen at higher column densities in the low-mass sample indicate there are is only one point in each column density bin. This highlights how rare high-column density systems are around low-mass galaxies.}
    \label{fig:vel_box}
    \end{figure}

    In Figure \ref{fig:vel_box}, we show the distribution of velocity centroids of the 416 detected absorption components associated with these 181 CGM systems in the rest-frame of the galaxy systemic redshift as a function of column density. The boxes show the quartiles of distribution in velocity while the whiskers show the extent of the distribution. Outliers beyond 1.5 times the innerquartile range are displayed as points. We see that systems with \NHI$< 10^{14.0}$ cm$^{-2}$ have a higher median and larger spread in velocity. The component velocity centroids do not exhibit any clear trends with impact parameter. Relevant to this discussion, we recall that absorption components are required to lie within $\pm$1000 km s$^{-1}$ of each other. The only velocity constraint imposed with respect to the galaxy is that at least one of these absorption components must lie within $\pm$ 500 km s$^{-1}$ of a \cgmsq galaxy systemic redshift in order to be classified as a CGM absorption system.
    
    We find that $>$ 53\% of the detected \ion{H}{1} components are located within $\pm$ 250 km s$^{-1}$ of the galaxy systemic redshift, independent of \ion{H}{1} column density. Furthermore, only 27\% of all absorption components lie at $|\delta v|$ $>$ 500 km s$^{-1}$. Finally, approximately 83\% of the total \ion{H}{1} column density in the CGM$^{2}$ survey lies within  $\pm$250 km s$^{-1}$ of the associated host galaxy, consistent with the results of COS-Halos \citep{tumlinson13}. Some absorption systems show a total extent $>$ 500 km s$^{-1}$, which likely captures both bulk motions of galaxies and possible ``missassociations" due to incompleteness our galaxy survey, especially at magnitudes $>$ 23. 
    
    Were absorption components distributed uniformly in velocity space relative to the galaxies, limited only by the 1000 km s$^{-1}$ window selection function, we would not expect to find such a high concentration of absorbers at low $\delta v$. To first order, Figure \ref{fig:vel_box} shows that our CGM survey is capturing gas that is mostly not exceeding the escape velocity of its host halo, where the escape velocity of an $M_{\star}$ $\approx$ 10$^{9.5}$ $M_{\odot}$ galaxy at $R_{\rm vir}$ is $\sim$250 km s$^{-1}$. There is very clearly a gravitational association between the galaxies and the absorption at higher column densities. The component velocity centroids do not exhibit a clear trend with impact parameter but we find that the median velocity for components at $\rho > R_{\rm vir}$ to be 233 km s$^{-1}$ with a standard deviation of 453 km s$^{-1}$ while components at $\rho < R_{\rm vir}$ have a median velocity of 185 km s$^{-1}$ with a standard deviation of 354 km s$^{-1}$. In this unique galaxy-absorber pairing framework, 80\% of the absorption components with \NHI $>$ $10^{15}$ cm$^{-2}$ lie within 1 $R_{\rm vir}$. We caution that for this kinematic analysis we paired specific galaxies with absorbers on the criterion that they were the closest in physical separation from the absorption sightline. Without this bias, the trend remains, however. 

   Lastly, Figure \ref{fig:vel_box} also shows the column-density dependence of the absorption component velocity distribution. Here, we see that absorption systems with \NHI $> 10^{14}$ cm$^{-2}$ concentrate more strongly at low $|\delta v|$ than weaker systems, and there is a clear column density dependence to the overall velocity concentration at $|\delta| v < 250$ km s$^{-1}$.  Counting by column density we find that 53\% of the total fitted column density lies within $\pm$ 250 km s$^{-1}$. In contrast, we find that only 44\% of the \NHI $< 10^{14}$ cm$^{-2}$ absorption components lie at $|\delta v| <$ 250 km s$^{-1}$, while 67\% of the \NHI $> 10^{14.0}$ cm$^{-2}$ absorption components lie at $|\delta v| < 250$ km s$^{-1}$. This concentration increases with increasing column density. For a limit of \NHI $= 10^{15}$ cm$^{-2}$, the percentages of components within $|\delta v| <$ 250 km s$^{-1}$ shift to 50\% and 86\% for the low and high column density thresholds, respectively. Similarly, the high velocity components tend to have low column density. 27\% of all absorption components lie at $|\delta v| > 500$ km s$^{-1}$, but less than 5\% of those components have \NHI $>10^{14}$ cm$^{-2}$. Generally, there is no systematic effect preventing high column density components from appearing at higher velocities, so this trend captures an important characteristic of the CGM.

    \section{HI-Galaxy Clustering: $R^{14}_{\rm CGM}$}\label{section:clustering}
    
    \subsection{Setup}
    In order to quantitatively describe the radial dependence of the CGM and estimate its extent, we perform an absorber-galaxy cross-correlation analysis. We aim to measure the excess probability of detecting an absorber given the proximity of a galaxy over the proximity-agnostic average rate. Our analysis of the HI-galaxy clustering is similar to the one developed by \cite{hennawi07} and follows more closely the analysis by \cite{prochaska19}. We define the 3D cross-correlation function, $\xi_{ag}(r)$ as
    
    \begin{equation} \label{xi}
        \xi_{ag}(r) = \left( \frac{r}{r_0}\right)^{-\gamma}.
    \end{equation}
    In order to determine the best fitting parameters, $r_{0}$ and $\gamma$, we define a likelihood function as
    \begin{equation} \label{eq_likelihood}
        \mathcal{L} = \prod_{i} P_{i}^{\rm hit}(r, z) \prod_{j} P_{j}^{\rm miss}(r, z),
    \end{equation} 
    where $P^{\rm hit}$ is defined to be the probability of detecting one or more \ion{H}{1} systems and $P^{\rm miss}$ is the probability of detecting none. This probability has both a radial and redshift dependence.  An absorber is considered a ``hit" if it falls within our window of $\delta v = \pm 500$ km s$^{-1}$ of a galaxy and we measure a column density above a threshold $N_{\rm HI}^{\rm thresh}$. 
    We define $P^{\rm miss}$ to be the probability of observing zero events from a Poisson distribution where the rate is the number of events expected from the average density of absorbers and our clustering term:
    
    \begin{equation}
        P^{\rm miss} = \exp{(-[1+ \chi_{\perp}(r)]\biggl< \frac{d\mathcal{N}}{dz} \biggr> \delta z)}.
    \end{equation}
    Here $\langle d\mathcal{N}/dz \rangle \delta z$ is the mean number of absorbers in a window of redshift $\delta z$. $[1+ \chi_{\perp}(r)]$ represents an excess in the number of absorbers due to clustering. This boost due to clustering can be expressed in terms of the 3-dimensional correlation function as
    \begin{equation} \label{chiperp}
    \begin{split}
        \chi_{\perp}(r) & = \frac{1}{V} \int_V \xi_{ag}(r) dV \\
        & \approx \frac{aH(z)}{2\delta v} \int_{-\delta v/[aH(z)]}^{-\delta v/[aH(z)]} \left(\frac{\sqrt{R_{\perp}^2 + R_{\parallel}^2}}{r_0}\right)^{-\gamma} dR_{\parallel}
    \end{split}
    \end{equation}
    where we are integrating Equation (\ref{xi}) along the length of a cylinder of length $2\delta v / a H(z)$. Here $a = 1 / (1 + z)$ and $H(z)$ is the Hubble parameter.
    The probability of a ``hit," $P^{\rm hit}$, is the complement of the probability of a ``miss": $P^{\rm hit} = 1 - P^{\rm miss}$. $P^{\rm hit}$ is equivalent to the covering fraction $f_c$.
     The covering fraction of a random sightline is then:
    \begin{equation}
        f_c = 1 - \exp{(\biggl< \frac{d\mathcal{N}}{dz} \biggr> \delta z)}.
    \end{equation}
    
    We take $\langle d\mathcal{N}/dz \rangle$ from \cite{danforth16}, who measured the occurrence of 5138 individual extragalactic absorption lines of \ion{H}{1} in 82 QSO/AGN spectra at redshifts $z_{\rm AGN} < 0.85$. The occurrence rate of \ion{H}{1} absorbers $d\mathcal{N}/dz$ is expressed with a functional form as follows:
    \begin{equation} \label{eq_dndz}
        \frac {d\mathcal{N}(\rm N_{\rm HI} \geq \rm N_{\rm HI}^{\rm thresh} ,z)}{dz} =
        C_0 (1+z)^{\gamma},
    \end{equation}
    with $C_0 = 16$ and $\gamma = 2.3$. This measurement is valid only for absorbers with $\rm N_{\rm HI}^{\rm thresh} \geq 10^{14}$ cm$^{-2}$, which coincides with the threshold in our definition of ``hits" and ``misses."
    
    In order to estimate $r_0$ and $\gamma$ in our 3D correlation function, Equation (\ref{xi}), we follow a Bayesian approach. The posterior probability function can be defined as

    \begin{multline}
        p(r_0, \gamma | \left\{k_i, r_i, z_i \right\}_{i=1}^{N})
        \propto p(r_0, \gamma) p({\bf k} | {\bf r}, {\bf z}, r_0, \gamma).
    \end{multline}
    where $k_i\in \{0, 1\}$ specifies whether system $i$ is a ``hit" or a ``miss" and $p({\bf k} | {\bf r}, {\bf z}, r_0, \gamma)$ is the likelihood function $\mathcal{L}$ defined in Equation (\ref{eq_likelihood}). We define the priors as follows:
    \begin{equation}
        p(r_0) = \left \{\begin{array}{ll}
                1/10 \,, & \mbox{if} \, 0 < r_0/\rm Mpc \rm < 10 \\
                0 \,, & \mbox{otherwise}
            \end{array}
            \right .
    \end{equation}
    and
    \begin{equation}
       p(\gamma) = \left \{\begin{array}{ll}
                \mathcal{N}(\mu=1.6, \sigma=1) \,, & \mbox{if} \,\gamma > 0 \\
                0 \,, & \mbox{otherwise}
            \end{array}
            \right .
    \end{equation}
    where $r_0$ is measured in $h^{-1}_{68}$ comoving Mpc and $\mathcal{N}(\mu, \sigma)$ is the normal distribution. These priors were chosen based on physical arguments and previous results on absorber-galaxy clustering  \citep[e.g.][]{tejos14}. We used the Markov Chain Monte Carlo (MCMC) sampler \texttt{emcee} \citep{emcee} to generate samples from the posterior probability distribution function over $r_0,\gamma$.

    The data were cut as in the previous analysis with a redshift cut of $z < 0.481$ and subdivided into the three equal sized mass samples as before with identical priors used for each sample. 
    
    \subsection{HI-Galaxy Clustering Results}
\begin{figure*}
    $
    \begin{array}{c c}
    \includegraphics[scale=0.3]{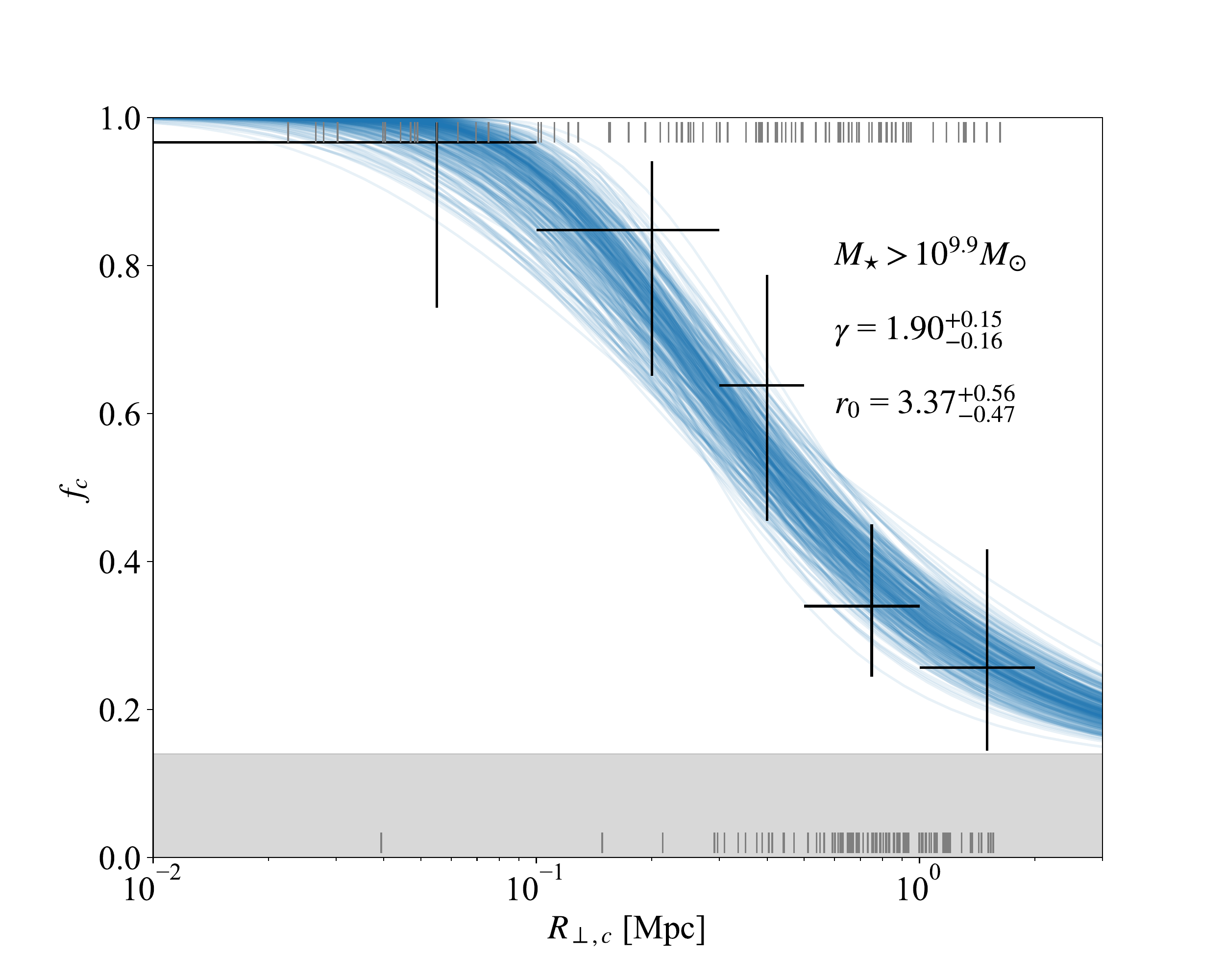} &
    \includegraphics[scale=0.3]{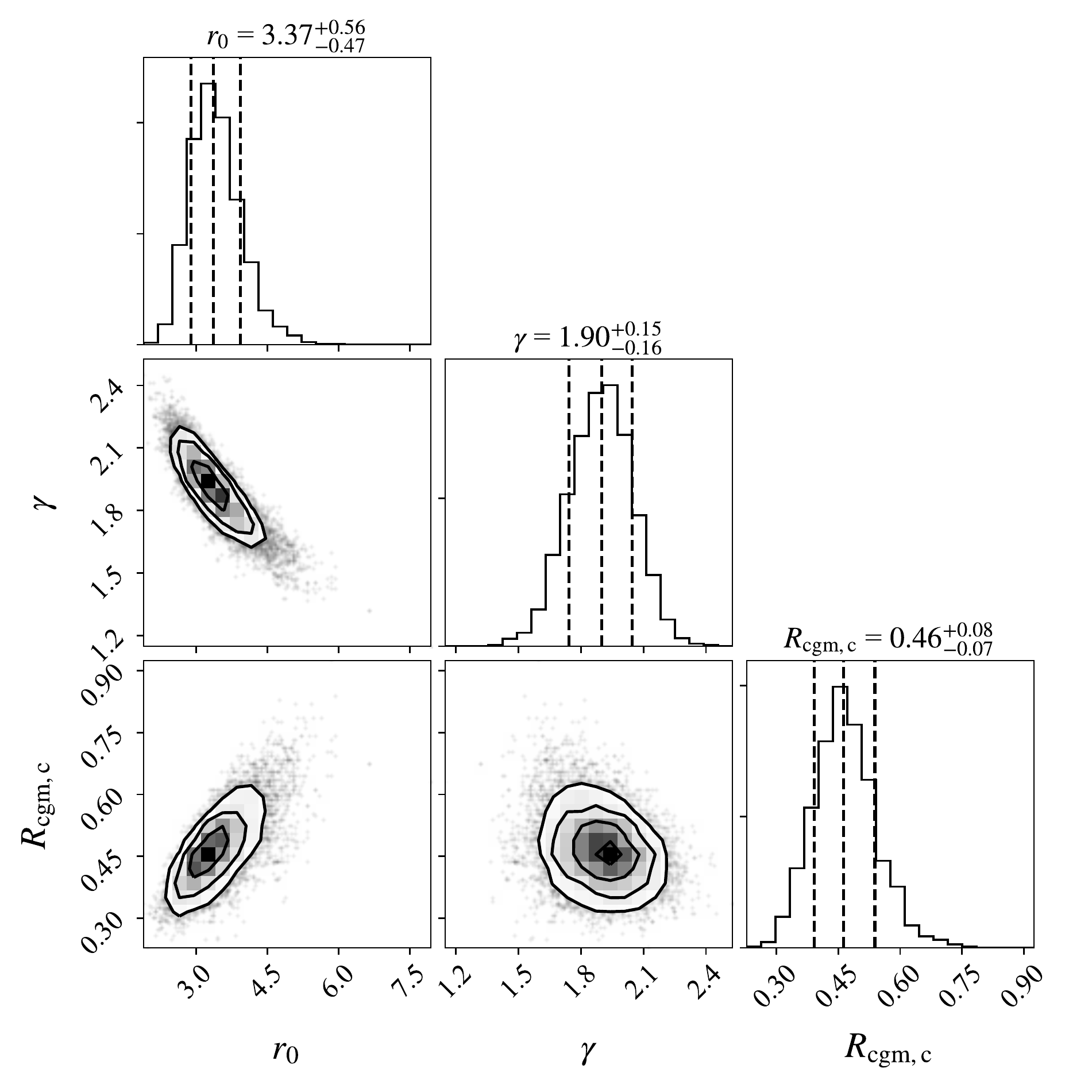} \\
    \includegraphics[scale=0.3]{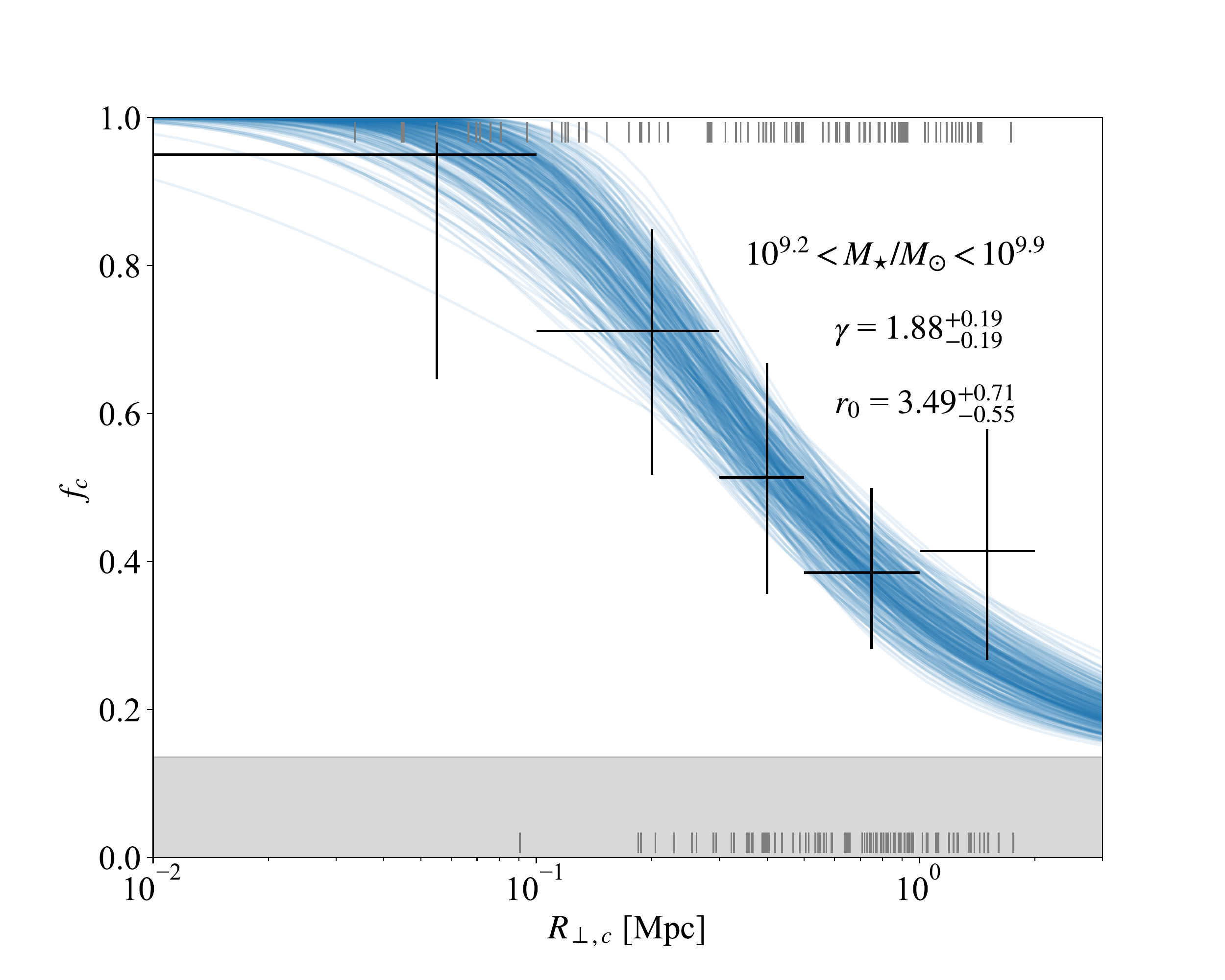} &
    \includegraphics[scale=0.3]{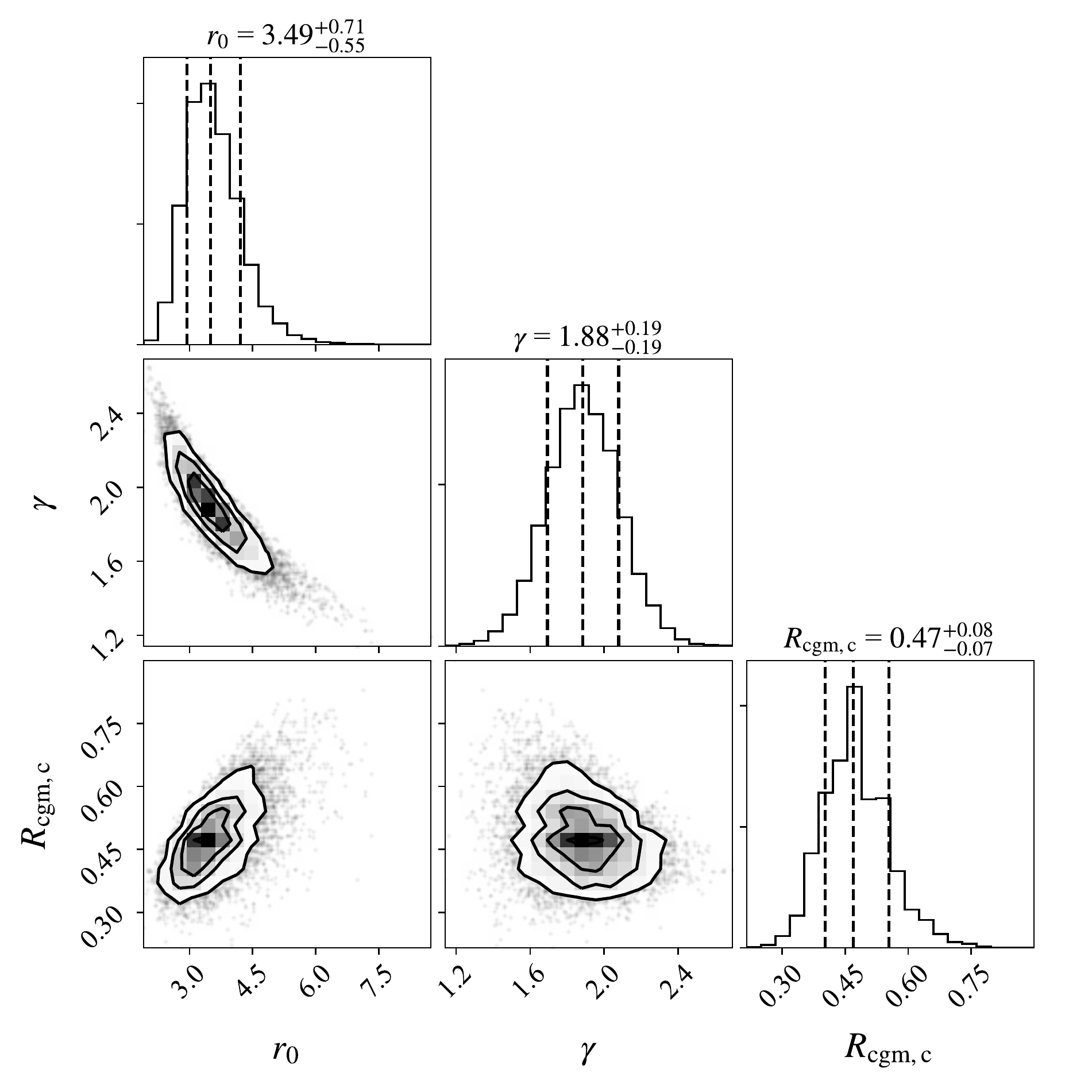} \\
    \includegraphics[scale=0.3]{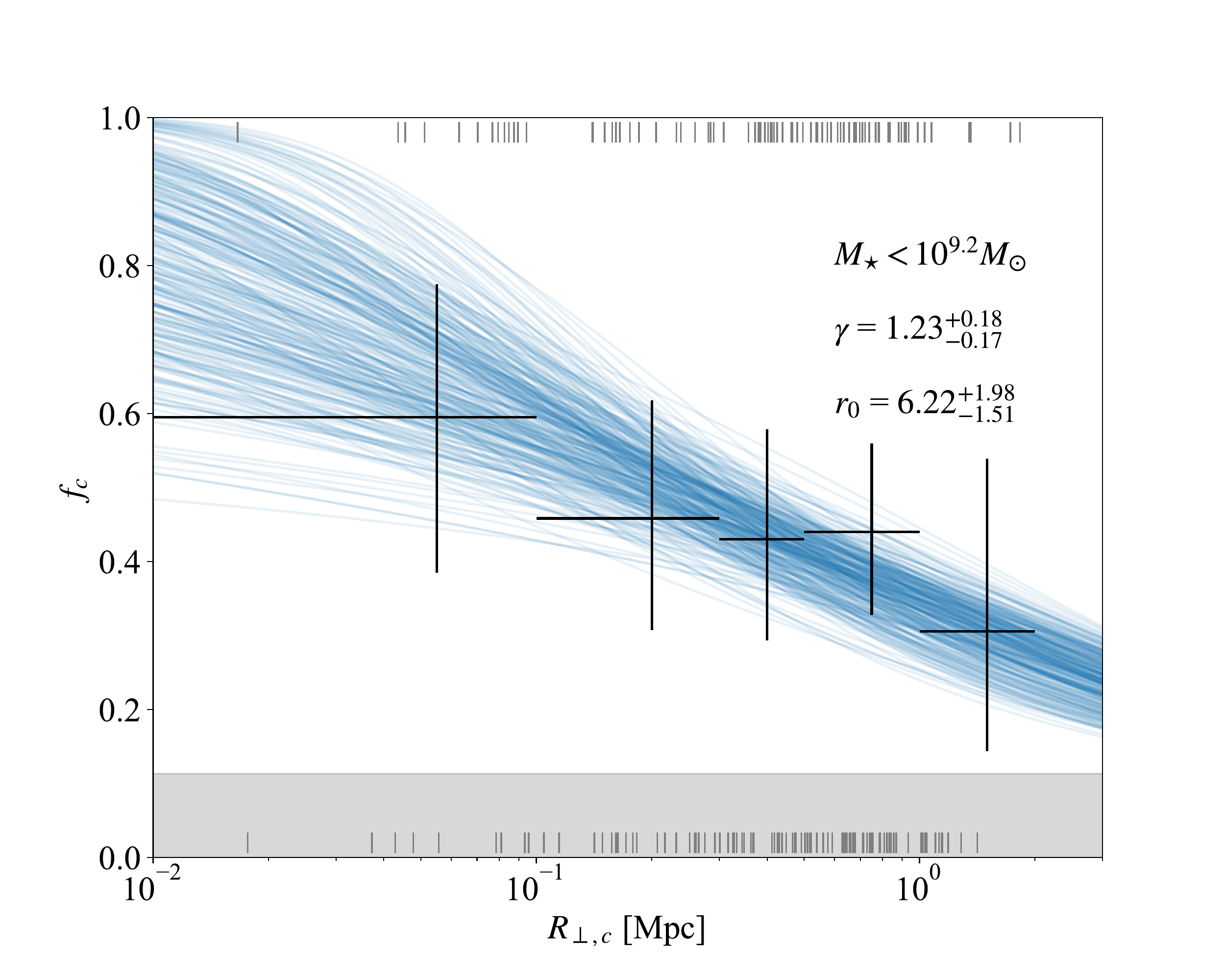} & 
    \includegraphics[scale=0.3]{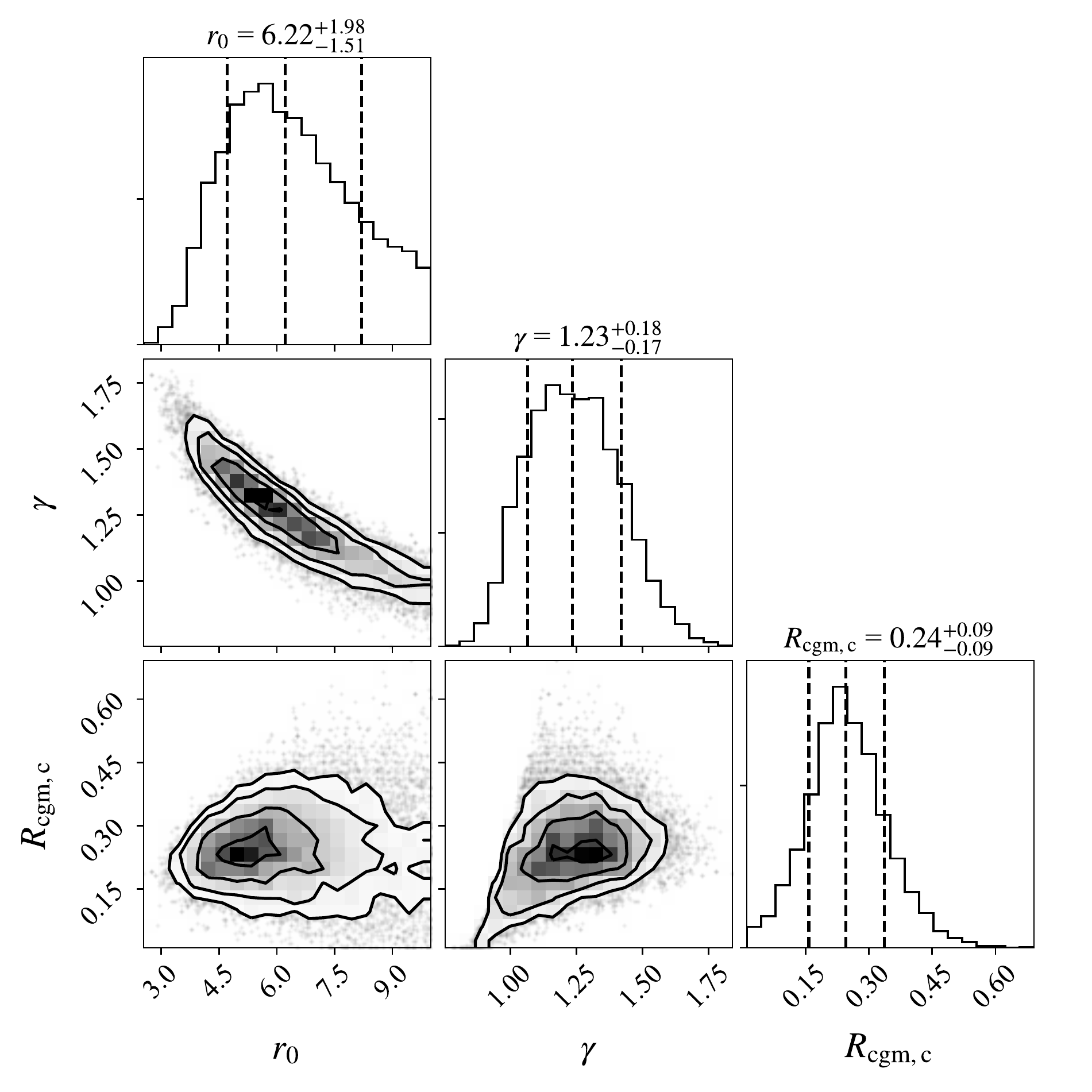} 
    \end{array}$
    %
    \caption{Left: covering fraction of \ion{H}{1} gas with \NHI $>10^{14}$ cm$^{-2}$ as a function of the physical impact parameter in comoving Mpc for galaxies in order of decreasing mass. The black crosses are binned evaluations of the covering fraction in each bin with 95\%  binomial confidence limits on covering fraction. The small grey ticks at the top and bottom of the figures indicate impact parameter for those systems that were \textit{hits} (top) and \textit{misses} (bottom). Beneath the black crosses are samples drawn from the posterior distribution showing the range in the $\gamma$--$r_0$ parameter space. The grey boxes on the bottom of the plots on the left indicate the covering fraction for sightlines taken at random, $\langle d\mathcal{N}/dz \rangle \delta z$. The plots on the right further illustrate the one and two dimensional projections of the posterior probability distributions of our parameters as well as indicate the median values for each. The highest mass sample contains 191 CGM systems at mean redshift of $\bar{z}=0.34$ and a random sightline covering fraction of $f_{c,\rm rand} = 0.14$. The intermediate-mass sample contains 191 CGM systems at a mean redshift of $\bar{z}=0.33$ and a random covering fraction of $f_{c,\rm rand} = 0.14$. The lowest mass sample contains 190 galaxies at a mean redshift of $\bar{z}=0.26$ and a random covering fraction of $f_{c,\rm rand} = 0.14$. Using the samples from the posterior distributions in $r_0$ and $\gamma$ we calculate $R_{\rm CGM,p}^{14} = 346_{-53}^{+57}$ kpc, $R_{\rm CGM, p}^{14} = 353_{-50}^{+64}$ kpc and  $R_{\rm CGM, p}^{14} = 177_{-65}^{+70}$ kpc in order of decreasing mass samples. The extent of the CGM drops off in the lowest mass sample.
    }
    \label{fig:fig_mcmc}
    \end{figure*}

    Figure \ref{fig:fig_mcmc} illustrates the results of the MCMC parameter estimation. The plots on the left show the covering fraction as a function of the perpendicular separation in comoving Mpc. 
    
    The grey boxes on the bottom of the plots on the left indicate the covering fraction for sightlines taken at random, $f_{c}^{rand} = 1 - \exp[\langle d\mathcal{N}/dz \rangle \delta z]$ at the mean redshift of the sample, $\bar{z} = 0.34$, $\bar{z}=0.34$ and $\bar{z}=0.26$ for the high, intermediate and low-mass samples, respectively. The plots on the right further illustrate the marginal distributions of the posteriors of our parameters as well as indicate the median values for each. The covering fraction due a random sightline is $f_{c,\rm rand} = 0.14$.
    
    \subsection{Estimating $R_{\rm CGM}^{14}$}
    In order to estimate a characteristic size of the \NHI $> 10^{14}$ cm$^{-2}$ CGM, $R_{\rm CGM}^{14}$, we devise a method in which we use the parameters to calculate the impact parameter at which the covering fraction ($f_c = P^{\rm hit}$) exceeds 0.5.  Within this impact parameter, a sightline has a greater than 50\% chance of exhibiting a \ion{H}{1} column with \NHI $>10^{14}$cm$^{-2}$. We can then estimate the posterior distribution of $R_{\rm CGM}$ by calculating it for each sample taken from the posterior distributions of $\gamma$ and $r_0$. Using the samples from the posterior distributions in $r_0$ and $\gamma$ we calculate $R_{\rm CGM,p}^{14} = 346_{-53}^{+57}$ kpc ($R_{\rm CGM,c}^{14} = 463_{-71}^{+76}$ comoving kpc), $R_{\rm CGM, p}^{14} = 353_{-50}^{+64}$ kpc ($R_{\rm CGM,c}^{14} = 469_{-66}^{+85}$ comoving kpc) and $R_{\rm CGM, p}^{14} = 177_{-65}^{+70}$ kpc ($R_{\rm CGM,c}^{14} = 222_{-81}^{+88}$ comoving kpc) in order of decreasing mass samples. The extent of the CGM remains relatively similar in size ($\sim 350$ physical kpc) except for the lowest mass sample. These correspond to $R_{\rm CGM}^{14} = 1.2^{+0.2}_{-0.2}$ $R_{\rm vir}$, $R_{\rm CGM}^{14} = 2.4_{-0.3}^{+0.4}$ $R_{\rm vir}$ and $R_{\rm CGM}^{14} = 1.6_{-0.6}^{+0.6}$ $R_{\rm vir}$ for the mass samples in order of decreasing mass, respectively, where $R_{\rm vir}$ was calculated using the mean redshift and mass of each sample. These estimates are in agreement with our qualitative empirical estimates from the previous analysis. These results are summarized in Table \ref{tab:results}.
    
    \begin{deluxetable}{C|C|C|C|C}
    \tablecaption{Results \label{tab:results}}
    \tablenum{5}
    \tablehead{
             \colhead{$M_{\star}$} & \colhead{$R_{\rm CGM,p}^{14}$} & \colhead{$R_{\rm CGM,c}^{14}$} & \colhead{$R_{\rm CGM}^{14}$} & \colhead{$\bar{z}$} \\
             \colhead{[$M_{\odot}$]} & \colhead{[kpc]} & \colhead{[kpc]} & \colhead{[$R_{\rm vir}$]} & \colhead{}
              }
    \colnumbers
    \startdata
    M_{\star} > 10^{9.9} & 346_{-53}^{+57} & 463_{-71}^{+76} & 1.3_{-0.2}^{+0.2} & 0.34 \\
    10^{9.2} < M_{\star} < 10^{9.9} & 353_{-50}^{+64} & 469_{-66}^{+85} & 2.4_{-0.4}^{+0.4} & 0.33 \\
    M_{\star} < 10^{9.2} & 177_{-65}^{+70} & 222_{-81}^{+88} & 1.6_{-0.6}^{+0.6} & 0.26 \\
    \enddata
    \tablecomments{Results of \ion{H}{1}-galaxy clustering analysis and $R_{\rm CGM}^{14}$: (1) Stellar Mass limits of each sample; (2) $R_{\rm CGM}^{14}$ in physical kpc; (3) $R_{\rm CGM}^{14}$ in physical kpc; (4) $R_{\rm CGM}^{14}$ normalized to the average virial radius of the sample; (5) Average redshift of each sample}
    \end{deluxetable}

\section{Comparison with Other Work}\label{discussion}
    \subsection{Previous Surveys}

    \begin{figure*}
    \begin{center}$
    \begin{array}{c c}
    \includegraphics[scale=0.35]{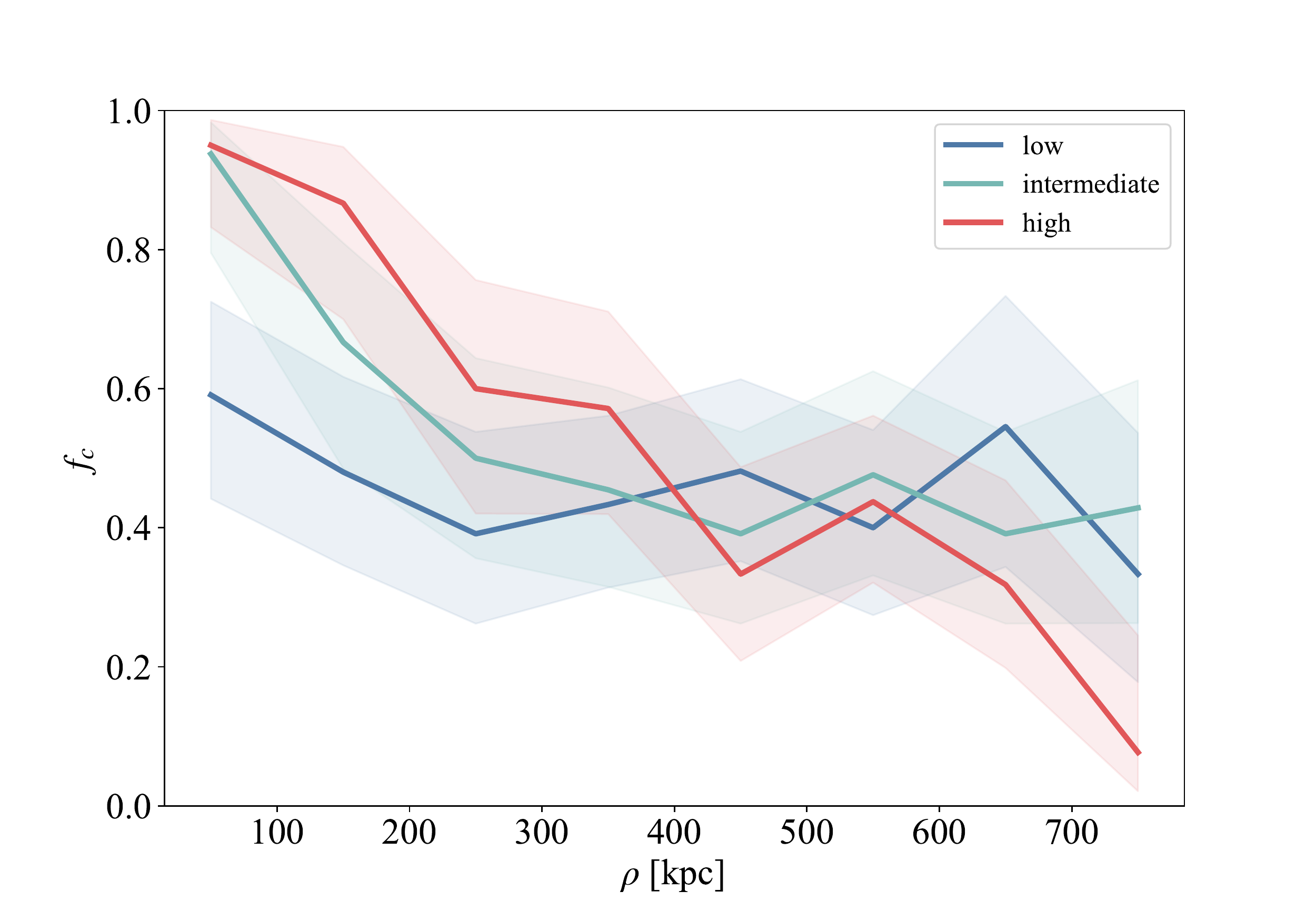}
     &
    \includegraphics[scale=0.35]{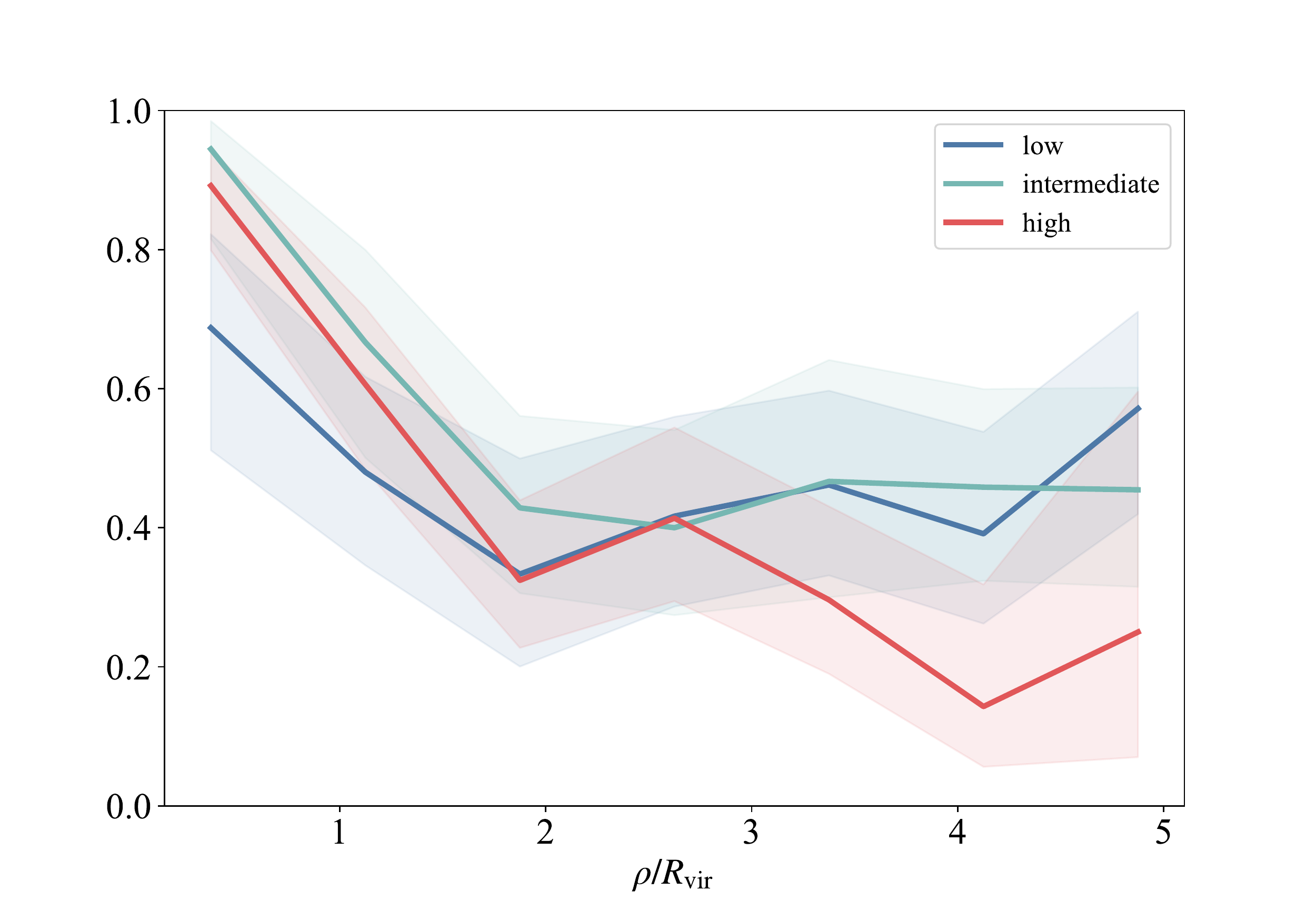}
    \end{array}$
    \end{center}
    %
    \caption{Covering fraction of \ion{H}{1} as a function of impact parameter $\rho$ (left) and $\rho/R_{\rm vir}$ (right) highlighting \NHI $> 10^{14}$ cm$^{-2}$ broken down into the three mass samples. The solid lines correspond to the covering fraction in bins of 100 physical kpc (left) and 1 $R_{\rm vir}$/ (right). The shaded region encodes the 1$\sigma$ error in the covering fraction measurement. The covering fraction decreases monotonically in either measure to $\sim300$~kpc or 2$R_{\rm vir}$. We find that the covering fraction in the lowest radial bins is consistent with $\sim100\%$ in each case except for the lowest mass sample. We note that the intermediate mass sample extends to further radius when looking at the normalized impact parameter (right) as compared to either of the other mass samples. \label{Fig:NHI_threshmass}}
    \end{figure*}

    \cgmsq contains the largest sample by a factor of 5 of CGM absorption systems within $\sim600$kpc of putative host galaxies. Our increased coverage is partly due to the fact that we probe a larger volume than previous surveys by confirming sub-L$*$ galaxies out to higher redshifts. For this reason, \cgmsq presents an excellent opportunity to study the HI-traced CGM at galaxy-absorber separations that exceed R$_{\rm vir}$ for a wide range of stellar masses. Here, we compare our results to those from other observational surveys of the CGM. 
    
    Figures \ref{fig:fig_mcmc} and \ref{Fig:NHI_threshmass} and Table \ref{tab:results} provide a summary of our covering fraction results for our mass segregated galaxy samples. The red, green, and blue-shaded curves in Figure \ref{Fig:NHI_threshmass} correspond to the samples cut at masses of $M_{\star, cut} = 10^{9.204}, 10^{9.888} M_{\odot}$ as described above. This figure allows us to better compare the covering fraction profiles of each mass sample to each other. As we saw in the last section, we find a mass dependence on the extent of the CGM; the covering fraction remains higher for larger galaxy mass when looking at the physical impact parameter. When we consider the impact parameter normalized to the average virial radii of each mass sample, we find the covering fraction remains elevated to larger radii ($\gtrsim 2R_{\rm vir}$) for the intermediate mass sample, while for the other mass samples the extent of the CGM is similar ($\gtrsim 1R_{\rm vir}$). We also find that the covering fraction of the lowest mass sample never exceeds 80\% while for both higher mass samples, the covering fraction is consistent with $\sim100\%$.

    Many previous studies have examined the covering fraction of \ion{H}{1} as a function of galaxy projected separation, although most of these surveys are limited to $\rho \lesssim 300$\,kpc. These surveys are all generally constructed in the same manner, with HST UV spectroscopy with COS and/or STIS and a spectroscopic galaxy catalog to connect galaxies and absorbers. 
   
    One of the largest such surveys was carried out by \cite{tejos14} who attempted to explore the connection between the IGM and galaxies by measuring the HI-galaxy cross-correlation at $z<1$ at distances between $\sim$1 and 10 Mpc. They used multiple ground based instruments to build a new spectroscopic survey of 2143 galaxies in 8 QSO fields with HST spectroscopy. They combined their catalog with existing catalogs to build a survey of $\sim$17500 galaxies. They did not limit their galaxy survey to only the nearest galaxies but used a statistical approach to examine the cross-correlation of galaxies and absorption systems. They found that the \ion{H}{1} \lya forest to be divided into two main categories: a population of low column density absorbers tracing the cosmic web and a higher-column density population that traces the dark matter halos in which galaxies reside. \cgmsq has the benefit of being more sensitive to fainter galaxies closer to the QSO itself allowing us to better constrain the extent of the CGM. However, their larger galaxy sample at larger separations (which explores different scales) presents a great opportunity to compare our cross-correlation analysis. Using the same 3D cross-correlation power law, Equation \ref{xi}, they found $r_0 = 3.8 \pm 0.2$ $h^{-1}_{70}$ Mpc and $\gamma =1.7 \pm 0.1$ in their sample of SF-galaxies. This sample is similar to our high-mass sample since we are dominated by SF galaxies and indeed, we find agreeme  nt consistent within 1$\sigma$.
    
    A mass dependent CGM was examined in \cite{bordoloi18} who used the 85 galaxies in the COS-Halos and COS-Dwarfs surveys at $z\sim0$ with $M_{\star}$ ranging from 8 to 11.6 $\log M_{\star}/M_{\odot}$. This sample was limited to impact parameters ($\rho < 160$ kpc). They found a mass and radius dependence of the strength of \ion{H}{1} absorption where the equivalent width of \ion{H}{1} increases with $M_{\star}$ and with decreasing impact parameter. Here we are primarily focused on the mass and radius dependence of the covering fraction, but find trends generally consistent with those observed by \cite{bordoloi18}.
    
    \cite{chen01} found \lya with column densities $N_{\rm HI} \gtrsim 10^{14}$ cm$^{-2}$ in 34/47 galaxies ($f_c \approx 0.7$) out to $\rho \simeq 330$~kpc. Their sample consists of $\sim L*$ HI-Galaxy pairs with $|\delta v| < 500$ km s$^{-1}$ spanning $0.1 <z<0.9$ ($\bar{z} = 0.36$). They also found a sharp decline thereafter. We find close agreement $f_c = 0.78^{+0.06}_{-0.89}$ (162/209) when applying the same criterion to our sample also seeing a sharp decline around 400 kpc. (Figure \ref{Fig:NHI_threshmass}).
    
    \cite{prochaska11b} used 14 QSO sightlines with previously published equivalent width ($W_0$) measurements of \lya to carry out a galaxy survey that targeted 37 $L > 0.01 L_{\star}$ galaxies at $\bar{z} = 0.18$. They connected absorbers and galaxies with $|\delta v| < 400$ km s$^{-1}$, although they show it makes no qualitative difference in their results using a larger velocity window of $|\delta v| < 600$ km s$^{-1}$. They found covering fractions of order unity ($\approx 90\%$) for \NHI $> 10^{13}$ cm$^{-2}$ gas out to $\rho = 300$~kpc. Comparing our covering fractions with \NHI $> 10^{13}$ cm$^{-2}$ and for galaxies with $M_{\star} > 10^{8.55} M_{\odot}$ to approximate their $L > 0.01 L_{\star}$ sample, we find $f_c = 0.85^{+0.04}_{-0.05}$ (128/150),  which is roughly consistent with their value.
  
    \cite{wakker09} conducted a large survey of the HI-galaxy connection at $z \lesssim 0.017$, consisting of 76 QSO sightlines and $\sim 20\,000$ local galaxies. They found covering fractions of 77\%  for Ly$\alpha$ absorbers $>$50~ m\AA  within $\rho < 400$~kpc and $|\delta v| < 400$ km s$^{-1}$ of $L>0.1 L_{\star}$ galaxies. If we limit our sample to $L>0.1 L_{\star}$ galaxies and use a Ly$\alpha$ absorber threshold of $\sim$50 m\AA~  (\NHI $\sim 10^{13}$ cm$^{-2}$), we find covering fractions, $f_c = 0.89^{+0.04}_{-0.05}$ (99/111) for galaxies within 400 kpc. This discrepancy could imply that covering fractions increase with redshift.
    
    Most recently, a large survey of \ion{H}{1} was carried out by \cite{keeney18} (K18).  Their survey consisted of 47 COS sightlines (COS GTO) with higher signal-to-noise (S/N $\sim$ 15-50 compared to our $\sim 10$) QSO spectra. Using ground based telescopes, they constructed a spectroscopic galaxy database of $\sim 9,000$ galaxies with the aim of $>90\%$ completeness to 1~Mpc down to $0.1L*$ at $z \lesssim 0.1$. Due to their higher S/N, they could consistently measure weaker absorption lines, down to $N_{\rm HI} \geq 10^{12.8}$ cm$^{-2}$.  Leveraging the high completeness of K18 at low redshifts, they were able to measure the \ion{H}{1} column densities out to $4 R_{\rm vir}$ with enough galaxies in this range (243) to make precise statements about the radial profile of HI. They find the covering fraction for $L < L*$ galaxies (corresponding to our low-mass sample) to feature a shallower decline than that of their $L > L*$ sample. Qualitatively, we show similar results. This difference in covering fraction behavior between high and low mass samples can readily be seen in Figures \ref{fig:covfrac_threshold} and \ref{Fig:NHI_threshmass}. The elevated covering fractions at large radii in the $L<0.1L*$ sample imply that contributions to \NHI $>10^{14}$ cm$^{-2}$ CGM gas are dominated by low-mass galaxies. This result is consistent with \cite{prochaska11b} in relation to \ion{H}{1} and with high-ionization metals such as \ion{O}{6}  \citep[]{tumlinson05, pratt18, prochaska19}.

    We note that the high mass sample drops to a very low covering fraction $f_c \simeq 0.20$ at 4$R_{\rm vir}$ while in the lower mass sample, $f_c$ remains elevated. At radii greater than 3$R_{\rm vir}$, we may be limited to only higher-$z$ high-mass galaxies due to the detector size but this detector size bias should not affect the lower mass samples.
   
    In addition, previous low-redshift studies have found that low column density gas (\NHI $= 10^{13 - 14}$ cm$^{-2}$) is likely uncorrelated with galaxy halos \citep{chen05, prochaska11b, danforth16} (but, see \cite{tejos14} who find that 50\% of weak lines can still be correlated with galaxies on 1-10 Mpc and \cite{tripp98} who show that the weakest absorbers are not randomly distributed). First, low column density material exhibits high covering fractions out to 1~Mpc which we can see in Figure \ref{fig:covfrac_threshold}. Second, low column density material exhibits less velocity correspondence with the systemic velocities of galaxies nearby in projection \citep{tumlinson13}, as we find a median velocity difference of 233 km s$^{-1}$ with a standard deviation 453 km s$^{-1}$ for $\rho > 1 R_{\rm vir}$ while we find a median velocity of 185 km s$^{-1}$ with a standard deviation of 354 km s$^{-1}$ for $\rho < 1 R_{\rm vir}$. Traditionally, such low column density material is attributed to the \lya forest, or to gas in a filament like structure \citep{tejos14}, physically distinct from the CGM. This was examined in greater depth by \cite{burchett20} who conclusively tie the diffuse IGM to the cosmic web. They find the \ion{H}{1} absorption signature decreases past $\rho > R_{\rm vir}$ and settles to the cosmic mean matter density. 
    
    Turning to higher redshifts, \cite{rudie12} use the ground based \textit{Keck Baryonic Structure Survey; KBSS} to investigate the $z \sim 2-3$ CGM surrounding 886 galaxies. Their sample contains 48 galaxies at $\bar{z} = 2.3$ within $\rho \lesssim 300$~kpc for which they measure a covering fraction of $f_c = 0.81 \pm 0.06$ for absorbers with \NHI $> 10^{14}$ cm$^{-2}$. By comparison, if we choose a mass range of $10^{10.4} < M_{\star}/M_{\odot} < 10^{11}$ to approximate their mass distribution \citep{erb06c} we find $f_c = 0.86^{+0.08}_{-0.14}$ (18/22) which is good agreement with their results. However, looking at the extended CGM $\rho < 1$Mpc, we find a discrepancy: we measure $f_c = 0.49^{+0.08}_{-0.08}$ (40/83) vs their $0.70 \pm 0.03$. These numbers are in agreement at the 2$\sigma$ level, however. Interestingly, the extended CGM may show a decrease in covering fraction as the universe evolves. This phenomenon may be due to the development of virial shocks that ionize the gas, as suggested by \cite{burchett18}. At $z=2.3$, 300kpc $\sim 2 R_{\rm vir}$ for a $7 \times 10^{10} M_{\odot}$ galaxy while at $z = 0.3$ (\cgmsq) 300~kpc $\sim 0.8 R_{\rm vir}$. Alternatively, this could simply be due to the fact that a column density of $N_{\rm HI} > 10^{14}$ cm$^{-2}$ traces lower density peaks at high-z.

    \subsection{Comparison with Hydrodynamical Simulations}
    
    We now turn to a brief comparison with hydrodynamical simulations. \cite{vandeVoort19} simulate a roughly $z \sim 0$ $L*$  galaxy using a new refinement technique to better resolve the CGM. They find an increase in the \ion{H}{1} column density and resultant covering fraction when the resolution is increased to resolve 1~kpc scales. We find their model to be in good agreement with our $N_{\rm HI} \geq 10^{14}$ cm$^{-2}$ covering fraction measurements when we limit our sample to $M_{\star} \sim 10^{10.5} M_{\odot}$ (see their Figure 3).  Comparing our column density measurements to \cite{vandeVoort19} and \cite{hummels19} we also see good agreement out to their limiting distance of 200~kpc and 100~kpc,  respectively (compare Figure \ref{Fig:NHI_v_rho} to \cite{vandeVoort19} Figure 2). Our high \ion{H}{1} covering fractions and column densities at $R<$ $R_{\rm vir}$ are in conflict with earlier simulations that consistently underpredict the column density of low-ions in the CGM \citep[e.g.,][]{hummels13, liang16, stinson12, shen12, ford13}, validating the work that has gone into creating these new high resolution techniques. To understand the extent of the CGM around a diverse sample of galaxies, we encourage future simulations extending to at least 4$R_{\rm vir}$, and covering a larger range of galaxy masses, down to 0.01$L*$.

\section{Summary and Conclusions} \label{summary}
We have reported the first results from the \cgmsq survey, a comprehensive survey of the $z<$1 CGM at least 5$\times$ larger than previous surveys such as COS-Halos at comparable redshifts. This paper has presented the detailed properties of the survey design and the procedures followed in the collection and processing of the data. We present an \ion{H}{1} study that combines high-resolution HST/COS UV spectra of 22 background QSOs with Gemini/GMOS spectra of 572 foreground galaxies having stellar masses 10$^{6}$ M$_{\odot}$ $<$ M$_{\star}$ $<$ 10$^{11.5}$ M$_{\odot}$ and z $<$ 0.481.  The S/N$\sim$10-12 of these COS spectra and access to the \ion{H}{1} Lyman limit enables us to constrain the \ion{H}{1} column densities and kinematics of associated CGM absorption, and to ultimately examine the extent of the CGM as a function of galaxy mass and physical separation from the QSO sightline. 

We find that high-column density circumgalactic material is associated with galaxies at high statistical significance out to 2$R_{\rm vir}$, whereas \ion{H}{1} absorption with \NHI $<$ $10^{14}$ cm$^{-2}$ is more broadly distributed in both impact parameter and velocity space and may not be associated directly with massive galaxies. Our kinematic analysis reveals that the detected strong \ion{H}{1} is most likely gravitationally associated with the most nearby galaxy, while weaker \ion{H}{1} components seen at absolute velocity offsets $\gtrsim$ $500$ km s$^{-1}$ may be instead associated with extended large scale structures. We find generally good agreement between our sample and the prior studies that have examined the CGM of low-redshift galaxies out to similar and larger separations. 

We define the cool CGM as the region surrounding a galaxy in which the probability of observing an absorber with \NHI $>$ 10$^{14.0}$ cm$^{-2}$ is $>$ 50\%. Our column density threshold, \NHI $>$ 10$^{14.0}$ cm$^{-2}$, is motivated by previous observational work that examines the statistical, large-scale (Mpc) correlations between galaxies and QSO absorption lines.  In essence, our definition of $R_{\rm CGM}^{14}$ demands that around a given galaxy, one is more likely than not to find material that has been empirically associated with galaxies. The picture that emerges is of a diffuse, CGM extending to $\sim350$ kpc around galaxies with stellar mass $\gtrsim 10^{9} M_{\odot}$ at $z \lesssim$ 0.5. For galaxies of lower stellar mass, the extent of the CGM is $<$ 200 kpc.  At all stellar masses,   the extent of this CGM exceeds a virial radius, especially for galaxies with intermediate masses where the $R_{\rm CGM}^{14}$ exceeds $2R_{\rm vir}$. Therefore, these results imply that using $R_{\rm vir}$ as a proxy for the characteristic edge of the CGM may significantly underestimate its true extent. The detailed nature of the mass dependence of the CGM will be examined in a forthcoming paper. Additional follow-up studies using \cgmsq data will consider transitions from a wide range of ionized metals and absorption-line profile analyses to characterize the ionization state, metallicity, kinematics, and mass of the CGM at low redshift.

\section{Acknowledgements}
MCW, JKW, and KT acknowledge the people of the Dkhw'Duw'Absh, the Duwamish Tribe, and other tribes on whose traditional lands we study and work at the University of Washington. MCW, JKW, and KT acknowledge support for this work from NSF-AST-1812521, and the contributions from the following University of Washington current and former undergraduate Astronomy majors who participated in ``Werk SQuAD'' from 2017-2020: Steven Bet, Douglas Branton, Joseph Breneman, Dustin Burnham, Olivia Petry Caplow-Munro, Justin Collins, Apurva Goel, Samuel Johnson, Courtney Klein, Dalton LaCoste, Christina Lindberg, Camellia Magness, Travis Mandeville, Karalyn Ostler, Magdalyn Paige, Locke Patton,  Brittany Platt, David Staker, Matthew Strasbourg, Sophia Taylor, Mercedes Thompson, and Leonardo Zhu. Additionally, the authors thank the anonymous referee for a productive report that improved the analysis. 

MCW would like to thank Daniel J. D'Orazio and David Fleming for useful discussion.

This work was based on observations obtained at the Gemini Observatory, which is operated by the Association of Universities for Research in Astronomy, Inc., under a cooperative agreement with the NSF on behalf of the Gemini partnership: the National Science Foundation (United States), National Research Council (Canada), CONICYT (Chile), Ministerio de Ciencia, Tecnolog\'{i}a e Innovaci\'{o}n Productiva (Argentina), Minist\'{e}rio da Ci\^{e}ncia, Tecnologia e Inova\c{c}\~{a}o (Brazil), and Korea Astronomy and Space Science Institute (Republic of Korea). The Gemini North telescope is located within the Maunakea Science Reserve and adjacent to the summit of Maunakea. We are grateful for the privilege of observing the Universe from a place that is unique in both its astronomical quality and its cultural significance.

This research has made use of the SVO Filter Profile Service (http://svo2.cab.inta-csic.es/theory/fps/) supported from the Spanish MINECO through grant AYA2017-84089.

This work is partly based on tools and data products produced by GAZPAR operated by CeSAM-LAM and IAP.

The Pan-STARRS1 Surveys (PS1) and the PS1 public science archive have been made possible through contributions by the Institute for Astronomy, the University of Hawaii, the Pan-STARRS Project Office, the Max-Planck Society and its participating institutes, the Max Planck Institute for Astronomy, Heidelberg and the Max Planck Institute for Extraterrestrial Physics, Garching, The Johns Hopkins University, Durham University, the University of Edinburgh, the Queen's University Belfast, the Harvard-Smithsonian Center for Astrophysics, the Las Cumbres Observatory Global Telescope Network Incorporated, the National Central University of Taiwan, the Space Telescope Science Institute, the National Aeronautics and Space Administration under Grant No. NNX08AR22G issued through the Planetary Science Division of the NASA Science Mission Directorate, the National Science Foundation Grant No. AST-1238877, the University of Maryland, Eotvos Lorand University (ELTE), the Los Alamos National Laboratory, and the Gordon and Betty Moore Foundation.

The Legacy Surveys consist of three individual and complementary projects: the Dark Energy Camera Legacy Survey (DECaLS; NOAO Proposal ID  2014B-0404; PIs: David Schlegel and Arjun Dey), the Beijing-Arizona Sky Survey (BASS; NOAO Proposal ID  2015A-0801; PIs: Zhou Xu and Xiaohui Fan), and the Mayall z-band Legacy Survey (MzLS; NOAO Proposal ID 2016A-0453; PI: Arjun Dey). DECaLS, BASS and MzLS together include data obtained, respectively, at the Blanco telescope, Cerro Tololo Inter-American Observatory, National Optical Astronomy Observatory (NOAO); the Bok telescope, Steward Observatory, University of Arizona; and the Mayall telescope, Kitt Peak National Observatory, NOAO. The Legacy Surveys project is honored to be permitted to conduct astronomical research on Iolkam Du'ag (Kitt Peak), a mountain with particular significance to the Tohono O'odham Nation.

NOAO is operated by the Association of Universities for Research in Astronomy (AURA) under a cooperative agreement with the National Science Foundation.

This project used data obtained with the Dark Energy Camera (DECam), which was constructed by the Dark Energy Survey (DES) collaboration. Funding for the DES Projects has been provided by the U.S. Department of Energy, the U.S. National Science Foundation, the Ministry of Science and Education of Spain, the Science and Technology Facilities Council of the United Kingdom, the Higher Education Funding Council for England, the National Center for Supercomputing Applications at the University of Illinois at Urbana-Champaign, the Kavli Institute of Cosmological Physics at the University of Chicago, Center for Cosmology and Astro-Particle Physics at the Ohio State University, the Mitchell Institute for Fundamental Physics and Astronomy at Texas A\&M University, Financiadora de Estudos e Projetos, Fundacao Carlos Chagas Filho de Amparo, Financiadora de Estudos e Projetos, Fundacao Carlos Chagas Filho de Amparo a Pesquisa do Estado do Rio de Janeiro, Conselho Nacional de Desenvolvimento Cientifico e Tecnologico and the Ministerio da Ciencia, Tecnologia e Inovacao, the Deutsche Forschungsgemeinschaft and the Collaborating Institutions in the Dark Energy Survey. The Collaborating Institutions are Argonne National Laboratory, the University of California at Santa Cruz, the University of Cambridge, Centro de Investigaciones Energeticas, Medioambientales y Tecnologicas-Madrid, the University of Chicago, University College London, the DES-Brazil Consortium, the University of Edinburgh, the Eidgenossische Technische Hochschule (ETH) Zurich, Fermi National Accelerator Laboratory, the University of Illinois at Urbana-Champaign, the Institut de Ciencies de l'Espai (IEEC/CSIC), the Institut de Fisica d'Altes Energies, Lawrence Berkeley National Laboratory, the Ludwig-Maximilians Universitat Munchen and the associated Excellence Cluster Universe, the University of Michigan, the National Optical Astronomy Observatory, the University of Nottingham, the Ohio State University, the University of Pennsylvania, the University of Portsmouth, SLAC National Accelerator Laboratory, Stanford University, the University of Sussex, and Texas A\&M University.

BASS is a key project of the Telescope Access Program (TAP), which has been funded by the National Astronomical Observatories of China, the Chinese Academy of Sciences (the Strategic Priority Research Program ``The Emergence of Cosmological Structures" Grant  XDB09000000), and the Special Fund for Astronomy from the Ministry of Finance. The BASS is also supported by the External Cooperation Program of Chinese Academy of Sciences (Grant  114A11KYSB20160057), and Chinese National Natural Science Foundation (Grant 11433005).

The Legacy Survey team makes use of data products from the Near-Earth Object Wide-field Infrared Survey Explorer (NEOWISE), which is a project of the Jet Propulsion Laboratory/California Institute of Technology. NEOWISE is funded by the National Aeronautics and Space Administration.

The Legacy Surveys imaging of the DESI footprint is supported by the Director, Office of Science, Office of High Energy Physics of the U.S. Department of Energy under Contract No. DE-AC02-05CH1123, by the National Energy Research Scientific Computing Center, a DOE Office of Science User Facility under the same contract; and by the U.S. National Science Foundation, Division of Astronomical Sciences under Contract No. AST-0950945 to NOAO.

\software{SEXTRACTOR \citep{sextractor96}, 
PYIGM \citep{pyigm},
PypeIt \citep[][]{pypeit20},
ASTROPY \citep[][]{astropy13, astropy18}, 
PyRAF \citep{pyraf}, 
linetools \citep{linetools}, 
REDROCK (https://github.com/desihub/redrock), \\
CIGALE \citep[][]{cigale11,boquien19},
seaborn \citep{seaborn},
pandas \citep[][]{pandas_software, pandas_paper},
corner \citep{corner},
emcee \citep{emcee}}

\bibliography{references.bib}

\end{document}